\documentclass[twocolumn,groupedaddress,amsmath,aps,longbibliography,nofootinbib]{revtex4-1}
\usepackage{hyperref}
\usepackage{graphicx}
\usepackage{amsmath,amsfonts}
\usepackage{dcolumn}
\usepackage{bm}
\usepackage{color}
\usepackage{multirow}
\usepackage{float}
\usepackage{url}
\usepackage{physics}
\usepackage{subfigure}
\usepackage[utf8]{inputenc}

\begin{document}

\title{Local distortions of the crystal structure and their influence on the electronic structure and superconductivity of the high-entropy alloy 
(TaNb)$_{0.67}$(HfZrTi)$_{0.33}$}

\author{Kinga Jasiewicz}
\affiliation{AGH University of Krakow, Faculty of Physics and Applied Computer Science, Aleja Mickiewicza 30, 30-059 Krakow, Poland}
\author{Janusz Tobola}
\affiliation{AGH University of Krakow, Faculty of Physics and Applied Computer Science, Aleja Mickiewicza 30, 30-059 Krakow, Poland}
\author{Bartlomiej Wiendlocha}
\email{wiendlocha@fis.agh.edu.pl}
\affiliation{AGH University of Krakow, Faculty of Physics and Applied Computer Science, Aleja Mickiewicza 30, 30-059 Krakow, Poland}

\date{\today}

\begin{abstract}
Local distortions of the crystal structure and their influence on the electronic structure, electron-phonon interaction, and superconductivity are theoretically studied in the superconducting high-entropy alloy (HEA) (TaNb)$_{0.67}$(HfZrTi)$_{0.33}$. Distortions of the crystal lattice are caused by the relaxation of atomic positions and are studied in the twelve models of supercells. The largest relative changes in the interatomic distances due to relaxation reach 8\%. On average, local distortions tend to lower the density of states at the Fermi level and significantly reduce the electron-phonon coupling parameter $\lambda$. As a result, the calculated superconducting critical temperature is reduced to about 50\% of the initial value, which shows the strong impact of structural disorder on superconductivity in this prototype high-entropy alloy.
With the reduced value of $\lambda$, the theoretical $T_c$ is closer to the experiment for typical values of the Columomb pseudopotential parameter $\mu^*$. The experimental $T_c$ can be reproduced by taking a slightly enhanced $\mu^* = 0.176$, which leaves little room for the possibility of suppression of superconductivity by disorder.
\end{abstract}

\maketitle

\section{Introduction}	

The history of high-entropy alloys goes back to 1979, when Brian Cantor and Alan Vincent successfully synthesized the multicomponent alloy Fe$_{0.20}$Ni$_{0.20}$Cr$_{0.20}$Co$_{0.20}$Mn$_{0.20}$ with a simple {\it fcc} structure \cite{murty2014,hea_book_2016, cantor_2004}. However, the term ''high-entropy alloys'' was first used by Yeh {\it et al.} \cite{yeh_2004}, along with the definition, according to which HEA are alloys that contain at least five elements with concentrations between 5 and 35\%. 
Their simple crystal structure (usually the basic "monoatomic" \textit{bcc} or \textit{fcc}) despite high level of chemical disorder was surprising, as it was widely accepted that the more elements involved in the synthesis of the materials, the lower the chance that the alloy could form a simple crystal structure, and the higher the probability of a metallic glass formation \cite{greer_1993}.
The literature review allows us to identify two families of HEA that have been the most investigated. The first group of alloys is based on CrCoFeNi, while the second group contains alloys built from Ta, Nb, Hf, Zr, and Ti atoms. The representative of the second family is Ta$_{0.34}$Nb$_{0.33}$Hf$_{0.08}$Zr$_{0.14}$Ti$_{0.11}$, the first superconducting high-entropy alloy \cite{kozejl_2014}, synthesized in 2014. Since then more superconducting HEA alloys have been reported, e.g. $bcc$ NbReHfZrTi \cite{hea_sup_1} 
layered REO$_{0.5}$F$_{0.5}$BiS$_2$ (RE - rare earth) \cite{hea_sup_2}, 
(ScZrNb)$_{1-x}$(RhPd)$_{x}$ with CsCl-type lattice \cite{hea_sup_3}, $\alpha$Mn-type (ZrNb)$_{1-x}$(MoReRu)$_x$, (HfTaWIr)$_{1-x}$Re$_x$ and (HfTaWPt)$_{1-x}$Re$_x$ \cite{hea_sup_4}, Re$_{0.35}$Os$_{0.35}$Mo$_{0.10}$W$_{0.10}$Zr$_{0.10}$~\cite{Motla_2023}, Nb$_3$(AlSnGeGaSi) \cite{hea_sup_5}, tetragonal Ta$_{10}$Mo$_{35-x}$Cr$_x$Re$_{35}$Ru$_{20}$ \cite{hea_sup_6}, CuAl$_2$ type (FeCoNiCuGa)Zr$_2$, Co$_{0.2}$Ni$_{0.1}$Cu$_{0.1}$Rh$_{0.3}$Ir$_{0.3}$Zr$_2$, \\Mo$_{0.11}$W$_{0.11}$V$_{0.11}$Re$_{0.34}$B$_{0.33}$ \cite{hea_sup_7,hea_sup_8,hea_sup_10}
or hexagonal Nb$_{10+2x}$Mo$_{35-x}$Ru$_{35-x}$Rh$_{10}$Pd$_{10}$ \cite{hea_sup_9}, Ru$_{0.35}$Os$_{0.35}$Mo$_{0.10}$W$_{0.10}$Zr$_{0.10}$ \cite{Motla_2023}.

However, among superconducting HEAs, the Ta-Nb-Hf-Zr-Ti family is attracting the most attention \cite{kozejl_2014, rohr_tnhzt_2016, guo-pressure, vrtnik_2017, gabani_2023_tnhzt, krnel_2022_tnhzt, zeng_2023_tnhzt, zhang_2020_tnhzt, hattori_2023_tnhzt, zeng_2023_tnhzt2, yuan_2018_tnhzt, kim_2022_tnhzt, hung_2020_tnhzt, kim_2020_tnhzt, prista_2023_tnhzt, zhang_2020_tnhzt2, hong_2022_tnhzt, leung_2022_tnhzt, heidelmann_2016_tnhzt, rohr_2018_tnhzt}.
Previous works show that (TaNb)$_{0.67}$(HfZrTi)$_{0.33}$ alloy (TNHZT), on which we focus in this work, is a conventional superconductor with \textit{bcc} crystal structure and lattice constant $a = 3.36$~\AA. 
The electron-phonon coupling is rather strong, with the electron-phonon coupling constant $\lambda\approx 1$. Two independent experiments \cite{rohr_tnhzt_2016,jasiewicz_2019} reported similar values for the electronic specific heat coefficient $\gamma$ (7.97 and 7.7 mJ/(mol~K$^2$)), Debye temperature (225 and 216 $K$) and the ratio $\Delta C (T_C)/\gamma T_C$ (1.89 and 1.93). 
Superconductivity in (TaNb)$_{0.67}$(HfZrTi)$_{0.33}$ appeared to be robust against extremely high pressures \cite{guo-pressure,cava2020}, as the critical temperature increases in the pressure range between 0 and 60 GPa, then becomes constant up to 100 GPa. 
Theoretical studies \cite{jasiewicz_2019} performed using the Korringa-Kohn-Rostoker method with coherent potential approximation (KKR-CPA)  \cite{bansil_1999,stopa_2004} showed that this pressure evolution is correctly described by the conventional electron-phonon mechanism of superconductivity. 
Furthermore, these calculations revealed the Lifshitz transition \cite{lifshitz_1960}, previously reported for Nb \cite{struzhkin_1997}, one of the main constituent atoms of (TaNb)$_{0.67}$(HfZrTi)$_{0.33}$.  
Another interesting and rather unexpected feature of TNHZT superconductors, revealed by calculations~\cite{jasiewicz_2016,jasiewicz_2019}, is their sharp bandstructure.
In a perfect-crystalline material, because of the Bloch theorem, electrons are not scattered by the lattice. Electrons keep their wavewectors $\mathbf{k}$, electronic states have an infinite lifetime $\tau$ and energy states have a definite eigenvalue $E(\mathbf{k})$. This defines the "sharp" electronic bandstructure with no band smearing.
In a substitutional alloy, on the other hand, because of the presence of the chemical disorder, electrons are scattered. This is, of course, a source of the residual resistivity of the alloys. 
When the scattering becomes strong, the electronic lifetime $\tau$ is short, and the resistivity increases. 
This influences the electronic bandstructure. The finite lifetime of electronic states leads to the smearing of electronic bands, which gain a bandwidth $\Gamma_{\bf k} = {\hbar}/{\tau_{\bf k}}$~\cite{butler_1985,wiendlocha_2021}.
Although TNHZT has a high level of substitutional disorder, with all five elements occupying a single site in the primitive cell, surprisingly calculations revealed that its electronic band structure is only slightly smeared ~\cite{jasiewicz_2016,jasiewicz_2019}, in contrast, for example, to the (ScZrNb)$_{1-x}$RhPd$_x$ alloy~\cite{gutowska2023}. This is probably a consequence of TNHZT being an alloy of two isoelectronic series of elements (groups 4 and 5 in the periodic table), which have similar Pauling electronegativities~\cite{gutowska2023}.
This shows that the substitutional (chemical) disorder has a relatively weak impact on the electronic properties of TNHZT.

In this work, we investigate the second type of disorder that is present in high-entropy alloys: the structural disorder. 
On average, HEAs form ordered simple crystal structures with random occupation of the crystal sites. However, locally, because of the different sizes and types of atoms, and the different arrangements of the nearest neighbors, atoms will move from ideal positions to minimize the total energy and relax the forces. Thus, local distortions of the ideal crystal structure will be formed.
These two types of disorder, substitutional and structural, are of course connected since the structural disorder is a direct consequence of the chemical disorder. Only in calculations are we able to decouple them and investigate their effects separately.
The formation of such local distortions and their influence on electronic structure, electron-phonon interaction, and superconductivity is studied here for the (TaNb)$_{0.67}$(HfZrTi)$_{0.33}$ alloy.

\section{Calculation details}

A set of twelve $3\times 3 \times 1$ supercells was constructed, based on the elementary cubic unit cell. 
Each supercell contains 18 atoms with a different nearest-neighbor arrangement and has no symmetry operations (space group P1). 
The lattice parameter $a$ was kept constant, equal to the experimental value.
The choice of supercell size allows to keep the original concentration ratio of the atoms of the (TaNb)$_{0.67}$(HfZrTi)$_{0.33}$ alloy. 
One of the supercells is shown in Fig.~\ref{fig_sc331_1}, details of all supercells are given in Supplemental Material~\cite{suppl}.

\begin{figure}[t!]
	\centering
	\includegraphics[width=\columnwidth]{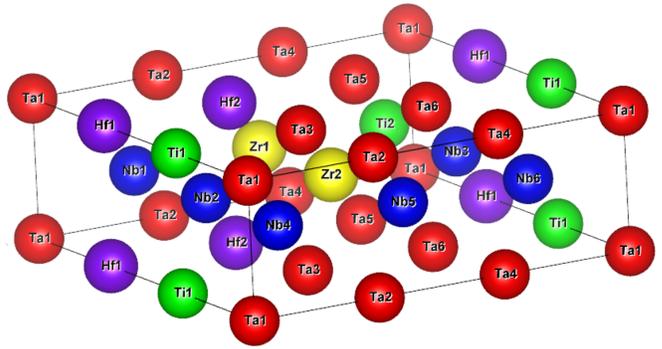}
	\caption{One of the studied $3\times3\times1$ supercells of (TaNb)$_{0.67}$(HfZrTi)$_{0.33}$ (model 1).}
	\label{fig_sc331_1}
\end{figure}

As a starting point, we have constructed a supercell with site occupations selected by a random number generator. This supercell is labeled as model 1. Models 2-8 were generated by changing the positions of isoelectronic atoms.
Model 9 was designed to have a larger cluster of Ta and Nb atoms.
Model 10, similarly to model 1, was constructed by another random assignment of the sites. Model 11 is based on model 10 with exchanged Ta, Ti, and Hf. Model 12 is equivalent to model 11 but with the exchanged positions of Hf and Zr.

The atomic positions in the investigated supercells were relaxed using the full potential LAPW method implemented in the {\sc wien2k} package \cite{wien2k2020,wien2k}, until the forces acting on all atoms were less than 2 mRy/a$_B$. 
From then on, to be able to compare the results obtained for supercells with those for the fully random ideal structure, all calculations were performed using the same Korringa-Kohn-Rostoker (KKR) method combined with the coherent potential approximation (CPA) \cite{bansil_1999,stopa_2004}, successfully used before to determine the electronic structure of the random TNHZT alloys~\cite{jasiewicz_2016,jasiewicz_2019}.
The CPA was actually used only in calculations for the fully random and undistorted alloy, where five atoms occupy a single site with appropriate concentrations, whereas in the supercells, each crystal site was occupied only by a single atom. 
The crystal potential in the spherical muffin-tin form was constructed using the local density approximation (LDA), with Perdew-Wang parameterization \cite{perdew_1992} in the semi-relativistic approach. 
The angular momentum cutoff was set to $l_{max}=3$. The Fermi level was precisely determined from the generalized Lloyd formula \cite{kaprzyk_1990}.  
The radii of the muffin-tin spheres, surrounding each atom, were set equal to $R_{\mathsf{MT}} = 2.5$ a$_B$ for all atoms. 
This value does not provide a maximum packing coefficient for the unit cell, preferential for spherical potential calculations; however, a reduction of $R_{\mathsf{MT}}$ was necessary to allow the atoms to move from their positions and to avoid overlapping of the spheres in relaxed supercells. 
The results obtained for the system without atomic distortions will be labeled as ''D(-)'', and for supercells after relaxation as ''D(+)''. 
The supercell results will be compared with the results of the KKR-CPA calculations for the fully disordered primitive cell, indicated as ''CPA''.
To obtain the appropriate reference values of the computed electronic properties for the fully disordered system, we have recalculated the electronic structure, the densities of states, the McMillan-Hopfield parameters and the electron-phonon coupling constant $\lambda$ using the same muffin tin radius of 2.5 a$_B$, since in our previous work~\cite{jasiewicz_2019} a larger $R_{\mathsf{MT}} = 2.75$ a$_B$ was used.
As a result of this different computational geometry, the McMillan-Hopfield parameters slightly increased, resulting in a 9\% larger $\lambda_{\rm CPA} = 1.21$ (it was 1.10 in~\cite{jasiewicz_2019}).

The electron-phonon interaction was studied using the Rigid Muffin Tin Approximation (RMTA), in which the electron-phonon coupling constant (EPC) is decoupled into electronic and lattice contributions:

\begin{equation}\label{eq_1}
\lambda = \sum_i \frac{\eta_i}{M_i\langle{\omega_i^2}\rangle}
\end{equation}
where $\eta_i$ is the $i$-th atom's McMillan-Hopfield parameter, calculated using the formula \cite{gaspari_1972,gomersall_1974,mazin_1990, kaprzyk_1996} :

\begin{equation}\label{eq:eta}
\eta_i =\!\sum_l \frac{(2l + 2)\,n_l(E_F)\,
	n_{l+1}(E_F)}{2(2l+1)(2l+3)N(E_F)} \left|\int_0^{R_{\mathsf{MT}}}\!\!r^2
R_l\frac{dV}{dr}R_{l+1} \right|^2\!.
\end{equation}
In the above equation $V(r)$ stands for the self-consistent potential at site $i$, $R_l(r)$ is a normalized regular solution of the radial Schr\"odinger equation, $n_l(E_F)$ is the $l$--th partial DOS at the Fermi level, $E_F$, and $N(E_F)$ is the total DOS per primitive cell. More comprehensive overview of the approximations involved in this methodology can be found in Refs.~\cite{mazin_1990,wiendlocha_2006} and references therein. 

In the current work, we focus on the changes in the electronic structure only. The phonon contribution, that is, the denominator in Eq.(\ref{eq_1}), as in the earlier works~\cite{jasiewicz_2016,jasiewicz_2019} is approximated by the product of average mass $\langle M \rangle$ and average square phonon frequency. The latter is calculated by assuming the Debye phonon spectrum, for which $\langle{\omega^2}\rangle = \frac{1}{2} \omega_D^2$ (see \cite{jasiewicz_2019}), and using the experimental Debye temperature $\theta_D$, $\hbar\omega_D = k_B\theta_D$.
Therefore, the final formula to calculate $\lambda$ is
\begin{equation}\label{eq_2}
\lambda =  \frac{ \sum_i\eta_i}{\frac{1}{2}\langle M \rangle \omega_D^2}.
\end{equation} 
Note that in the case of the supercell, the sum runs over all atoms, whereas in the case of CPA calculations in the n-times smaller primitive cell in \cite{jasiewicz_2016,jasiewicz_2019} $\eta_i$ are multiplied by the atomic concentrations. The factor that maintains the consistency of both approaches is the total density of states per computational cell, which is in the denominator in Eq.~(\ref{eq:eta}) and increases n-times with the size of the supercell, playing an equal role as the atomic concentration factor.
Here, in each case where CPA results are compared to supercell calculations, densities of states and McMillan-Hopfield parameters are adequately scaled to take into account the 18 times larger number of atoms of the supercell, compared to the single-atomic primitive {\it bcc} cell.

\section{Results and discussion}

\subsection{Lattice distortions}

\begin{figure}[b]
\centering
		\includegraphics[width=\columnwidth]{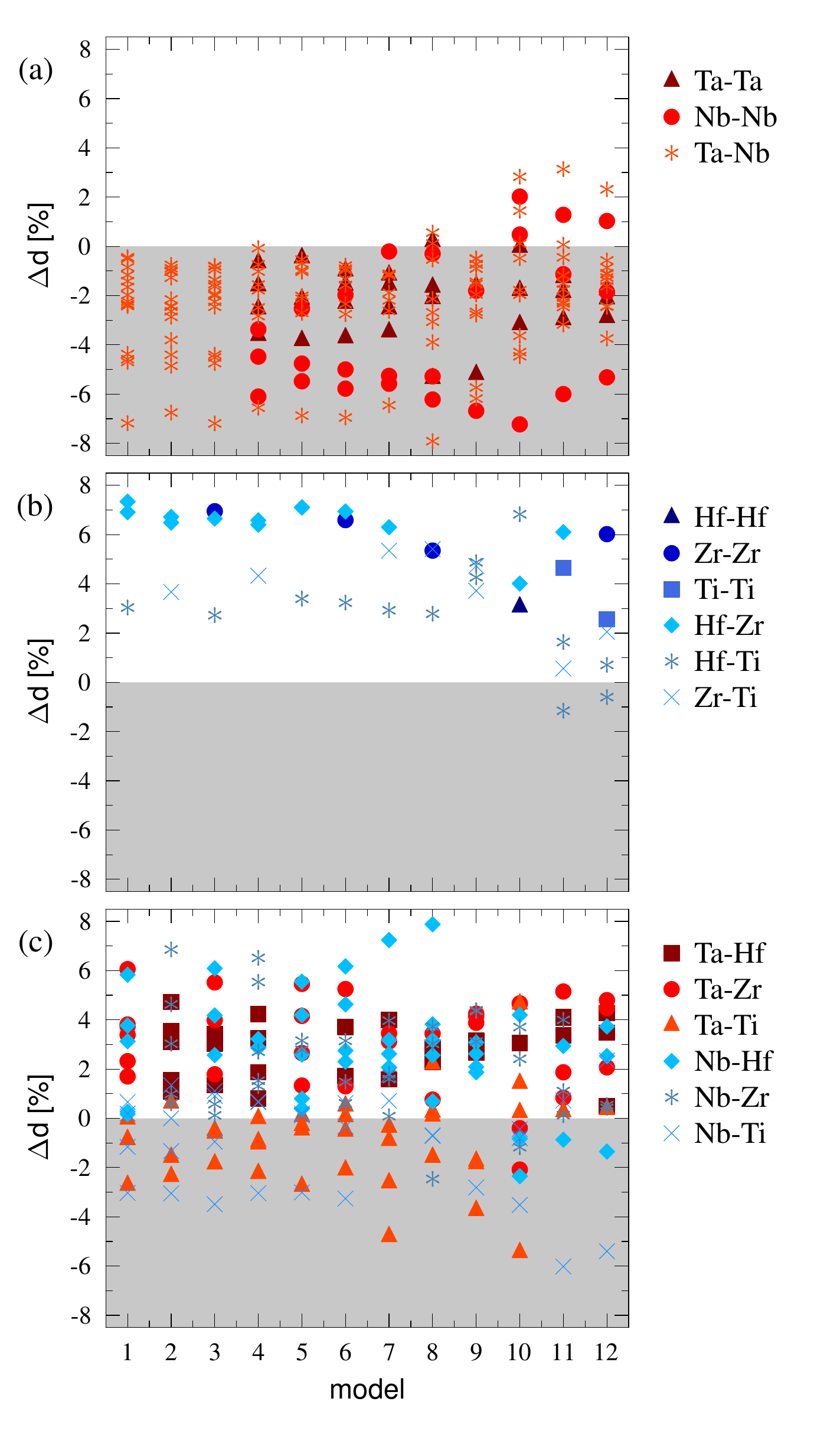}
		\caption{Relative changes in interatomic distances calculated for twelve $3\times3\times1$ supercells with different models of atom arrangements.}
		\label{fig_interat_dist}
\end{figure}

Figure~\ref{fig_interat_dist} presents the relative changes in interatomic distances $\Delta d$ in all models, due to the relaxation of the structure,  calculated using the formula:

\begin{equation}\label{eq_3}
\Delta d = \frac{d^{AB}_{D(+)}-d^{AB}_{D(-)}}{d^{AB}_{D(-)}}\times  100\%
\end{equation}

They are divided into three sets according to the number of valence electrons of the A and B atoms, forming the A-B pair:
(a) both A and B are pentavalent elements (Ta, Nb);
(b) both A and B are tetravalent elements (Hf, Zr, Ti);
(c) A and B have different valency.

\begin{figure*}[t!]
\centering
	\includegraphics[width=0.95\textwidth]{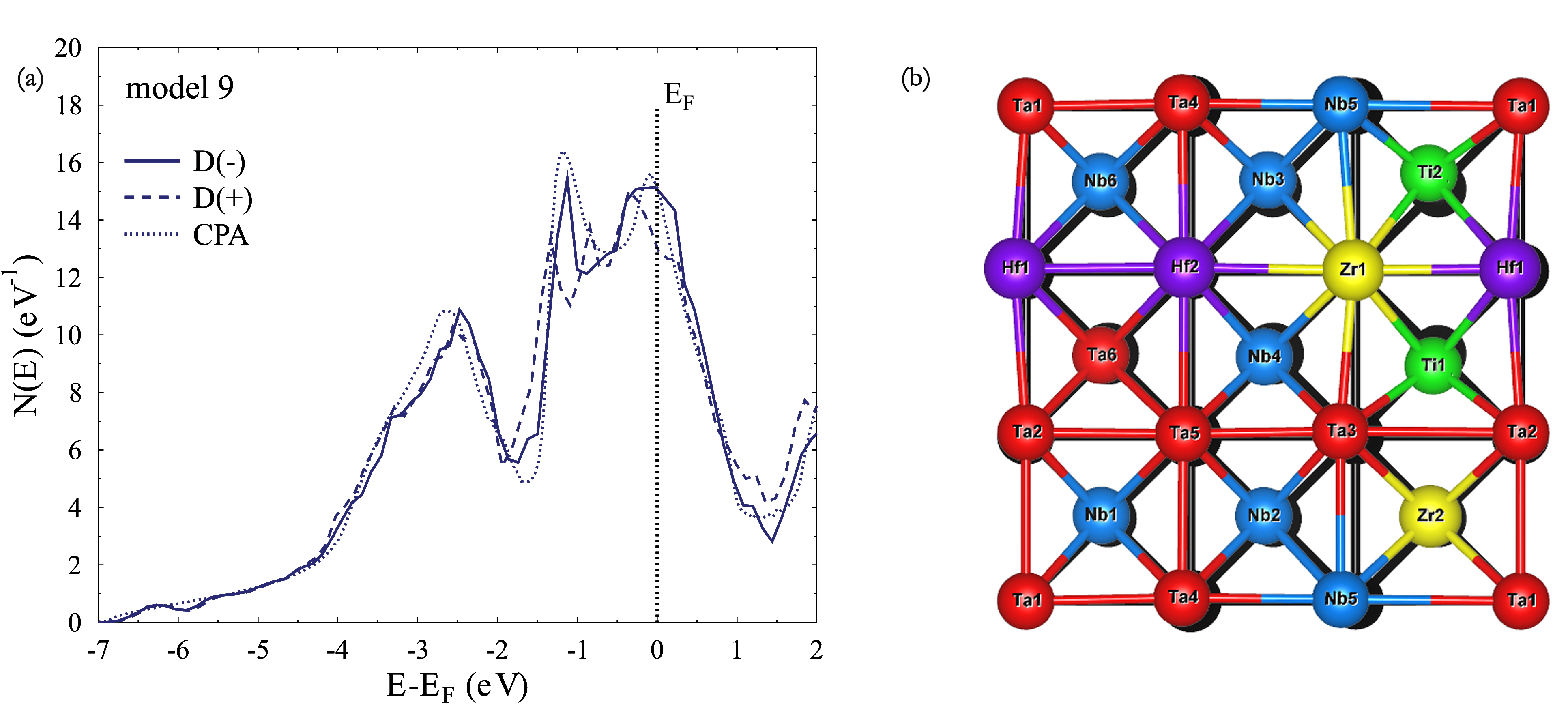}
	\caption{(a) Total electronic densities of states, obtained for the three structural variants: one of the supercell models (no. 9) without distortions, labeled "D(-)"; the same supercell with distortions, labeled as  "D(+)"; and KKR-CPA result for the full disorder with random site occupations, labeled "CPA". (b) Projection of the distorted supercell no. 9 on the $xy$ plane. In the background, the undistorted positions of the atoms are marked in black.}
	\label{fig_dos}
\end{figure*}

Considerable local lattice distortions are observed as the absolute values of $\Delta d$ reach 8\%. 
{As a result of the relaxation process, each supercell reduces its total energy, and the average gain per atom is between 2 and 4 mRy, depending on the supercell model.
Several trends in the changes of the interatomic distances can be distinguished. }
When the A and B atoms are pentavalent Ta and/or Nb, the values of $\Delta d$ show a strong tendency to be negative [Fig.~\ref{fig_interat_dist}(a)].
Only the niobium atoms in models 10-12 increase the distance between each other. The same behavior is observed for Ta-Nb pairs. 
This can be understood as both Ta and Nb in their elemental structures form {\it bcc}-type structures with a lower lattice parameter (Ta: 3.3013~\AA ~\cite{ta_lattice_const} and Nb: 3.3004~\AA ~\cite{nb_lattice_const}) than in the HEA structure (3.36~\AA).

\begin{figure*}[htb]
\centering
	\includegraphics[width=0.95\textwidth]{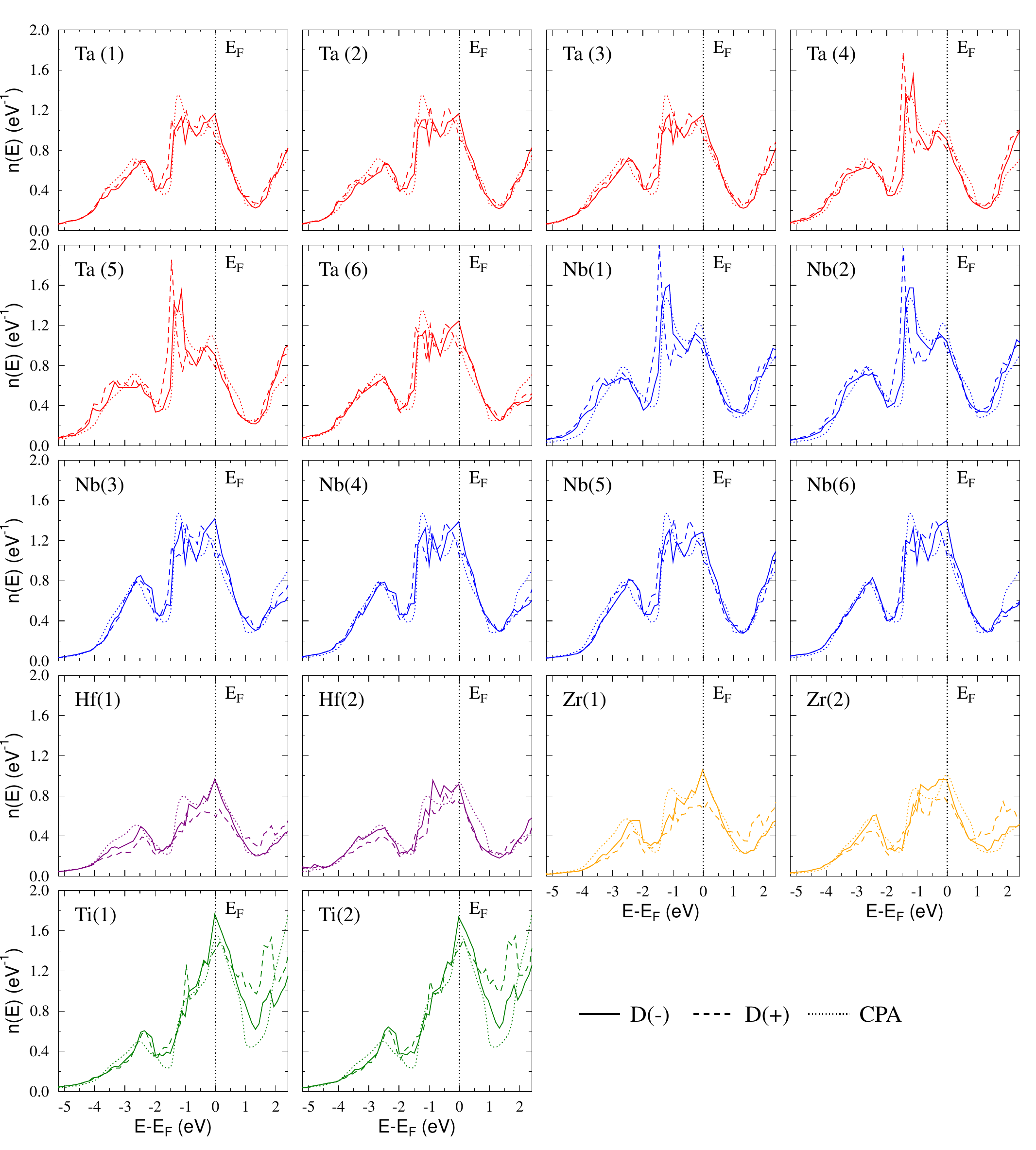}
	\caption{Electronic densities of states for each of the atoms, obtained for model 9  without distortions, labeled as D(-); with distortions, labeled as  D(+); and KKR-CPA result for the full disorder with random site occupations, labeled CPA. For CPA, the curves for the same atom type are identical.}
	\label{fig_pdos}
\end{figure*}

Atoms with four valence electrons, that is, Hf, Zr and Ti, which, as elements, crystallize in a hexagonal {\it hcp} structure~\cite{hf_lattice_constant,zr_lattice_constant,ti_lattice_constant},
exhibit a strong tendency to move away from each other, as seen in Fig.~\ref{fig_interat_dist}(b).
There is only one Hf-Ti pair in models 11 and 12, with a small negative value of $\Delta d$. 
For Hf and Zr this also correlates with the tendency to increase the bond lengths (between 2.999 and 3.11~\AA) closer to those in their elemental structures, since in both cases the nearest-neighbor atomic distances in the {\it hcp} structures are larger (3.093~\AA ~for Hf and 3.152~\AA ~in Zr) than in the HEA structure (2.91~\AA). In metallic elemental Ti, the nearest-neighbor distance is similar to that in HEA (2.869~\AA), however the distances between the Ti atoms in the hexagonal planes are also larger (2.95~\AA).

Pairs of pentavalent and tetravalent atoms do not show a unique trend for changes in interatomic distances; however, in more cases positive values of $\Delta d$ are seen in Fig.~\ref{fig_interat_dist}(c). 
Especially for
Ta-Hf pairs, where only positive $\Delta d$ are noticed. 
The strongest distortions (highest $\Delta d$) are found for atoms located in clusters where three atoms with the largest atomic radius (hafnium and zirconium) are the nearest neighbors. 
The fact that elemental Hf, Zr, and Ti, as well as their binary alloys, crystallize in the \textit{hcp} structure is correlated with the strong local distortion effect and the existence of hexagonal phase precipitations reported in a similar TaNbHfZrTi alloy~\cite{stepanov_2018,chen_2019}.

\subsection{Densities of states}

Figure \ref{fig_dos}(a) shows the effect of atomic relaxation on the total densities of states $N(E)$ for model 9.
Three curves are plotted, the DOS of the D(-) and D(+) structures, as well as of the fully disordered structure, obtained using the coherent potential approximation and labeled as "CPA".
Similar sets of figures for the remaining supercell models are included in the Supplemental Material~\cite{suppl}, Figs. S1-S12.
Figure \ref{fig_dos}(b) presents the projection of the relaxed supercell of model 9 on the $xy$ plane, with the initial (not relaxed) structure in the background. The local distortions are well seen here. 
This particular supercell, after relaxation, had one of the lowest total energies among the structures studied. This is likely related to the presence of the Ta- and Nb-rich clusters.

The general shape of the total DOS curve calculated for the D(-) supercell and the fully disordered variant is similar. As the D(+) DOS curve shows, changes in interatomic distances due to relaxation have the strongest impact on the DOS in the energy range from about 2 eV below the Fermi level to 0.5 eV above $E_F$. 
The relaxed structure has a significantly lower $N(E_F)$ value, 14.2\% for this particular supercell model. 

\begin{figure}[hb!]
\centering
	\includegraphics[width=0.94\columnwidth]{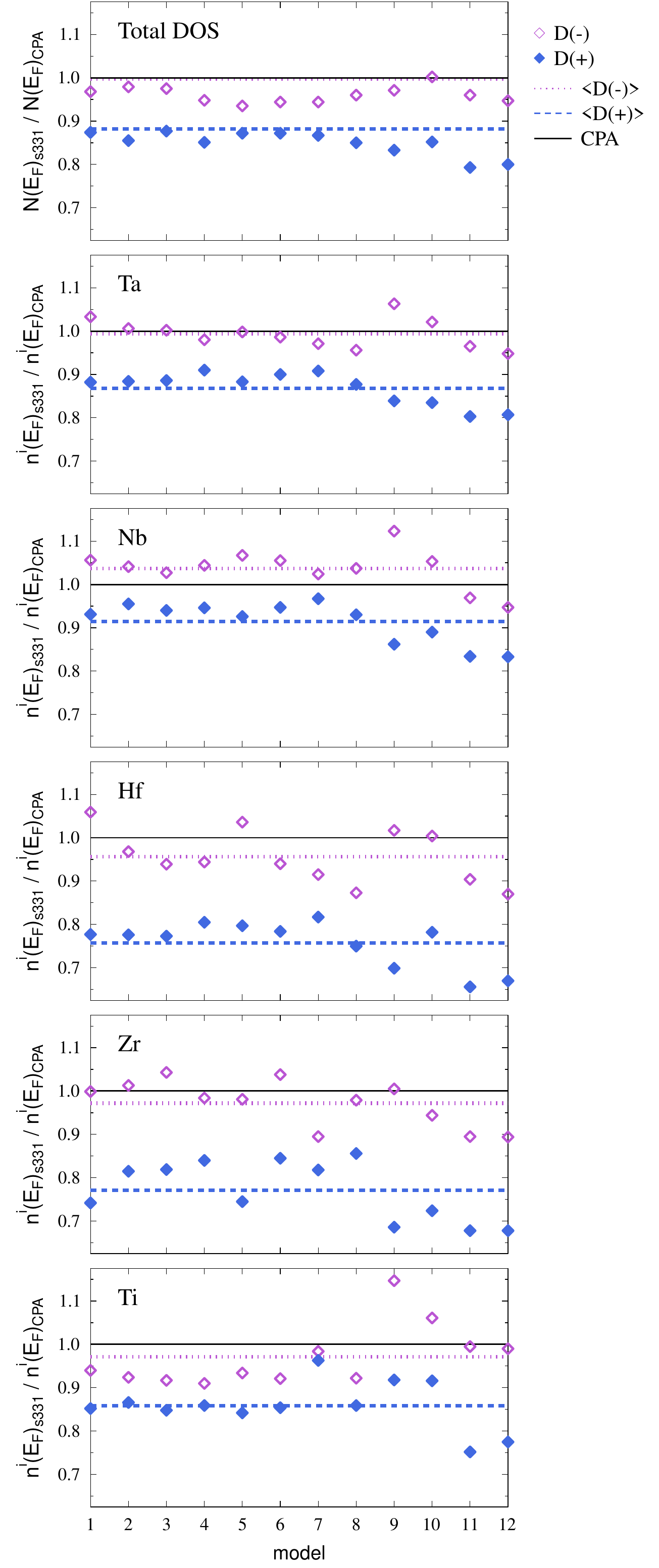}
	\caption{Total and atomic densities of states at Fermi level (average value for a given atom type in different supercell models), calculated for each model of {D(-) (undistorted, open symbols) and D(+) (distorted, solid symbols) supercells}. Average values over all models are marked with a {dotted} line (undistorted supercells, $\langle D(-)\rangle$) and dashed line (distorted supercells, $\langle D(+)\rangle$). All values given as relative to the full-disordered CPA result, marked at $1.0$ with a continuous line.}
	\label{fig_nef_atom}
\end{figure}

Analyzing the atomic densities of states of D(-) supercell $n(E)$, presented in Fig. \ref{fig_pdos}, a correlation can be observed between the shape of the Ta/Nb DOS curve in the vicinity of the Fermi level and the first coordination sphere of the atom. 
Ta(4) and Ta(5), as well as Nb (1) and Nb (2), are surrounded only by the isoelectronic atoms. The main feature of their density of states is a large DOS peak with maximum at ~1.2 eV below the Fermi level. 
Also, $n(E_F)$ does not exceed 0.42 eV (Ta) and 0.48 eV (Nb), which is a relatively low value compared to $n(E_F)$ of the group V atoms which are coordinated by at least two Hf/Zr/Ti atoms in the first coordination sphere (numbers are $0.51-0.55$ eV$^{-1}$ for Ta atoms and $0.56-0.60$ eV$^{-1}$ for Nb atoms). 
This behavior may favor the clustering of these atoms. 

The details of changes in the densities of states are slightly different for other models, as can be seen in the Supplemental Material~\cite{suppl},  but for all of them we observe a similar effect of the lowering of $N(E_F)$.

Figure \ref{fig_nef_atom} summarizes the calculations of the densities of states by showing the values of the total [$N(E_F)$] and partial atomic [$n(E_F)$] DOS at the Fermi level in all supercell models. 
Results for both variants, distorted and undistorted, are shown, relative to the values obtained in the fully random CPA calculations, which, of course, do not include distortions caused by the atomic relaxation process. 
As all 18 atoms were not equivalent in supercell calculations, so to extract essential information, we have averaged the values of $n^i(E_F)$ for each type of atom in the given supercell.

The CPA reference results are represented by a continuous line at $y = 1.0$. 
First, we focus on the undistorted D(-) results, marked with empty symbols. For all of the atoms, as a function of the supercell model number, they fluctuate very close to the CPA reference line. 
The averages of the 12 configurations, $\langle D(-)\rangle$, marked with a dotted line, agree well with the CPA results: the smallest difference of 0.6\% is seen for Ta and the largest for Hf is $4.4\%$. 
The top panel of Fig. \ref{fig_nef_atom} shows the variation of the total $N(E_F)$ value. Here, the differences between the models are not large and also here the average value $\langle D(-)\rangle$ agrees well with the CPA result, being only 0.3 \% smaller.
This is an important result, as it allows us to make the statement that the number and choice of supercell models is statistically sufficient to draw conclusions on the average behavior of this disordered alloy, as in the CPA calculations, the results are averaged over all possible atomic configurations \cite{kaprzyk_1996}.

Now we move to disordered structures. As mentioned above, atomic relaxation in all cases leads to a decrease in the densities of states, both total $N(E_F)$ and atomic $n^i(E_F)$.
The relative changes in the average $n^i(E_F)$ are 11.5\% for Ti, 11.9\% for Nb, 12.7\% for Ta and 20.8\% for Hf and Zr. 
For the total density of states, the average decrease is 11.6\% for all the models considered\footnote{
Due to distortions, there is a small relative transfer of states to the interstitial region. Its participation in total DOS increases, which slightly compensates for the drops observed for atomic $n^i(E_F)$ values.}.

\begin{figure}[ht!]
\centering
	\includegraphics[width=\columnwidth]{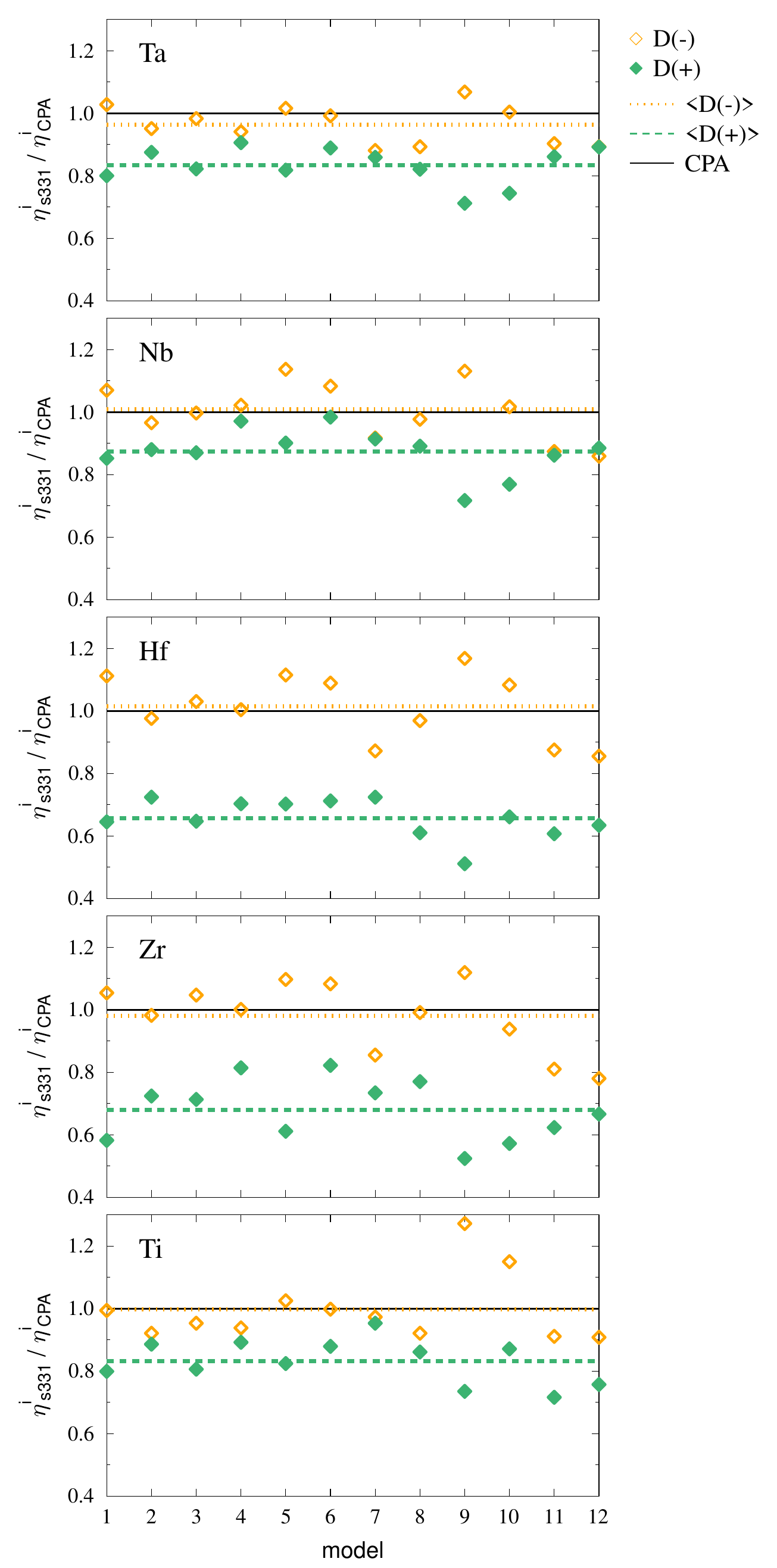}
	\caption{McMillan-Hopfield parameters (average value for a given atom type in different supercell models) calculated for each model of  {D(-) (undistorted, open symbols) and D(+) (distorted, solid symbols) supercells}. Average values over all models are marked with a {dotted} line (undistorted supercells, $\langle D(-)\rangle$) and dashed line (distorted supercells, $\langle D(+)\rangle$). All values given as relative to the full-disordered CPA result, marked at $1.0$ with a continuous line.}
	\label{fig_mcmh_atom}
\end{figure}

\subsection{McMillan-Hopfield parameters}

In a similar way to the densities of states presented in Fig.~\ref{fig_nef_atom}, Fig. \ref{fig_mcmh_atom} presents the McMillan-Hopfield parameters $\eta$ calculated for each type of atom in the different supercells.

Here, the differences between the results obtained for the different models are larger than in the case of densities of states.
Nevertheless, for the undistorted D(-) results, 
as in the case of densities of states, the averages over all 12 configurations, $\langle D(-)\rangle$, marked with a dotted line, agree very well with the CPA results for all atoms. The largest difference is observed for Ta and is about 3.5\%.

Local distortions created by atomic relaxation
generally lead to a decrease in McMillan-Hopfield parameters (full points in Fig. \ref{fig_mcmh_atom}).
This is correlated with the decrease in the densities of states at the Fermi level, discussed above.
The average values for the distorted systems, $\langle D(+)\rangle$, are plotted using the dashed line. 
The strongest effect is observed for Hf and Zr, where the decrease is about 35\% and 31\%, respectively. 
The third tetravalent element, Ti, has a $\langle D(+)\rangle$ 17\% lower than in the undistorted structure, and the effect is the weakest for Ta and Nb, 13.5\% and 13.3\%, respectively.

Taking a closer look at the results for single supercells, the largest differences due to relaxation were obtained for Hf and Zr atoms in model 9, where the difference reaches 56\% and 53\%, while the smallest differences are found for Ta atoms in model 12 (0.2\%).

\subsection{Electron-phonon coupling}

\begin{figure}[t]
\centering
	\includegraphics[width=\columnwidth]{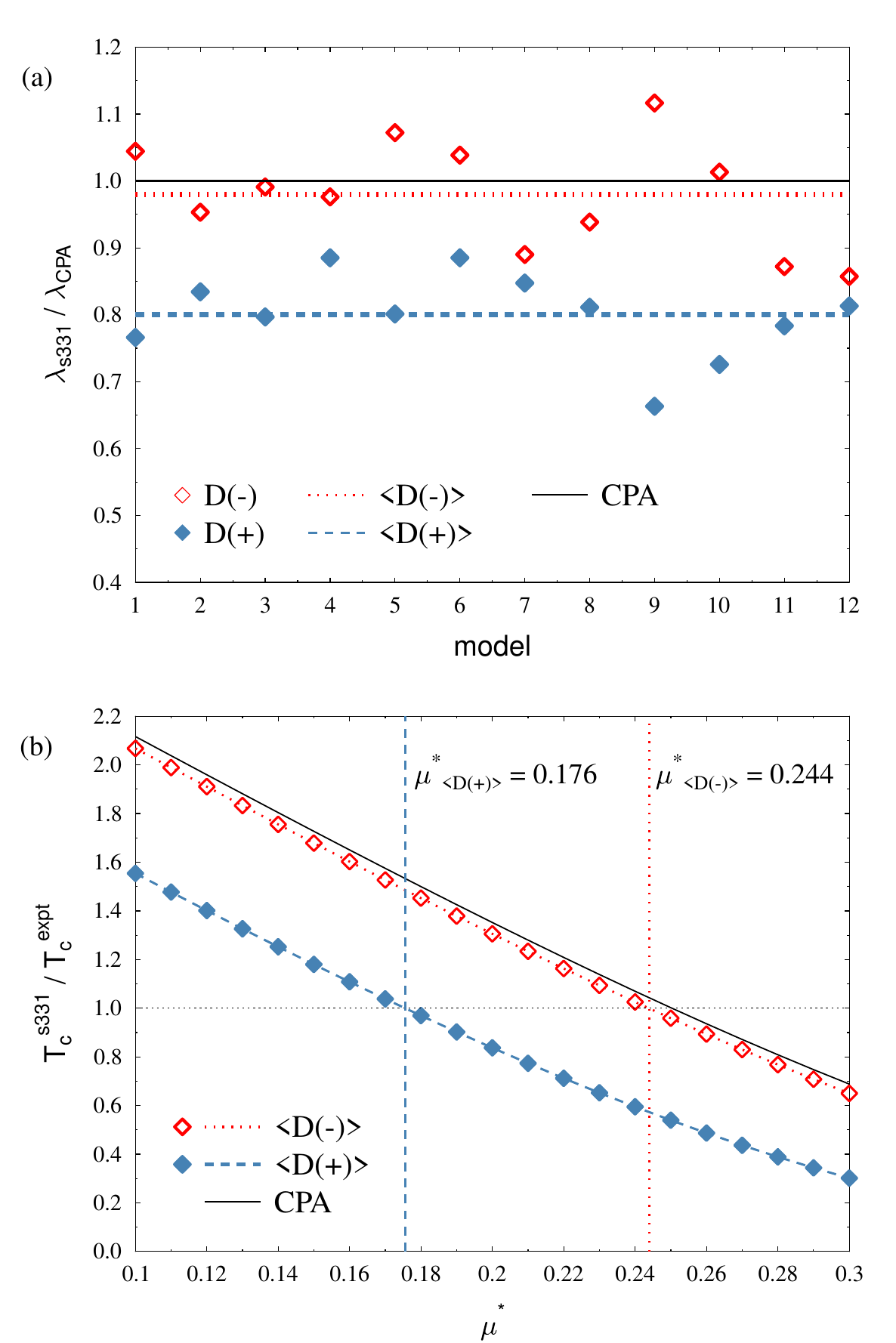}
	\caption{(a) Values of the electron-phonon coupling constants $\lambda$ calculated for each model of  {D(-) (undistorted, open symbols) and D(+) (distorted, solid symbols) supercells. Dotted line} represents the average value over all models for undistorted supercells $\langle D(-)\rangle$, dashed lines represent the average value over all models for distorted supercells $\langle D(+)\rangle$. All values are given relative to the full-disordered CPA result, marked at $1.0$ with a continuous line.
 (b) Critical temperature, calculated using average undistorted $\langle D(-)\rangle$ and distorted $\langle D(+)\rangle$ values of $\lambda$, as a function of the Coulomb pseudopotential $\mu^*$. The results are normalized by the experimental $T_c = 7.7$~K. Vertical lines mark the values of $\mu^*$ for which the experimental $T_c$ is obtained: $\mu^*_{\langle D(+)\rangle} = 0.176$ for $\lambda_{\langle D(+)\rangle} = 0.98$ and $\mu^*_{\langle D(-)\rangle} = 0.244$ for $\lambda_{\langle D(-)\rangle} = 1.19$.
 }
	\label{fig_epc_tc}
\end{figure}

Now we can proceed to the discussion of the effect of local distortions on the electron-phonon coupling parameter $\lambda$. As we have mentioned before, the phonon contribution, via Eq.~(\ref{eq_1}), is estimated using the Debye model, the average atomic mass, and the experimental Debye temperature. Thus, our analysis models only how the distortions affect $\lambda$ through the changes in the electronic contribution, described by the McMillan-Hopfield parameters. 
Figure~\ref{fig_epc_tc}(a) presents the calculated values of $\lambda$ for distorted and undistorted supercells. Again, to capture the relative effect of relaxation on $\lambda$, all results are divided by the fully random undistorted CPA value ($\lambda_{\rm CPA} = 1.21$), which marks a reference $y=1.0$ line.
As in the case of the McMillan-Hopfield parameters discussed above, the calculated values of $\lambda$ for different undistorted supercells fluctuate around the reference CPA result, with a spread of approximately $\pm 15$\%. This shows that a single supercell calculation is not sufficient to give a conclusive prediction of the value of $\lambda$ for this alloy.
However, after averaging the results for 12 models, the average {$\lambda_{\langle D(-)\rangle} = 1.19$} is almost identical to $\lambda_{\rm CPA}$, again confirming the statistical significance of our probe. 
The direct consequence of the relaxation process, due to the decrease in the McMillan-Hopfield parameters, is a reduction of $\lambda$ for the distorted supercells. On average, $\lambda$ decreases by almost 20\%. Referring to the absolute values, this gives a reduction from {1.19 to $\lambda_{\langle D(+)\rangle} = 0.98.$}
Such a decrease will strongly influence the critical temperature due to the exponential dependence of $T_c(\lambda)$ in the
McMillan formula~\cite{mcmillan}:
\begin{equation}
T_c=\frac{\theta_D}{1.45}
\exp\left[ 
\frac{-1.04(1+\lambda)}{\lambda-\mu^*(1+0.62\lambda)}\right].
\label{eq:tc}
\end{equation}
Figure \ref{fig_epc_tc}(b) presents the calculated $T_c$ as a function of the Coulomb pseudopotential $\mu^*$ parameter, in eq.(\ref{eq:tc}) the
experimental Debye temperature of $\theta_D = 216$ K \cite{jasiewicz_2019} was used.
The critical temperature is normalized by the experimental value of $T_c^{\rm expt} = 7.7$ K to visualize the relative change.
The average values of undistorted {$\lambda_{\langle D(-)\rangle} = 1.19$ and distorted $\lambda_{\langle D(+)\rangle} = 0.98$}  were used. As we can see, for the same pseudopotential Coulomb parameters, taking into account the lattice distortions reduces $T_c$ in more than 50\%. Thus, the effect of structural disorder on $T_c$ is very strong. Moreover, the introduction of distortions partially solves the problem with unusually high values of $\mu^*$ required to reproduce the experimental critical temperatures in calculations. 
As reported in our KKR-CPA calculations for fully disordered systems in ideal cubic {\it bcc} structures \cite{jasiewicz_2016,jasiewicz_2019}, $\mu^* > 0.20$ was needed to achieve agreement between experimental and theoretical $T_c$. In the present calculations, for undistorted structures, the experimental $T_c$ is reproduced for a large $\mu^* = 0.244$, while for the average value of $\lambda = 0.98$, obtained for distorted structures, $\mu^* = 0.176$ is sufficient\footnote{If, as reference, we take the $\lambda_{\rm CPA} = 1.10$ value, obtained in \cite{jasiewicz_2019} for the larger $R_{\rm MT} = 2.75$ a$_B$ for which the maximal primitive cell volume is filled with muffin-tin spheres, while keeping the same 20\% reduction of $\lambda$ due to distortions, even a lower
$\mu^* = 0.15$ becomes sufficient to reproduce the experimental $T_c$.}. 
This shows that structural disorder, that is, the deviation of the lattice from the ideal cubic symmetry, is an important factor for determining the electron-phonon coupling in high-entropy alloys, and it needs to be taken into account to accurately explain their superconducting properties.

\subsection{Discussion of the effect of disorder on superconductivity}
An issue of particular interest in the study of superconducting HEAs is to what extent disorder affects the properties of the material, in particular its superconductivity. 
It can be investigated on many levels, but we can certainly distinguish two most basic questions. The first is the question of the influence of disorder on the electronic and phonon structures and the electron-phonon interaction, manifested by the influence on the value of the $\lambda$ parameter. The second is the problem of influence of disorder on the formation of the superconducting phase, manifested by the value of the critical temperature $T_c$ exhibited by the alloy at a given value of $\lambda$.
Anderson's theorem ~\cite{anderson_59} states that conventional superconductivity is robust with respect to weak disorder due to nonmagnetic impurities. 
In such a case, we expect that $T_c$ will be well reproduced by the McMillan formula (\ref{eq:tc}) using the value of $\lambda$, calculated based on electronic and phonon structures, with a typical Coulomb pseudopotential parameter $\mu^* = 0.10 - 0.15$.
However, in strongly disordered cases, superconductivity can be suppressed~\cite{anderson_degradation,weakly_localized_regime}, as was observed in A-15 superconductors~\cite{anderson_degradation}, thin films~\cite{destruction_films} or some highly disordered metals~\cite{destruction_granular}.
In this case, the competition between the strong scattering of electrons on the disorder, and the formation of Cooper pairs and superconducting phase, effectively reduces $T_c$ at a given $\lambda$. This effect is usually captured as an increase in the Coulomb pseudopotential value $\mu^*$, required to reproduce experimental $T_c$~\cite{anderson_degradation}.
Apart from the influence on $T_c$, other thermodynamic properties of the superconducting phase may be affected; for example,
the specific heat jump at the superconducting transition $\Delta C/\gamma T_c$ ($\gamma$ is the electronic specific heat coefficient) can be reduced below the conventional BCS value of 1.43~\cite{heat-prb}.
Recently, an anomalous broadening of the specific heat jump at $T_c$ was observed in high-entropy alloy-type superconductor $Tr$Zr$_2$ ($Tr$ - transition metals) \cite{Kasem_2021}, which may be related to disorder.
Furthermore, very strong electron scattering was found in (ScZrNb)$_{1-x}$RhPd$_x$ superconductors~\cite{gutowska2023}. 
In these alloys, the critical temperature was found to decrease with increasing $x$ in the alloy composition, which is not explained by calculations of the electronic contribution to $\lambda$, as they suggest the opposite behavior. 
The suppression of superconductivity in this series of alloys may be related to the increase in the strength of electron scattering due to disorder, as $T_c$ and the electronic life time $\tau(x)$ have a similar composition dependence, both of which decrease rapidly with $x$~\cite{gutowska2023}.

To discuss the effects of disorder on the superconductivity of (TaNb)$_{0.67}$(HfZrTi)$_{0.33}$ let us list the important findings made so far, related to the interplay of disorder, electron-phonon coupling and superconductivity:
(i) chemical disorder does not lead to strong electron scattering, as the disorder-induced electronic band smearing effect is weak~\cite{jasiewicz_2019};
(ii) structural disorder reduces the electronic contribution to $\lambda$ in about 20\%, as shown in this work;
(iii) there is no information available on how the disorder influences the phonon spectrum, but as the mass disorder is quite significant here, we expect it to be rather strong\footnote{Note that in our calculations of $\lambda$ the phonon contribution is modeled using the experimental Debye temperature; thus it takes into account the possible influence of disorder on phonons without specifying how strong it is.};
(iv) the experimental critical temperature is reproduced by the McMillan formula using slightly enhanced value of $\mu^* > 0.15$.
On the basis of that we can conclude that disorder (mostly structural) has a significant impact on the electron-phonon coupling parameter $\lambda$. As far as $T_c$ is concerned, 
taking into account local distortions in the alloy reduced the parameter $\mu^*$ required to reproduce its experimental $T_c$ closer to the "conventional" range.
Thus, the formation of the superconducting phase is probably only slightly suppressed by disorder.

\section{Summary}
Local distortions of the crystal structure and their influence on the electronic structure and superconductivity of (TaNb)$_{0.67}$(HfZrTi)$_{0.33}$ high-entropy alloy
have been theoretically studied. A set of twelve supercell models occurs to be a statistically sufficient probe to conclude on the properties of these random alloys, as the averaged densities of states and McMillan-Hopfield parameters of the undistorted structures agree very well with the coherent potential approximation results, in which all computed quantities are automatically averaged over all possible configurations.
Local distortions from the ideal cubic lattice appear as a result of forces relaxations, and the interatomic distances change significantly, up to 8\%. 
As a consequence, the total density of states at the Fermi level, $N(E_F)$, and the McMillan-Hopfield parameters $\eta$ decrease, lowering the electronic contribution to the electron-phonon coupling constant $\lambda$ in 20\%. This significantly affects the superconducting critical temperature, which is reduced by about 50\%. As a consequence, the Coulomb pseudopotential parameter $\mu^*$ required to reproduce the experimental $T_c$ in the calculations is much closer to the conventional range of 0.10 - 0.15, observed for the intermetallic systems. 
This suggests that the disorder has only a weak suppressing effect on the thermodynamics of the superconducting phase in (TaNb)$_{0.67}$(HfZrTi)$_{0.33}$.

\section*{Acknowledgements}
KJ and BW were supported by the National Science Centre (Poland), project no. 2017/26/E/ST3/00119.

\bibliography{references}

\begin{thebibliography}{71}%
\makeatletter
\providecommand \@ifxundefined [1]{%
 \@ifx{#1\undefined}
}%
\providecommand \@ifnum [1]{%
 \ifnum #1\expandafter \@firstoftwo
 \else \expandafter \@secondoftwo
 \fi
}%
\providecommand \@ifx [1]{%
 \ifx #1\expandafter \@firstoftwo
 \else \expandafter \@secondoftwo
 \fi
}%
\providecommand \natexlab [1]{#1}%
\providecommand \enquote  [1]{``#1''}%
\providecommand \bibnamefont  [1]{#1}%
\providecommand \bibfnamefont [1]{#1}%
\providecommand \citenamefont [1]{#1}%
\providecommand \href@noop [0]{\@secondoftwo}%
\providecommand \href [0]{\begingroup \@sanitize@url \@href}%
\providecommand \@href[1]{\@@startlink{#1}\@@href}%
\providecommand \@@href[1]{\endgroup#1\@@endlink}%
\providecommand \@sanitize@url [0]{\catcode `\\12\catcode `\$12\catcode `\&12\catcode `\#12\catcode `\^12\catcode `\_12\catcode `\%12\relax}%
\providecommand \@@startlink[1]{}%
\providecommand \@@endlink[0]{}%
\providecommand \url  [0]{\begingroup\@sanitize@url \@url }%
\providecommand \@url [1]{\endgroup\@href {#1}{\urlprefix }}%
\providecommand \urlprefix  [0]{URL }%
\providecommand \Eprint [0]{\href }%
\providecommand \doibase [0]{http://dx.doi.org/}%
\providecommand \selectlanguage [0]{\@gobble}%
\providecommand \bibinfo  [0]{\@secondoftwo}%
\providecommand \bibfield  [0]{\@secondoftwo}%
\providecommand \translation [1]{[#1]}%
\providecommand \BibitemOpen [0]{}%
\providecommand \bibitemStop [0]{}%
\providecommand \bibitemNoStop [0]{.\EOS\space}%
\providecommand \EOS [0]{\spacefactor3000\relax}%
\providecommand \BibitemShut  [1]{\csname bibitem#1\endcsname}%
\let\auto@bib@innerbib\@empty
\bibitem [{\citenamefont {Murty}\ \emph {et~al.}(2014)\citenamefont {Murty}, \citenamefont {Yeh},\ and\ \citenamefont {Ranganathan}}]{murty2014}%
  \BibitemOpen
  \bibfield  {author} {\bibinfo {author} {\bibfnamefont {Bhagevatula~Satyanarayana}\ \bibnamefont {Murty}}, \bibinfo {author} {\bibfnamefont {Jien-Wei}\ \bibnamefont {Yeh}}, \ and\ \bibinfo {author} {\bibfnamefont {Srinivasa}\ \bibnamefont {Ranganathan}},\ }\href {\doibase 10.1016/C2013-0-14235-3} {\emph {\bibinfo {title} {High-entropy alloys}}}\ (\bibinfo  {publisher} {Elsevier},\ \bibinfo {year} {2014})\BibitemShut {NoStop}%
\bibitem [{\citenamefont {Gao}\ \emph {et~al.}(2016)\citenamefont {Gao}, \citenamefont {Yeh}, \citenamefont {Liaw},\ and\ \citenamefont {Zhang}}]{hea_book_2016}%
  \BibitemOpen
  \bibfield  {author} {\bibinfo {author} {\bibfnamefont {M.~C.}\ \bibnamefont {Gao}}, \bibinfo {author} {\bibfnamefont {J.~W.}\ \bibnamefont {Yeh}}, \bibinfo {author} {\bibfnamefont {P.~K.}\ \bibnamefont {Liaw}}, \ and\ \bibinfo {author} {\bibfnamefont {Y.}~\bibnamefont {Zhang}},\ }\href@noop {} {\emph {\bibinfo {title} {{High entropy alloys. Fundamentals and Applications}}}}\ (\bibinfo  {publisher} {Springer International Publishing},\ \bibinfo {address} {Switzerland},\ \bibinfo {year} {2016})\BibitemShut {NoStop}%
\bibitem [{\citenamefont {Cantor}\ \emph {et~al.}(2004)\citenamefont {Cantor}, \citenamefont {Chang}, \citenamefont {Knight},\ and\ \citenamefont {Vincent}}]{cantor_2004}%
  \BibitemOpen
  \bibfield  {author} {\bibinfo {author} {\bibfnamefont {B.}~\bibnamefont {Cantor}}, \bibinfo {author} {\bibfnamefont {I.~T.~H.}\ \bibnamefont {Chang}}, \bibinfo {author} {\bibfnamefont {P.}~\bibnamefont {Knight}}, \ and\ \bibinfo {author} {\bibfnamefont {A.~J.~B.}\ \bibnamefont {Vincent}},\ }\bibfield  {title} {\enquote {\bibinfo {title} {Microstructural development in equiatomic multicomponent alloys},}\ }\href {\doibase 10.1016/j.msea.2003.10.257} {\bibfield  {journal} {\bibinfo  {journal} {Materials Science and Engineering: A}\ }\textbf {\bibinfo {volume} {375-377}},\ \bibinfo {pages} {213 -- 218} (\bibinfo {year} {2004})}\BibitemShut {NoStop}%
\bibitem [{\citenamefont {Yeh}\ \emph {et~al.}(2004)\citenamefont {Yeh}, \citenamefont {Chen}, \citenamefont {Lin}, \citenamefont {Gan}, \citenamefont {Chin}, \citenamefont {Shun}, \citenamefont {Tsau},\ and\ \citenamefont {Chang}}]{yeh_2004}%
  \BibitemOpen
  \bibfield  {author} {\bibinfo {author} {\bibfnamefont {J.~W.}\ \bibnamefont {Yeh}}, \bibinfo {author} {\bibfnamefont {S.~K.}\ \bibnamefont {Chen}}, \bibinfo {author} {\bibfnamefont {S.~J.}\ \bibnamefont {Lin}}, \bibinfo {author} {\bibfnamefont {J.~Y.}\ \bibnamefont {Gan}}, \bibinfo {author} {\bibfnamefont {T.~S.}\ \bibnamefont {Chin}}, \bibinfo {author} {\bibfnamefont {T.~T.}\ \bibnamefont {Shun}}, \bibinfo {author} {\bibfnamefont {C.~H.}\ \bibnamefont {Tsau}}, \ and\ \bibinfo {author} {\bibfnamefont {S.~Y.}\ \bibnamefont {Chang}},\ }\bibfield  {title} {\enquote {\bibinfo {title} {{Nanostructured High Entropy Alloys with Multiple Principal Elements: Novel Alloy Design Concepts and Outcomes}},}\ }\href {\doibase 10.1002/adem.200300567} {\bibfield  {journal} {\bibinfo  {journal} {Advanced Engineering Materials}\ }\textbf {\bibinfo {volume} {6}},\ \bibinfo {pages} {299--303} (\bibinfo {year} {2004})}\BibitemShut {NoStop}%
\bibitem [{\citenamefont {Greer}(1993)}]{greer_1993}%
  \BibitemOpen
  \bibfield  {author} {\bibinfo {author} {\bibfnamefont {A.~L.}\ \bibnamefont {Greer}},\ }\bibfield  {title} {\enquote {\bibinfo {title} {Confusion by design},}\ }\href {\doibase 10.1038/366303a0} {\bibfield  {journal} {\bibinfo  {journal} {Nature}\ }\textbf {\bibinfo {volume} {366}},\ \bibinfo {pages} {303--304} (\bibinfo {year} {1993})}\BibitemShut {NoStop}%
\bibitem [{\citenamefont {Ko\ifmmode~\check{z}\else \v{z}\fi{}elj}\ \emph {et~al.}(2014)\citenamefont {Ko\ifmmode~\check{z}\else \v{z}\fi{}elj}, \citenamefont {Vrtnik}, \citenamefont {Jelen}, \citenamefont {Jazbec}, \citenamefont {Jagli\ifmmode \check{c}\else \v{c}\fi{}i\ifmmode~\acute{c}\else \'{c}\fi{}}, \citenamefont {Maiti}, \citenamefont {Feuerbacher}, \citenamefont {Steurer},\ and\ \citenamefont {Dolin\ifmmode~\check{s}\else \v{s}\fi{}ek}}]{kozejl_2014}%
  \BibitemOpen
  \bibfield  {author} {\bibinfo {author} {\bibfnamefont {P.}~\bibnamefont {Ko\ifmmode~\check{z}\else \v{z}\fi{}elj}}, \bibinfo {author} {\bibfnamefont {S.}~\bibnamefont {Vrtnik}}, \bibinfo {author} {\bibfnamefont {A.}~\bibnamefont {Jelen}}, \bibinfo {author} {\bibfnamefont {S.}~\bibnamefont {Jazbec}}, \bibinfo {author} {\bibfnamefont {Z.}~\bibnamefont {Jagli\ifmmode \check{c}\else \v{c}\fi{}i\ifmmode~\acute{c}\else \'{c}\fi{}}}, \bibinfo {author} {\bibfnamefont {S.}~\bibnamefont {Maiti}}, \bibinfo {author} {\bibfnamefont {M.}~\bibnamefont {Feuerbacher}}, \bibinfo {author} {\bibfnamefont {W.}~\bibnamefont {Steurer}}, \ and\ \bibinfo {author} {\bibfnamefont {J.}~\bibnamefont {Dolin\ifmmode~\check{s}\else \v{s}\fi{}ek}},\ }\bibfield  {title} {\enquote {\bibinfo {title} {{Discovery of a Superconducting High Entropy Alloy}},}\ }\href {\doibase 10.1103/PhysRevLett.113.107001} {\bibfield  {journal} {\bibinfo  {journal} {Phys. Rev. Lett.}\ }\textbf {\bibinfo {volume} {113}},\ \bibinfo {pages} {107001} (\bibinfo
  {year} {2014})}\BibitemShut {NoStop}%
\bibitem [{\citenamefont {Marik}\ \emph {et~al.}(2018)\citenamefont {Marik}, \citenamefont {Varghese}, \citenamefont {Sajilesh}, \citenamefont {Singh},\ and\ \citenamefont {Singh}}]{hea_sup_1}%
  \BibitemOpen
  \bibfield  {author} {\bibinfo {author} {\bibfnamefont {Sourav}\ \bibnamefont {Marik}}, \bibinfo {author} {\bibfnamefont {Maneesha}\ \bibnamefont {Varghese}}, \bibinfo {author} {\bibfnamefont {K.P.}\ \bibnamefont {Sajilesh}}, \bibinfo {author} {\bibfnamefont {Deepak}\ \bibnamefont {Singh}}, \ and\ \bibinfo {author} {\bibfnamefont {R.P.}\ \bibnamefont {Singh}},\ }\bibfield  {title} {\enquote {\bibinfo {title} {{Superconductivity in equimolar Nb-Re-Hf-Zr-Ti high entropy alloy}},}\ }\href {\doibase 10.1016/j.jallcom.2018.08.039} {\bibfield  {journal} {\bibinfo  {journal} {Journal of Alloys and Compounds}\ }\textbf {\bibinfo {volume} {769}},\ \bibinfo {pages} {1059--1063} (\bibinfo {year} {2018})}\BibitemShut {NoStop}%
\bibitem [{\citenamefont {Sogabe}\ \emph {et~al.}(2018)\citenamefont {Sogabe}, \citenamefont {Goto},\ and\ \citenamefont {Mizuguchi}}]{hea_sup_2}%
  \BibitemOpen
  \bibfield  {author} {\bibinfo {author} {\bibfnamefont {Ryota}\ \bibnamefont {Sogabe}}, \bibinfo {author} {\bibfnamefont {Yosuke}\ \bibnamefont {Goto}}, \ and\ \bibinfo {author} {\bibfnamefont {Yoshikazu}\ \bibnamefont {Mizuguchi}},\ }\bibfield  {title} {\enquote {\bibinfo {title} {{Superconductivity in {REO}$_{0.5}${F}$_{0.5}${Bi}{S}$_2$ with high entropy-alloy-type blocking layers}},}\ }\href {\doibase 10.7567/apex.11.053102} {\bibfield  {journal} {\bibinfo  {journal} {Applied Physics Express}\ }\textbf {\bibinfo {volume} {11}},\ \bibinfo {pages} {053102} (\bibinfo {year} {2018})}\BibitemShut {NoStop}%
\bibitem [{\citenamefont {Stolze}\ \emph {et~al.}(2018{\natexlab{a}})\citenamefont {Stolze}, \citenamefont {Tao}, \citenamefont {von Rohr}, \citenamefont {Kong},\ and\ \citenamefont {Cava}}]{hea_sup_3}%
  \BibitemOpen
  \bibfield  {author} {\bibinfo {author} {\bibfnamefont {Karoline}\ \bibnamefont {Stolze}}, \bibinfo {author} {\bibfnamefont {Jing}\ \bibnamefont {Tao}}, \bibinfo {author} {\bibfnamefont {Fabian~O.}\ \bibnamefont {von Rohr}}, \bibinfo {author} {\bibfnamefont {Tai}\ \bibnamefont {Kong}}, \ and\ \bibinfo {author} {\bibfnamefont {Robert~J.}\ \bibnamefont {Cava}},\ }\bibfield  {title} {\enquote {\bibinfo {title} {{{Sc–Zr–Nb–Rh–Pd and Sc–Zr–Nb–Ta–Rh–Pd} High Entropy Alloy Superconductors on a {CsCl}-Type Lattice}},}\ }\href {\doibase 10.1021/acs.chemmater.7b04578} {\bibfield  {journal} {\bibinfo  {journal} {Chemistry of Materials}\ }\textbf {\bibinfo {volume} {30}},\ \bibinfo {pages} {906--914} (\bibinfo {year} {2018}{\natexlab{a}})}\BibitemShut {NoStop}%
\bibitem [{\citenamefont {Stolze}\ \emph {et~al.}(2018{\natexlab{b}})\citenamefont {Stolze}, \citenamefont {Cevallos}, \citenamefont {Kong},\ and\ \citenamefont {Cava}}]{hea_sup_4}%
  \BibitemOpen
  \bibfield  {author} {\bibinfo {author} {\bibfnamefont {Karoline}\ \bibnamefont {Stolze}}, \bibinfo {author} {\bibfnamefont {F.~Alex}\ \bibnamefont {Cevallos}}, \bibinfo {author} {\bibfnamefont {Tai}\ \bibnamefont {Kong}}, \ and\ \bibinfo {author} {\bibfnamefont {Robert~J.}\ \bibnamefont {Cava}},\ }\bibfield  {title} {\enquote {\bibinfo {title} {{High entropy alloy superconductors on an $\sigma$-Mn lattice}},}\ }\href {\doibase 10.1039/C8TC03337D} {\bibfield  {journal} {\bibinfo  {journal} {J. Mater. Chem. C}\ }\textbf {\bibinfo {volume} {6}},\ \bibinfo {pages} {10441--10449} (\bibinfo {year} {2018}{\natexlab{b}})}\BibitemShut {NoStop}%
\bibitem [{\citenamefont {Motla}\ \emph {et~al.}(2023)\citenamefont {Motla}, \citenamefont {Arushi}, \citenamefont {Jangid}, \citenamefont {Meena}, \citenamefont {Kushwaha},\ and\ \citenamefont {Singh}}]{Motla_2023}%
  \BibitemOpen
  \bibfield  {author} {\bibinfo {author} {\bibfnamefont {K}~\bibnamefont {Motla}}, \bibinfo {author} {\bibnamefont {Arushi}}, \bibinfo {author} {\bibfnamefont {S}~\bibnamefont {Jangid}}, \bibinfo {author} {\bibfnamefont {P~K}\ \bibnamefont {Meena}}, \bibinfo {author} {\bibfnamefont {R~K}\ \bibnamefont {Kushwaha}}, \ and\ \bibinfo {author} {\bibfnamefont {R~P}\ \bibnamefont {Singh}},\ }\bibfield  {title} {\enquote {\bibinfo {title} {Superconducting properties of new hexagonal and noncentrosymmetric cubic high entropy alloys},}\ }\href {\doibase 10.1088/1361-6668/acfac5} {\bibfield  {journal} {\bibinfo  {journal} {Superconductor Science and Technology}\ }\textbf {\bibinfo {volume} {36}},\ \bibinfo {pages} {115024} (\bibinfo {year} {2023})}\BibitemShut {NoStop}%
\bibitem [{\citenamefont {Yamashita}\ \emph {et~al.}(2021)\citenamefont {Yamashita}, \citenamefont {Matsuda},\ and\ \citenamefont {Mizuguchi}}]{hea_sup_5}%
  \BibitemOpen
  \bibfield  {author} {\bibinfo {author} {\bibfnamefont {Aichi}\ \bibnamefont {Yamashita}}, \bibinfo {author} {\bibfnamefont {Tatsuma~D.}\ \bibnamefont {Matsuda}}, \ and\ \bibinfo {author} {\bibfnamefont {Yoshikazu}\ \bibnamefont {Mizuguchi}},\ }\bibfield  {title} {\enquote {\bibinfo {title} {{Synthesis of new high-entropy alloy-type Nb$_3$(Al, Sn, Ge, Ga, Si) superconductors}},}\ }\href {\doibase 10.1016/j.jallcom.2021.159233} {\bibfield  {journal} {\bibinfo  {journal} {Journal of Alloys and Compounds}\ }\textbf {\bibinfo {volume} {868}},\ \bibinfo {pages} {159233} (\bibinfo {year} {2021})}\BibitemShut {NoStop}%
\bibitem [{\citenamefont {Liu}\ \emph {et~al.}(2021)\citenamefont {Liu}, \citenamefont {Wu}, \citenamefont {Cui}, \citenamefont {Zhu}, \citenamefont {Xiao}, \citenamefont {Wu}, \citenamefont {han Cao},\ and\ \citenamefont {Ren}}]{hea_sup_6}%
  \BibitemOpen
  \bibfield  {author} {\bibinfo {author} {\bibfnamefont {Bin}\ \bibnamefont {Liu}}, \bibinfo {author} {\bibfnamefont {Jifeng}\ \bibnamefont {Wu}}, \bibinfo {author} {\bibfnamefont {Yanwei}\ \bibnamefont {Cui}}, \bibinfo {author} {\bibfnamefont {Qinqing}\ \bibnamefont {Zhu}}, \bibinfo {author} {\bibfnamefont {Guorui}\ \bibnamefont {Xiao}}, \bibinfo {author} {\bibfnamefont {Siqi}\ \bibnamefont {Wu}}, \bibinfo {author} {\bibfnamefont {Guang}\ \bibnamefont {han Cao}}, \ and\ \bibinfo {author} {\bibfnamefont {Zhi}\ \bibnamefont {Ren}},\ }\bibfield  {title} {\enquote {\bibinfo {title} {{Superconductivity and paramagnetism in Cr-containing tetragonal high-entropy alloys}},}\ }\href {\doibase 10.1016/j.jallcom.2021.159293} {\bibfield  {journal} {\bibinfo  {journal} {Journal of Alloys and Compounds}\ }\textbf {\bibinfo {volume} {869}},\ \bibinfo {pages} {159293} (\bibinfo {year} {2021})}\BibitemShut {NoStop}%
\bibitem [{\citenamefont {Kasem}\ \emph {et~al.}(2022)\citenamefont {Kasem}, \citenamefont {Arima}, \citenamefont {Ikeda}, \citenamefont {Yamashita},\ and\ \citenamefont {Mizuguchi}}]{hea_sup_7}%
  \BibitemOpen
  \bibfield  {author} {\bibinfo {author} {\bibfnamefont {Md~Riad}\ \bibnamefont {Kasem}}, \bibinfo {author} {\bibfnamefont {Hiroto}\ \bibnamefont {Arima}}, \bibinfo {author} {\bibfnamefont {Yoichi}\ \bibnamefont {Ikeda}}, \bibinfo {author} {\bibfnamefont {Aichi}\ \bibnamefont {Yamashita}}, \ and\ \bibinfo {author} {\bibfnamefont {Yoshikazu}\ \bibnamefont {Mizuguchi}},\ }\bibfield  {title} {\enquote {\bibinfo {title} {{Superconductivity of high-entropy-alloy-type transition-metal zirconide (Fe,Co,Ni,Cu,Ga)Zr$_2$}},}\ }\href {\doibase 10.1088/2515-7639/ac8e34} {\bibfield  {journal} {\bibinfo  {journal} {Journal of Physics: Materials}\ }\textbf {\bibinfo {volume} {5}},\ \bibinfo {pages} {045001} (\bibinfo {year} {2022})}\BibitemShut {NoStop}%
\bibitem [{\citenamefont {Mizuguchi}\ \emph {et~al.}(2021)\citenamefont {Mizuguchi}, \citenamefont {Kasem},\ and\ \citenamefont {Matsuda}}]{hea_sup_8}%
  \BibitemOpen
  \bibfield  {author} {\bibinfo {author} {\bibfnamefont {Yoshikazu}\ \bibnamefont {Mizuguchi}}, \bibinfo {author} {\bibfnamefont {Md.~Riad}\ \bibnamefont {Kasem}}, \ and\ \bibinfo {author} {\bibfnamefont {Tatsuma~D.}\ \bibnamefont {Matsuda}},\ }\bibfield  {title} {\enquote {\bibinfo {title} {{Superconductivity in CuAl$_2$-type Co$_{0.2}$Ni$_{0.1}$Cu$_{0.1}$Rh$_{0.3}$Ir$_{0.3}$Zr$_2$ with a high-entropy-alloy transition metal site}},}\ }\href {\doibase 10.1080/21663831.2020.1860147} {\bibfield  {journal} {\bibinfo  {journal} {Materials Research Letters}\ }\textbf {\bibinfo {volume} {9}},\ \bibinfo {pages} {141--147} (\bibinfo {year} {2021})}\BibitemShut {NoStop}%
\bibitem [{\citenamefont {Motla}\ \emph {et~al.}(2022)\citenamefont {Motla}, \citenamefont {Soni}, \citenamefont {Meena},\ and\ \citenamefont {Singh}}]{hea_sup_10}%
  \BibitemOpen
  \bibfield  {author} {\bibinfo {author} {\bibfnamefont {Kapil}\ \bibnamefont {Motla}}, \bibinfo {author} {\bibfnamefont {V}~\bibnamefont {Soni}}, \bibinfo {author} {\bibfnamefont {P~K}\ \bibnamefont {Meena}}, \ and\ \bibinfo {author} {\bibfnamefont {R~P}\ \bibnamefont {Singh}},\ }\bibfield  {title} {\enquote {\bibinfo {title} {{Boron based new high entropy alloy superconductor Mo$_{0.11}$W$_{0.11}$V$_{0.11}$Re$_{0.34}$B$_{0.33}$}},}\ }\href {\doibase 10.1088/1361-6668/ac5bea} {\bibfield  {journal} {\bibinfo  {journal} {Superconductor Science and Technology}\ }\textbf {\bibinfo {volume} {35}},\ \bibinfo {pages} {074002} (\bibinfo {year} {2022})}\BibitemShut {NoStop}%
\bibitem [{\citenamefont {Liu}\ \emph {et~al.}(2020)\citenamefont {Liu}, \citenamefont {Wu}, \citenamefont {Cui}, \citenamefont {Zhu}, \citenamefont {Xiao}, \citenamefont {Wu}, \citenamefont {Cao},\ and\ \citenamefont {Ren}}]{hea_sup_9}%
  \BibitemOpen
  \bibfield  {author} {\bibinfo {author} {\bibfnamefont {Bin}\ \bibnamefont {Liu}}, \bibinfo {author} {\bibfnamefont {Jifeng}\ \bibnamefont {Wu}}, \bibinfo {author} {\bibfnamefont {Yanwei}\ \bibnamefont {Cui}}, \bibinfo {author} {\bibfnamefont {Qinqing}\ \bibnamefont {Zhu}}, \bibinfo {author} {\bibfnamefont {Guorui}\ \bibnamefont {Xiao}}, \bibinfo {author} {\bibfnamefont {Siqi}\ \bibnamefont {Wu}}, \bibinfo {author} {\bibfnamefont {Guanghan}\ \bibnamefont {Cao}}, \ and\ \bibinfo {author} {\bibfnamefont {Zhi}\ \bibnamefont {Ren}},\ }\bibfield  {title} {\enquote {\bibinfo {title} {{Superconductivity in hexagonal Nb-Mo-Ru-Rh-Pd high-entropy alloys}},}\ }\href {\doibase 10.1016/j.scriptamat.2020.03.004} {\bibfield  {journal} {\bibinfo  {journal} {Scripta Materialia}\ }\textbf {\bibinfo {volume} {182}},\ \bibinfo {pages} {109--113} (\bibinfo {year} {2020})}\BibitemShut {NoStop}%
\bibitem [{\citenamefont {von Rohr}\ \emph {et~al.}(2016)\citenamefont {von Rohr}, \citenamefont {Winiarski}, \citenamefont {Tao}, \citenamefont {Klimczuk},\ and\ \citenamefont {Cava}}]{rohr_tnhzt_2016}%
  \BibitemOpen
  \bibfield  {author} {\bibinfo {author} {\bibfnamefont {Fabian}\ \bibnamefont {von Rohr}}, \bibinfo {author} {\bibfnamefont {Micha{\l}~J.}\ \bibnamefont {Winiarski}}, \bibinfo {author} {\bibfnamefont {Jing}\ \bibnamefont {Tao}}, \bibinfo {author} {\bibfnamefont {Tomasz}\ \bibnamefont {Klimczuk}}, \ and\ \bibinfo {author} {\bibfnamefont {Robert~Joseph}\ \bibnamefont {Cava}},\ }\bibfield  {title} {\enquote {\bibinfo {title} {{Effect of electron count and chemical complexity in the {Ta-Nb-Hf-Zr-Ti} high entropy alloy superconductor}},}\ }\href {\doibase 10.1073/pnas.1615926113} {\bibfield  {journal} {\bibinfo  {journal} {Proceedings of the National Academy of Sciences}\ }\textbf {\bibinfo {volume} {113}},\ \bibinfo {pages} {E7144--E7150} (\bibinfo {year} {2016})}\BibitemShut {NoStop}%
\bibitem [{\citenamefont {Guo}\ \emph {et~al.}(2017)\citenamefont {Guo}, \citenamefont {Wang}, \citenamefont {von Rohr}, \citenamefont {Wang}, \citenamefont {Cai}, \citenamefont {Zhou}, \citenamefont {Yang}, \citenamefont {Li}, \citenamefont {Jiang}, \citenamefont {Wu}, \citenamefont {Cava},\ and\ \citenamefont {Sun}}]{guo-pressure}%
  \BibitemOpen
  \bibfield  {author} {\bibinfo {author} {\bibfnamefont {J.}~\bibnamefont {Guo}}, \bibinfo {author} {\bibfnamefont {H.}~\bibnamefont {Wang}}, \bibinfo {author} {\bibfnamefont {Fabian}\ \bibnamefont {von Rohr}}, \bibinfo {author} {\bibfnamefont {Zhe}\ \bibnamefont {Wang}}, \bibinfo {author} {\bibfnamefont {Shu}\ \bibnamefont {Cai}}, \bibinfo {author} {\bibfnamefont {Yazhou}\ \bibnamefont {Zhou}}, \bibinfo {author} {\bibfnamefont {Ke}~\bibnamefont {Yang}}, \bibinfo {author} {\bibfnamefont {Aiguo}\ \bibnamefont {Li}}, \bibinfo {author} {\bibfnamefont {Sheng}\ \bibnamefont {Jiang}}, \bibinfo {author} {\bibfnamefont {Qi}~\bibnamefont {Wu}}, \bibinfo {author} {\bibfnamefont {Robert~J.}\ \bibnamefont {Cava}}, \ and\ \bibinfo {author} {\bibfnamefont {Liling}\ \bibnamefont {Sun}},\ }\bibfield  {title} {\enquote {\bibinfo {title} {{Robust zero resistance in a superconducting high entropy alloy at pressures up to 190 {GP}a}},}\ }\href {\doibase 10.1073/pnas.1716981114} {\bibfield  {journal} {\bibinfo  {journal}
  {Proceedings of the National Academy of Sciences}\ }\textbf {\bibinfo {volume} {114}},\ \bibinfo {pages} {13144--13147} (\bibinfo {year} {2017})}\BibitemShut {NoStop}%
\bibitem [{\citenamefont {Vrtnik}\ \emph {et~al.}(2017)\citenamefont {Vrtnik}, \citenamefont {Koželj}, \citenamefont {Meden}, \citenamefont {Maiti}, \citenamefont {Steurer}, \citenamefont {Feuerbacher},\ and\ \citenamefont {Dolinšek}}]{vrtnik_2017}%
  \BibitemOpen
  \bibfield  {author} {\bibinfo {author} {\bibfnamefont {S.}~\bibnamefont {Vrtnik}}, \bibinfo {author} {\bibfnamefont {P.}~\bibnamefont {Koželj}}, \bibinfo {author} {\bibfnamefont {A.}~\bibnamefont {Meden}}, \bibinfo {author} {\bibfnamefont {S.}~\bibnamefont {Maiti}}, \bibinfo {author} {\bibfnamefont {W.}~\bibnamefont {Steurer}}, \bibinfo {author} {\bibfnamefont {M.}~\bibnamefont {Feuerbacher}}, \ and\ \bibinfo {author} {\bibfnamefont {J.}~\bibnamefont {Dolinšek}},\ }\bibfield  {title} {\enquote {\bibinfo {title} {{Superconductivity in thermally annealed Ta-Nb-Hf-Zr-Ti high-entropy alloys}},}\ }\href {\doibase 10.1016/j.jallcom.2016.11.417} {\bibfield  {journal} {\bibinfo  {journal} {Journal of Alloys and Compounds}\ }\textbf {\bibinfo {volume} {695}},\ \bibinfo {pages} {3530--3540} (\bibinfo {year} {2017})}\BibitemShut {NoStop}%
\bibitem [{\citenamefont {Gabáni}\ \emph {et~al.}(2023)\citenamefont {Gabáni}, \citenamefont {Cedervall}, \citenamefont {Ek}, \citenamefont {Pristáš}, \citenamefont {Orendáč}, \citenamefont {Bačkai}, \citenamefont {Onufriienko}, \citenamefont {Gažo},\ and\ \citenamefont {Flachbart}}]{gabani_2023_tnhzt}%
  \BibitemOpen
  \bibfield  {author} {\bibinfo {author} {\bibfnamefont {Slavomír}\ \bibnamefont {Gabáni}}, \bibinfo {author} {\bibfnamefont {Johan}\ \bibnamefont {Cedervall}}, \bibinfo {author} {\bibfnamefont {Gustav}\ \bibnamefont {Ek}}, \bibinfo {author} {\bibfnamefont {Gabriel}\ \bibnamefont {Pristáš}}, \bibinfo {author} {\bibfnamefont {Matúš}\ \bibnamefont {Orendáč}}, \bibinfo {author} {\bibfnamefont {Július}\ \bibnamefont {Bačkai}}, \bibinfo {author} {\bibfnamefont {Oleksandr}\ \bibnamefont {Onufriienko}}, \bibinfo {author} {\bibfnamefont {Emil}\ \bibnamefont {Gažo}}, \ and\ \bibinfo {author} {\bibfnamefont {Karol}\ \bibnamefont {Flachbart}},\ }\bibfield  {title} {\enquote {\bibinfo {title} {{Search for superconductivity in hydrides of TiZrNb, TiZrNbHf and TiZrNbHfTa equimolar alloys}},}\ }\href {\doibase 10.1016/j.physb.2022.414414} {\bibfield  {journal} {\bibinfo  {journal} {Physica B: Condensed Matter}\ }\textbf {\bibinfo {volume} {648}},\ \bibinfo {pages} {414414} (\bibinfo {year} {2023})}\BibitemShut
  {NoStop}%
\bibitem [{\citenamefont {Krnel}\ \emph {et~al.}(2022)\citenamefont {Krnel}, \citenamefont {Jelen}, \citenamefont {Vrtnik}, \citenamefont {Luzar}, \citenamefont {Gačnik}, \citenamefont {Koželj}, \citenamefont {Wencka}, \citenamefont {Meden}, \citenamefont {Hu}, \citenamefont {Guo},\ and\ \citenamefont {Dolinšek}}]{krnel_2022_tnhzt}%
  \BibitemOpen
  \bibfield  {author} {\bibinfo {author} {\bibfnamefont {Mitja}\ \bibnamefont {Krnel}}, \bibinfo {author} {\bibfnamefont {Andreja}\ \bibnamefont {Jelen}}, \bibinfo {author} {\bibfnamefont {Stanislav}\ \bibnamefont {Vrtnik}}, \bibinfo {author} {\bibfnamefont {Jože}\ \bibnamefont {Luzar}}, \bibinfo {author} {\bibfnamefont {Darja}\ \bibnamefont {Gačnik}}, \bibinfo {author} {\bibfnamefont {Primož}\ \bibnamefont {Koželj}}, \bibinfo {author} {\bibfnamefont {Magdalena}\ \bibnamefont {Wencka}}, \bibinfo {author} {\bibfnamefont {Anton}\ \bibnamefont {Meden}}, \bibinfo {author} {\bibfnamefont {Qiang}\ \bibnamefont {Hu}}, \bibinfo {author} {\bibfnamefont {Sheng}\ \bibnamefont {Guo}}, \ and\ \bibinfo {author} {\bibfnamefont {Janez}\ \bibnamefont {Dolinšek}},\ }\bibfield  {title} {\enquote {\bibinfo {title} {{The Effect of Scandium on the Structure, Microstructure and Superconductivity of Equimolar Sc-Hf-Nb-Ta-Ti-Zr Refractory High-Entropy Alloys}},}\ }\href {\doibase 10.3390/ma15031122} {\bibfield  {journal}
  {\bibinfo  {journal} {Materials}\ }\textbf {\bibinfo {volume} {15}} (\bibinfo {year} {2022}),\ 10.3390/ma15031122}\BibitemShut {NoStop}%
\bibitem [{\citenamefont {{Zeng}}\ \emph {et~al.}(2023)\citenamefont {{Zeng}}, \citenamefont {{Hu}}, \citenamefont {{Boubeche}}, \citenamefont {{Li}}, \citenamefont {{Li}}, \citenamefont {{Yu}}, \citenamefont {{Wang}}, \citenamefont {{Zhang}}, \citenamefont {{Jin}}, \citenamefont {{Yao}},\ and\ \citenamefont {{Luo}}}]{zeng_2023_tnhzt}%
  \BibitemOpen
  \bibfield  {author} {\bibinfo {author} {\bibfnamefont {Lingyong}\ \bibnamefont {{Zeng}}}, \bibinfo {author} {\bibfnamefont {Xunwu}\ \bibnamefont {{Hu}}}, \bibinfo {author} {\bibfnamefont {Mebrouka}\ \bibnamefont {{Boubeche}}}, \bibinfo {author} {\bibfnamefont {Kuan}\ \bibnamefont {{Li}}}, \bibinfo {author} {\bibfnamefont {Longfu}\ \bibnamefont {{Li}}}, \bibinfo {author} {\bibfnamefont {Peifeng}\ \bibnamefont {{Yu}}}, \bibinfo {author} {\bibfnamefont {Kangwang}\ \bibnamefont {{Wang}}}, \bibinfo {author} {\bibfnamefont {Chao}\ \bibnamefont {{Zhang}}}, \bibinfo {author} {\bibfnamefont {Kui}\ \bibnamefont {{Jin}}}, \bibinfo {author} {\bibfnamefont {Dao-Xin}\ \bibnamefont {{Yao}}}, \ and\ \bibinfo {author} {\bibfnamefont {Huixia}\ \bibnamefont {{Luo}}},\ }\bibfield  {title} {\enquote {\bibinfo {title} {{Extremely strong coupling s-wave superconductivity in the medium-entropy alloy TiHfNbTa}},}\ }\href {\doibase 10.1007/s11433-023-2113-6} {\bibfield  {journal} {\bibinfo  {journal} {Science China Physics,
  Mechanics, and Astronomy}\ }\textbf {\bibinfo {volume} {66}},\ \bibinfo {eid} {277412} (\bibinfo {year} {2023})}\BibitemShut {NoStop}%
\bibitem [{\citenamefont {Zhang}\ \emph {et~al.}(2020{\natexlab{a}})\citenamefont {Zhang}, \citenamefont {Peng}, \citenamefont {Li}, \citenamefont {Liu}, \citenamefont {Zhang}, \citenamefont {Wu}, \citenamefont {Yang}, \citenamefont {Greenberg}, \citenamefont {Prakapenka}, \citenamefont {Hui}, \citenamefont {Wang},\ and\ \citenamefont {Yang}}]{zhang_2020_tnhzt}%
  \BibitemOpen
  \bibfield  {author} {\bibinfo {author} {\bibfnamefont {Kai}\ \bibnamefont {Zhang}}, \bibinfo {author} {\bibfnamefont {Shang}\ \bibnamefont {Peng}}, \bibinfo {author} {\bibfnamefont {Nana}\ \bibnamefont {Li}}, \bibinfo {author} {\bibfnamefont {Xuqiang}\ \bibnamefont {Liu}}, \bibinfo {author} {\bibfnamefont {Mingjian}\ \bibnamefont {Zhang}}, \bibinfo {author} {\bibfnamefont {Yi-Dong}\ \bibnamefont {Wu}}, \bibinfo {author} {\bibfnamefont {Yanping}\ \bibnamefont {Yang}}, \bibinfo {author} {\bibfnamefont {Eran}\ \bibnamefont {Greenberg}}, \bibinfo {author} {\bibfnamefont {Vitali~B.}\ \bibnamefont {Prakapenka}}, \bibinfo {author} {\bibfnamefont {Xidong}\ \bibnamefont {Hui}}, \bibinfo {author} {\bibfnamefont {Yandong}\ \bibnamefont {Wang}}, \ and\ \bibinfo {author} {\bibfnamefont {Wenge}\ \bibnamefont {Yang}},\ }\bibfield  {title} {\enquote {\bibinfo {title} {{Tuning to more compressible phase in TiZrHfNb high entropy alloy by pressure}},}\ }\href {\doibase 10.1063/1.5136022} {\bibfield  {journal} {\bibinfo
  {journal} {Applied Physics Letters}\ }\textbf {\bibinfo {volume} {116}},\ \bibinfo {pages} {031901} (\bibinfo {year} {2020}{\natexlab{a}})}\BibitemShut {NoStop}%
\bibitem [{\citenamefont {Hattori}\ \emph {et~al.}(2023)\citenamefont {Hattori}, \citenamefont {Watanabe}, \citenamefont {Nishizaki}, \citenamefont {Hiraoka}, \citenamefont {Kakihara}, \citenamefont {Hoshi}, \citenamefont {Mizuguchi},\ and\ \citenamefont {Kitagawa}}]{hattori_2023_tnhzt}%
  \BibitemOpen
  \bibfield  {author} {\bibinfo {author} {\bibfnamefont {Takuma}\ \bibnamefont {Hattori}}, \bibinfo {author} {\bibfnamefont {Yuto}\ \bibnamefont {Watanabe}}, \bibinfo {author} {\bibfnamefont {Terukazu}\ \bibnamefont {Nishizaki}}, \bibinfo {author} {\bibfnamefont {Koki}\ \bibnamefont {Hiraoka}}, \bibinfo {author} {\bibfnamefont {Masato}\ \bibnamefont {Kakihara}}, \bibinfo {author} {\bibfnamefont {Kazuhisa}\ \bibnamefont {Hoshi}}, \bibinfo {author} {\bibfnamefont {Yoshikazu}\ \bibnamefont {Mizuguchi}}, \ and\ \bibinfo {author} {\bibfnamefont {Jiro}\ \bibnamefont {Kitagawa}},\ }\bibfield  {title} {\enquote {\bibinfo {title} {{Metallurgy, superconductivity, and hardness of a new high-entropy alloy superconductor Ti-Hf-Nb-Ta-Re}},}\ }\href {\doibase 10.1016/j.jalmes.2023.100020} {\bibfield  {journal} {\bibinfo  {journal} {Journal of Alloys and Metallurgical Systems}\ }\textbf {\bibinfo {volume} {3}},\ \bibinfo {pages} {100020} (\bibinfo {year} {2023})}\BibitemShut {NoStop}%
\bibitem [{\citenamefont {Zeng}\ \emph {et~al.}(2023)\citenamefont {Zeng}, \citenamefont {Wang}, \citenamefont {Song}, \citenamefont {Lin}, \citenamefont {Guo}, \citenamefont {Luo}, \citenamefont {Guo}, \citenamefont {Li}, \citenamefont {Yu}, \citenamefont {Zhang}, \citenamefont {Guo}, \citenamefont {Ma}, \citenamefont {Hou},\ and\ \citenamefont {Luo}}]{zeng_2023_tnhzt2}%
  \BibitemOpen
  \bibfield  {author} {\bibinfo {author} {\bibfnamefont {Lingyong}\ \bibnamefont {Zeng}}, \bibinfo {author} {\bibfnamefont {Zequan}\ \bibnamefont {Wang}}, \bibinfo {author} {\bibfnamefont {Jing}\ \bibnamefont {Song}}, \bibinfo {author} {\bibfnamefont {Gaoting}\ \bibnamefont {Lin}}, \bibinfo {author} {\bibfnamefont {Ruixin}\ \bibnamefont {Guo}}, \bibinfo {author} {\bibfnamefont {Si-Chun}\ \bibnamefont {Luo}}, \bibinfo {author} {\bibfnamefont {Shu}\ \bibnamefont {Guo}}, \bibinfo {author} {\bibfnamefont {Kuan}\ \bibnamefont {Li}}, \bibinfo {author} {\bibfnamefont {Peifeng}\ \bibnamefont {Yu}}, \bibinfo {author} {\bibfnamefont {Chao}\ \bibnamefont {Zhang}}, \bibinfo {author} {\bibfnamefont {Wei-Ming}\ \bibnamefont {Guo}}, \bibinfo {author} {\bibfnamefont {Jie}\ \bibnamefont {Ma}}, \bibinfo {author} {\bibfnamefont {Yusheng}\ \bibnamefont {Hou}}, \ and\ \bibinfo {author} {\bibfnamefont {Huixia}\ \bibnamefont {Luo}},\ }\bibfield  {title} {\enquote {\bibinfo {title} {{Discovery of the High-Entropy Carbide Ceramic
  Topological Superconductor Candidate (Ti$_{0.2}$Zr$_{0.2}$Nb$_{0.2}$Hf$_{0.2}$Ta$_{0.2}$)C}},}\ }\href {\doibase https://doi.org/10.1002/adfm.202301929} {\bibfield  {journal} {\bibinfo  {journal} {Advanced Functional Materials}\ }\textbf {\bibinfo {volume} {33}},\ \bibinfo {pages} {2301929} (\bibinfo {year} {2023})},\ \Eprint {http://arxiv.org/abs/https://onlinelibrary.wiley.com/doi/pdf/10.1002/adfm.202301929} {https://onlinelibrary.wiley.com/doi/pdf/10.1002/adfm.202301929} \BibitemShut {NoStop}%
\bibitem [{\citenamefont {Yuan}\ \emph {et~al.}(2018)\citenamefont {Yuan}, \citenamefont {Wu}, \citenamefont {Luo}, \citenamefont {Wang}, \citenamefont {Liang}, \citenamefont {Yang}, \citenamefont {Wang}, \citenamefont {Liu},\ and\ \citenamefont {Lu}}]{yuan_2018_tnhzt}%
  \BibitemOpen
  \bibfield  {author} {\bibinfo {author} {\bibfnamefont {Yuan}\ \bibnamefont {Yuan}}, \bibinfo {author} {\bibfnamefont {Yuan}\ \bibnamefont {Wu}}, \bibinfo {author} {\bibfnamefont {Huiqian}\ \bibnamefont {Luo}}, \bibinfo {author} {\bibfnamefont {Zhaosheng}\ \bibnamefont {Wang}}, \bibinfo {author} {\bibfnamefont {Xue}\ \bibnamefont {Liang}}, \bibinfo {author} {\bibfnamefont {Zhi}\ \bibnamefont {Yang}}, \bibinfo {author} {\bibfnamefont {Hui}\ \bibnamefont {Wang}}, \bibinfo {author} {\bibfnamefont {Xiongjun}\ \bibnamefont {Liu}}, \ and\ \bibinfo {author} {\bibfnamefont {Zhaoping}\ \bibnamefont {Lu}},\ }\bibfield  {title} {\enquote {\bibinfo {title} {{Superconducting Ti$_{15}$Zr$_{15}$Nb$_{35}$Ta$_{35}$ High-Entropy Alloy With Intermediate Electron-Phonon Coupling}},}\ }\href {\doibase 10.3389/fmats.2018.00072} {\bibfield  {journal} {\bibinfo  {journal} {Frontiers in Materials}\ }\textbf {\bibinfo {volume} {5}} (\bibinfo {year} {2018}),\ 10.3389/fmats.2018.00072}\BibitemShut {NoStop}%
\bibitem [{\citenamefont {Kim}\ \emph {et~al.}(2022)\citenamefont {Kim}, \citenamefont {Hidayati}, \citenamefont {Jung}, \citenamefont {Salawu}, \citenamefont {Kim}, \citenamefont {Yun},\ and\ \citenamefont {Rhyee}}]{kim_2022_tnhzt}%
  \BibitemOpen
  \bibfield  {author} {\bibinfo {author} {\bibfnamefont {Jin~Hee}\ \bibnamefont {Kim}}, \bibinfo {author} {\bibfnamefont {Rahmatul}\ \bibnamefont {Hidayati}}, \bibinfo {author} {\bibfnamefont {Soon-Gil}\ \bibnamefont {Jung}}, \bibinfo {author} {\bibfnamefont {Yusuff~Adeyemi}\ \bibnamefont {Salawu}}, \bibinfo {author} {\bibfnamefont {Heon-Jung}\ \bibnamefont {Kim}}, \bibinfo {author} {\bibfnamefont {Jae~Hyun}\ \bibnamefont {Yun}}, \ and\ \bibinfo {author} {\bibfnamefont {Jong-Soo}\ \bibnamefont {Rhyee}},\ }\bibfield  {title} {\enquote {\bibinfo {title} {{Enhancement of critical current density and strong vortex pinning in high entropy alloy superconductor ${\mathrm{Ta}}_{1/6}{\mathrm{Nb}}_{2/6}{\mathrm{Hf}}_{1/6}{\mathrm{Zr}}_{1/6}{\mathrm{Ti}}_{1/6}$ synthesized by spark plasma sintering}},}\ }\href {\doibase 10.1016/j.actamat.2022.117971} {\bibfield  {journal} {\bibinfo  {journal} {Acta Materialia}\ }\textbf {\bibinfo {volume} {232}},\ \bibinfo {pages} {117971} (\bibinfo {year} {2022})}\BibitemShut {NoStop}%
\bibitem [{\citenamefont {Hung}\ \emph {et~al.}(2020)\citenamefont {Hung}, \citenamefont {Kawasaki}, \citenamefont {Han}, \citenamefont {Lábár},\ and\ \citenamefont {Gubicza}}]{hung_2020_tnhzt}%
  \BibitemOpen
  \bibfield  {author} {\bibinfo {author} {\bibfnamefont {Pham~Tran}\ \bibnamefont {Hung}}, \bibinfo {author} {\bibfnamefont {Megumi}\ \bibnamefont {Kawasaki}}, \bibinfo {author} {\bibfnamefont {Jae-Kyung}\ \bibnamefont {Han}}, \bibinfo {author} {\bibfnamefont {János~L.}\ \bibnamefont {Lábár}}, \ and\ \bibinfo {author} {\bibfnamefont {Jenő}\ \bibnamefont {Gubicza}},\ }\bibfield  {title} {\enquote {\bibinfo {title} {{Thermal stability of a nanocrystalline HfNbTiZr multi-principal element alloy processed by high-pressure torsion}},}\ }\href {\doibase 10.1016/j.matchar.2020.110550} {\bibfield  {journal} {\bibinfo  {journal} {Materials Characterization}\ }\textbf {\bibinfo {volume} {168}},\ \bibinfo {pages} {110550} (\bibinfo {year} {2020})}\BibitemShut {NoStop}%
\bibitem [{\citenamefont {Kim}\ \emph {et~al.}(2020)\citenamefont {Kim}, \citenamefont {Lee}, \citenamefont {Yun}, \citenamefont {Rawat}, \citenamefont {Jung}, \citenamefont {Choi}, \citenamefont {You}, \citenamefont {Kim},\ and\ \citenamefont {Rhyee}}]{kim_2020_tnhzt}%
  \BibitemOpen
  \bibfield  {author} {\bibinfo {author} {\bibfnamefont {Gareoung}\ \bibnamefont {Kim}}, \bibinfo {author} {\bibfnamefont {Min-Ho}\ \bibnamefont {Lee}}, \bibinfo {author} {\bibfnamefont {Jae~Hyun}\ \bibnamefont {Yun}}, \bibinfo {author} {\bibfnamefont {Pooja}\ \bibnamefont {Rawat}}, \bibinfo {author} {\bibfnamefont {Soon-Gil}\ \bibnamefont {Jung}}, \bibinfo {author} {\bibfnamefont {Woongjin}\ \bibnamefont {Choi}}, \bibinfo {author} {\bibfnamefont {Tae-Soo}\ \bibnamefont {You}}, \bibinfo {author} {\bibfnamefont {Sung~Jin}\ \bibnamefont {Kim}}, \ and\ \bibinfo {author} {\bibfnamefont {Jong-Soo}\ \bibnamefont {Rhyee}},\ }\bibfield  {title} {\enquote {\bibinfo {title} {{Strongly correlated and strongly coupled s-wave superconductivity of the high entropy alloy ${\mathrm{Ta}}_{1/6}{\mathrm{Nb}}_{2/6}{\mathrm{Hf}}_{1/6}{\mathrm{Zr}}_{1/6}{\mathrm{Ti}}_{1/6}$ compound}},}\ }\href {\doibase https://doi.org/10.1016/j.actamat.2020.01.007} {\bibfield  {journal} {\bibinfo  {journal} {Acta Materialia}\ }\textbf {\bibinfo
  {volume} {186}},\ \bibinfo {pages} {250--256} (\bibinfo {year} {2020})}\BibitemShut {NoStop}%
\bibitem [{\citenamefont {Prist\'a\ifmmode~\check{s}\else \v{s}\fi{}}\ \emph {et~al.}(2023)\citenamefont {Prist\'a\ifmmode~\check{s}\else \v{s}\fi{}}, \citenamefont {Ba\ifmmode~\check{c}\else \v{c}\fi{}kai}, \citenamefont {Orend\'a\ifmmode~\check{c}\else \v{c}\fi{}}, \citenamefont {Gab\'ani}, \citenamefont {Ko\ifmmode~\check{s}\else \v{s}\fi{}uth}, \citenamefont {Kuzmiak}, \citenamefont {Szab\'o}, \citenamefont {Ga\ifmmode~\check{z}\else \v{z}\fi{}o}, \citenamefont {Franz}, \citenamefont {Hirn}, \citenamefont {Gruber}, \citenamefont {Mitterer}, \citenamefont {Vorobiov},\ and\ \citenamefont {Flachbart}}]{prista_2023_tnhzt}%
  \BibitemOpen
  \bibfield  {author} {\bibinfo {author} {\bibfnamefont {Gabriel}\ \bibnamefont {Prist\'a\ifmmode~\check{s}\else \v{s}\fi{}}}, \bibinfo {author} {\bibfnamefont {J\'ulius}\ \bibnamefont {Ba\ifmmode~\check{c}\else \v{c}\fi{}kai}}, \bibinfo {author} {\bibfnamefont {Mat\'u\ifmmode \check{s}\else~\v{s}\fi{}}\ \bibnamefont {Orend\'a\ifmmode~\check{c}\else \v{c}\fi{}}}, \bibinfo {author} {\bibfnamefont {Slavom\'{\i}r}\ \bibnamefont {Gab\'ani}}, \bibinfo {author} {\bibfnamefont {Filip}\ \bibnamefont {Ko\ifmmode~\check{s}\else \v{s}\fi{}uth}}, \bibinfo {author} {\bibfnamefont {Marek}\ \bibnamefont {Kuzmiak}}, \bibinfo {author} {\bibfnamefont {Pavol}\ \bibnamefont {Szab\'o}}, \bibinfo {author} {\bibfnamefont {Emil}\ \bibnamefont {Ga\ifmmode~\check{z}\else \v{z}\fi{}o}}, \bibinfo {author} {\bibfnamefont {Robert}\ \bibnamefont {Franz}}, \bibinfo {author} {\bibfnamefont {Sabrina}\ \bibnamefont {Hirn}}, \bibinfo {author} {\bibfnamefont {Georg~C.}\ \bibnamefont {Gruber}}, \bibinfo {author} {\bibfnamefont {Christian}\
  \bibnamefont {Mitterer}}, \bibinfo {author} {\bibfnamefont {Serhii}\ \bibnamefont {Vorobiov}}, \ and\ \bibinfo {author} {\bibfnamefont {Karol}\ \bibnamefont {Flachbart}},\ }\bibfield  {title} {\enquote {\bibinfo {title} {{Superconductivity in medium- and high-entropy alloy thin films: Impact of thickness and external pressure}},}\ }\href {\doibase 10.1103/PhysRevB.107.024505} {\bibfield  {journal} {\bibinfo  {journal} {Phys. Rev. B}\ }\textbf {\bibinfo {volume} {107}},\ \bibinfo {pages} {024505} (\bibinfo {year} {2023})}\BibitemShut {NoStop}%
\bibitem [{\citenamefont {Zhang}\ \emph {et~al.}(2020{\natexlab{b}})\citenamefont {Zhang}, \citenamefont {Winter}, \citenamefont {Witteveen}, \citenamefont {Moehl}, \citenamefont {Xiao}, \citenamefont {Krogh}, \citenamefont {Schilling},\ and\ \citenamefont {von Rohr}}]{zhang_2020_tnhzt2}%
  \BibitemOpen
  \bibfield  {author} {\bibinfo {author} {\bibfnamefont {Xiaofu}\ \bibnamefont {Zhang}}, \bibinfo {author} {\bibfnamefont {Natascha}\ \bibnamefont {Winter}}, \bibinfo {author} {\bibfnamefont {Catherine}\ \bibnamefont {Witteveen}}, \bibinfo {author} {\bibfnamefont {Thomas}\ \bibnamefont {Moehl}}, \bibinfo {author} {\bibfnamefont {Yuan}\ \bibnamefont {Xiao}}, \bibinfo {author} {\bibfnamefont {Fabio}\ \bibnamefont {Krogh}}, \bibinfo {author} {\bibfnamefont {Andreas}\ \bibnamefont {Schilling}}, \ and\ \bibinfo {author} {\bibfnamefont {Fabian~O.}\ \bibnamefont {von Rohr}},\ }\bibfield  {title} {\enquote {\bibinfo {title} {{Preparation and characterization of high-entropy alloy ${(\mathrm{TaNb})}_{1\ensuremath{-}x}{(\mathrm{ZrHfTi})}_{x}$ superconducting films}},}\ }\href {\doibase 10.1103/PhysRevResearch.2.013375} {\bibfield  {journal} {\bibinfo  {journal} {Phys. Rev. Res.}\ }\textbf {\bibinfo {volume} {2}},\ \bibinfo {pages} {013375} (\bibinfo {year} {2020}{\natexlab{b}})}\BibitemShut {NoStop}%
\bibitem [{\citenamefont {Hong}\ \emph {et~al.}(2022)\citenamefont {Hong}, \citenamefont {Jang}, \citenamefont {Jung}, \citenamefont {Han}, \citenamefont {Kim}, \citenamefont {Hidayati}, \citenamefont {Rhyee},\ and\ \citenamefont {Park}}]{hong_2022_tnhzt}%
  \BibitemOpen
  \bibfield  {author} {\bibinfo {author} {\bibfnamefont {Vuong Thi~Anh}\ \bibnamefont {Hong}}, \bibinfo {author} {\bibfnamefont {Harim}\ \bibnamefont {Jang}}, \bibinfo {author} {\bibfnamefont {Soon-Gil}\ \bibnamefont {Jung}}, \bibinfo {author} {\bibfnamefont {Yoonseok}\ \bibnamefont {Han}}, \bibinfo {author} {\bibfnamefont {Jin~Hee}\ \bibnamefont {Kim}}, \bibinfo {author} {\bibfnamefont {Rahmatul}\ \bibnamefont {Hidayati}}, \bibinfo {author} {\bibfnamefont {Jong-Soo}\ \bibnamefont {Rhyee}}, \ and\ \bibinfo {author} {\bibfnamefont {Tuson}\ \bibnamefont {Park}},\ }\bibfield  {title} {\enquote {\bibinfo {title} {{Probing superconducting gap of the high-entropy alloy ${\mathrm{Ta}}_{1/6}{\mathrm{Nb}}_{2/6}{\mathrm{Hf}}_{1/6}{\mathrm{Zr}}_{1/6}{\mathrm{Ti}}_{1/6}$ via Andreev reflection spectroscopy}},}\ }\href {\doibase 10.1103/PhysRevB.106.024504} {\bibfield  {journal} {\bibinfo  {journal} {Phys. Rev. B}\ }\textbf {\bibinfo {volume} {106}},\ \bibinfo {pages} {024504} (\bibinfo {year} {2022})}\BibitemShut
  {NoStop}%
\bibitem [{\citenamefont {Leung}\ \emph {et~al.}(2022)\citenamefont {Leung}, \citenamefont {Zhang}, \citenamefont {von Rohr}, \citenamefont {Lortz},\ and\ \citenamefont {J{\"a}ck}}]{leung_2022_tnhzt}%
  \BibitemOpen
  \bibfield  {author} {\bibinfo {author} {\bibfnamefont {Casey~KW}\ \bibnamefont {Leung}}, \bibinfo {author} {\bibfnamefont {Xiaofu}\ \bibnamefont {Zhang}}, \bibinfo {author} {\bibfnamefont {Fabian}\ \bibnamefont {von Rohr}}, \bibinfo {author} {\bibfnamefont {Rolf}\ \bibnamefont {Lortz}}, \ and\ \bibinfo {author} {\bibfnamefont {Berthold}\ \bibnamefont {J{\"a}ck}},\ }\bibfield  {title} {\enquote {\bibinfo {title} {Evidence for isotropic s-wave superconductivity in high-entropy alloys},}\ }\href@noop {} {\bibfield  {journal} {\bibinfo  {journal} {Scientific reports}\ }\textbf {\bibinfo {volume} {12}},\ \bibinfo {pages} {12773} (\bibinfo {year} {2022})}\BibitemShut {NoStop}%
\bibitem [{\citenamefont {Heidelmann}\ \emph {et~al.}(2016)\citenamefont {Heidelmann}, \citenamefont {Feuerbacher}, \citenamefont {Ma},\ and\ \citenamefont {Grabowski}}]{heidelmann_2016_tnhzt}%
  \BibitemOpen
  \bibfield  {author} {\bibinfo {author} {\bibfnamefont {Markus}\ \bibnamefont {Heidelmann}}, \bibinfo {author} {\bibfnamefont {Michael}\ \bibnamefont {Feuerbacher}}, \bibinfo {author} {\bibfnamefont {Duancheng}\ \bibnamefont {Ma}}, \ and\ \bibinfo {author} {\bibfnamefont {Blazej}\ \bibnamefont {Grabowski}},\ }\bibfield  {title} {\enquote {\bibinfo {title} {{Structural anomaly in the high-entropy alloy ZrNbTiTaHf}},}\ }\href {\doibase https://doi.org/10.1016/j.intermet.2015.08.013} {\bibfield  {journal} {\bibinfo  {journal} {Intermetallics}\ }\textbf {\bibinfo {volume} {68}},\ \bibinfo {pages} {11--15} (\bibinfo {year} {2016})}\BibitemShut {NoStop}%
\bibitem [{\citenamefont {von Rohr}\ and\ \citenamefont {Cava}(2018)}]{rohr_2018_tnhzt}%
  \BibitemOpen
  \bibfield  {author} {\bibinfo {author} {\bibfnamefont {Fabian~O.}\ \bibnamefont {von Rohr}}\ and\ \bibinfo {author} {\bibfnamefont {Robert~J.}\ \bibnamefont {Cava}},\ }\bibfield  {title} {\enquote {\bibinfo {title} {{Isoelectronic substitutions and aluminium alloying in the Ta-Nb-Hf-Zr-Ti high-entropy alloy superconductor}},}\ }\href {\doibase 10.1103/PhysRevMaterials.2.034801} {\bibfield  {journal} {\bibinfo  {journal} {Phys. Rev. Mater.}\ }\textbf {\bibinfo {volume} {2}},\ \bibinfo {pages} {034801} (\bibinfo {year} {2018})}\BibitemShut {NoStop}%
\bibitem [{\citenamefont {Jasiewicz}\ \emph {et~al.}(2019)\citenamefont {Jasiewicz}, \citenamefont {Wiendlocha}, \citenamefont {G\'ornicka}, \citenamefont {Gofryk}, \citenamefont {Gazda}, \citenamefont {Klimczuk},\ and\ \citenamefont {Tobola}}]{jasiewicz_2019}%
  \BibitemOpen
  \bibfield  {author} {\bibinfo {author} {\bibfnamefont {K.}~\bibnamefont {Jasiewicz}}, \bibinfo {author} {\bibfnamefont {B.}~\bibnamefont {Wiendlocha}}, \bibinfo {author} {\bibfnamefont {K.}~\bibnamefont {G\'ornicka}}, \bibinfo {author} {\bibfnamefont {K.}~\bibnamefont {Gofryk}}, \bibinfo {author} {\bibfnamefont {M.}~\bibnamefont {Gazda}}, \bibinfo {author} {\bibfnamefont {T.}~\bibnamefont {Klimczuk}}, \ and\ \bibinfo {author} {\bibfnamefont {J.}~\bibnamefont {Tobola}},\ }\bibfield  {title} {\enquote {\bibinfo {title} {{Pressure effects on the electronic structure and superconductivity of (TaNb)$_{0.67}$(HfZrTi)$_{0.33}$ high entropy alloy}},}\ }\href {\doibase 10.1103/PhysRevB.100.184503} {\bibfield  {journal} {\bibinfo  {journal} {Phys. Rev. B}\ }\textbf {\bibinfo {volume} {100}},\ \bibinfo {pages} {184503} (\bibinfo {year} {2019})}\BibitemShut {NoStop}%
\bibitem [{\citenamefont {Huang}\ \emph {et~al.}(2020)\citenamefont {Huang}, \citenamefont {Guo}, \citenamefont {Zhang}, \citenamefont {Stolze}, \citenamefont {Cai}, \citenamefont {Liu}, \citenamefont {Weng}, \citenamefont {Lu}, \citenamefont {Wu}, \citenamefont {Xiang}, \citenamefont {Cava},\ and\ \citenamefont {Sun}}]{cava2020}%
  \BibitemOpen
  \bibfield  {author} {\bibinfo {author} {\bibfnamefont {Cheng}\ \bibnamefont {Huang}}, \bibinfo {author} {\bibfnamefont {Jing}\ \bibnamefont {Guo}}, \bibinfo {author} {\bibfnamefont {Jianfeng}\ \bibnamefont {Zhang}}, \bibinfo {author} {\bibfnamefont {Karoline}\ \bibnamefont {Stolze}}, \bibinfo {author} {\bibfnamefont {Shu}\ \bibnamefont {Cai}}, \bibinfo {author} {\bibfnamefont {Kai}\ \bibnamefont {Liu}}, \bibinfo {author} {\bibfnamefont {Hongming}\ \bibnamefont {Weng}}, \bibinfo {author} {\bibfnamefont {Zhongyi}\ \bibnamefont {Lu}}, \bibinfo {author} {\bibfnamefont {Qi}~\bibnamefont {Wu}}, \bibinfo {author} {\bibfnamefont {Tao}\ \bibnamefont {Xiang}}, \bibinfo {author} {\bibfnamefont {Robert~J.}\ \bibnamefont {Cava}}, \ and\ \bibinfo {author} {\bibfnamefont {Liling}\ \bibnamefont {Sun}},\ }\bibfield  {title} {\enquote {\bibinfo {title} {{RSAVS superconductors: Materials with a superconducting state that is robust against large volume shrinkage}},}\ }\href {\doibase 10.1103/PhysRevMaterials.4.071801}
  {\bibfield  {journal} {\bibinfo  {journal} {Phys. Rev. Mater.}\ }\textbf {\bibinfo {volume} {4}},\ \bibinfo {pages} {071801} (\bibinfo {year} {2020})}\BibitemShut {NoStop}%
\bibitem [{\citenamefont {Bansil}\ \emph {et~al.}(1999)\citenamefont {Bansil}, \citenamefont {Kaprzyk}, \citenamefont {Mijnarends},\ and\ \citenamefont {Tobo\l{}a}}]{bansil_1999}%
  \BibitemOpen
  \bibfield  {author} {\bibinfo {author} {\bibfnamefont {A.}~\bibnamefont {Bansil}}, \bibinfo {author} {\bibfnamefont {S.}~\bibnamefont {Kaprzyk}}, \bibinfo {author} {\bibfnamefont {P.~E.}\ \bibnamefont {Mijnarends}}, \ and\ \bibinfo {author} {\bibfnamefont {J.}~\bibnamefont {Tobo\l{}a}},\ }\bibfield  {title} {\enquote {\bibinfo {title} {{Electronic structure and magnetism of {Fe}$_{3-x}${V}$_x${X} {(X = Si, Ga, and Al)} alloys by the KKR-CPA method}},}\ }\href {\doibase 10.1103/PhysRevB.60.13396} {\bibfield  {journal} {\bibinfo  {journal} {Phys. Rev. B}\ }\textbf {\bibinfo {volume} {60}},\ \bibinfo {pages} {13396--13412} (\bibinfo {year} {1999})}\BibitemShut {NoStop}%
\bibitem [{\citenamefont {Stopa}\ \emph {et~al.}(2004)\citenamefont {Stopa}, \citenamefont {Kaprzyk},\ and\ \citenamefont {Tobo{\"l}a}}]{stopa_2004}%
  \BibitemOpen
  \bibfield  {author} {\bibinfo {author} {\bibfnamefont {T.}~\bibnamefont {Stopa}}, \bibinfo {author} {\bibfnamefont {S.}~\bibnamefont {Kaprzyk}}, \ and\ \bibinfo {author} {\bibfnamefont {J}~\bibnamefont {Tobo{\"l}a}},\ }\bibfield  {title} {\enquote {\bibinfo {title} {{Linear aspects of the {Korringa{\textendash}Kohn{\textendash}Rostoker} formalism}},}\ }\href {\doibase 10.1088/0953-8984/16/28/012} {\bibfield  {journal} {\bibinfo  {journal} {Journal of Physics: Condensed Matter}\ }\textbf {\bibinfo {volume} {16}},\ \bibinfo {pages} {4921--4933} (\bibinfo {year} {2004})}\BibitemShut {NoStop}%
\bibitem [{\citenamefont {Lifshitz}(1960)}]{lifshitz_1960}%
  \BibitemOpen
  \bibfield  {author} {\bibinfo {author} {\bibfnamefont {I.M.}\ \bibnamefont {Lifshitz}},\ }\bibfield  {title} {\enquote {\bibinfo {title} {Anomalies of electron characteristics in the high pressure region},}\ }\href@noop {} {\bibfield  {journal} {\bibinfo  {journal} {Zhur. Eksptl'. i Teoret. Fiz.}\ }\textbf {\bibinfo {volume} {38}} (\bibinfo {year} {1960})}\BibitemShut {NoStop}%
\bibitem [{\citenamefont {Struzhkin}\ \emph {et~al.}(1997)\citenamefont {Struzhkin}, \citenamefont {Timofeev}, \citenamefont {Hemley},\ and\ \citenamefont {Mao}}]{struzhkin_1997}%
  \BibitemOpen
  \bibfield  {author} {\bibinfo {author} {\bibfnamefont {Viktor~V.}\ \bibnamefont {Struzhkin}}, \bibinfo {author} {\bibfnamefont {Yuri~A.}\ \bibnamefont {Timofeev}}, \bibinfo {author} {\bibfnamefont {Russell~J.}\ \bibnamefont {Hemley}}, \ and\ \bibinfo {author} {\bibfnamefont {Ho-kwang}\ \bibnamefont {Mao}},\ }\bibfield  {title} {\enquote {\bibinfo {title} {{Superconducting {T$_c$} and Electron-Phonon Coupling in {Nb} to 132 {GP}a: Magnetic Susceptibility at Megabar Pressures}},}\ }\href {\doibase 10.1103/PhysRevLett.79.4262} {\bibfield  {journal} {\bibinfo  {journal} {Phys. Rev. Lett.}\ }\textbf {\bibinfo {volume} {79}},\ \bibinfo {pages} {4262--4265} (\bibinfo {year} {1997})}\BibitemShut {NoStop}%
\bibitem [{\citenamefont {Jasiewicz}\ \emph {et~al.}(2016)\citenamefont {Jasiewicz}, \citenamefont {Wiendlocha}, \citenamefont {Korbe{\'n}}, \citenamefont {Kaprzyk},\ and\ \citenamefont {Tobola}}]{jasiewicz_2016}%
  \BibitemOpen
  \bibfield  {author} {\bibinfo {author} {\bibfnamefont {K.}~\bibnamefont {Jasiewicz}}, \bibinfo {author} {\bibfnamefont {B.}~\bibnamefont {Wiendlocha}}, \bibinfo {author} {\bibfnamefont {P.}~\bibnamefont {Korbe{\'n}}}, \bibinfo {author} {\bibfnamefont {S.}~\bibnamefont {Kaprzyk}}, \ and\ \bibinfo {author} {\bibfnamefont {J.}~\bibnamefont {Tobola}},\ }\bibfield  {title} {\enquote {\bibinfo {title} {{Superconductivity of Ta$_{34}$Nb$_{33}$Hf$_8$Zr$_{14}$Ti$_{11}$ high entropy alloy from first principles calculations}},}\ }\href {\doibase 10.1002/pssr.201600056} {\bibfield  {journal} {\bibinfo  {journal} {Physica Status Solidi (RRL) - Rapid Research Letters}\ }\textbf {\bibinfo {volume} {10}},\ \bibinfo {pages} {415--419} (\bibinfo {year} {2016})}\BibitemShut {NoStop}%
\bibitem [{\citenamefont {Butler}(1985)}]{butler_1985}%
  \BibitemOpen
  \bibfield  {author} {\bibinfo {author} {\bibfnamefont {W.~H.}\ \bibnamefont {Butler}},\ }\bibfield  {title} {\enquote {\bibinfo {title} {{Theory of electronic transport in random alloys: Korringa-Kohn-Rostoker coherent-potential approximation}},}\ }\href {\doibase 10.1103/PhysRevB.31.3260} {\bibfield  {journal} {\bibinfo  {journal} {Phys. Rev. B}\ }\textbf {\bibinfo {volume} {31}},\ \bibinfo {pages} {3260--3277} (\bibinfo {year} {1985})}\BibitemShut {NoStop}%
\bibitem [{\citenamefont {Wiendlocha}\ \emph {et~al.}(2021)\citenamefont {Wiendlocha}, \citenamefont {Misra}, \citenamefont {Dauscher}, \citenamefont {Lenoir},\ and\ \citenamefont {Candolfi}}]{wiendlocha_2021}%
  \BibitemOpen
  \bibfield  {author} {\bibinfo {author} {\bibfnamefont {Bartlomiej}\ \bibnamefont {Wiendlocha}}, \bibinfo {author} {\bibfnamefont {Shantanu}\ \bibnamefont {Misra}}, \bibinfo {author} {\bibfnamefont {Anne}\ \bibnamefont {Dauscher}}, \bibinfo {author} {\bibfnamefont {Bertrand}\ \bibnamefont {Lenoir}}, \ and\ \bibinfo {author} {\bibfnamefont {Christophe}\ \bibnamefont {Candolfi}},\ }\bibfield  {title} {\enquote {\bibinfo {title} {Residual resistivity as an independent indicator of resonant levels in semiconductors},}\ }\href {\doibase 10.1039/D1MH00416F} {\bibfield  {journal} {\bibinfo  {journal} {Mater. Horiz.}\ }\textbf {\bibinfo {volume} {8}},\ \bibinfo {pages} {1735--1743} (\bibinfo {year} {2021})}\BibitemShut {NoStop}%
\bibitem [{\citenamefont {Gutowska}\ \emph {et~al.}(2023)\citenamefont {Gutowska}, \citenamefont {Kawala},\ and\ \citenamefont {Wiendlocha}}]{gutowska2023}%
  \BibitemOpen
  \bibfield  {author} {\bibinfo {author} {\bibfnamefont {Sylwia}\ \bibnamefont {Gutowska}}, \bibinfo {author} {\bibfnamefont {Alicja}\ \bibnamefont {Kawala}}, \ and\ \bibinfo {author} {\bibfnamefont {Bartlomiej}\ \bibnamefont {Wiendlocha}},\ }\bibfield  {title} {\enquote {\bibinfo {title} {{Superconductivity near the Mott-Ioffe-Regel limit in the high-entropy alloy superconductor ${(\mathrm{ScZrNb})}_{1\ensuremath{-}x}{(\mathrm{RhPd})}_{x}$ with a CsCl-type lattice}},}\ }\href {\doibase 10.1103/PhysRevB.108.064507} {\bibfield  {journal} {\bibinfo  {journal} {Phys. Rev. B}\ }\textbf {\bibinfo {volume} {108}},\ \bibinfo {pages} {064507} (\bibinfo {year} {2023})}\BibitemShut {NoStop}%
\bibitem [{sup()}]{suppl}%
  \BibitemOpen
  \href@noop {} {}\bibinfo {note} {{See Supplemental Material for Table I with the atomic positions of all atoms in all 12 undistorted supercells. Figures S1-S12 with the total and partial densities of states of the 12 supercells before and after relaxation and projections of the distorted supercells on the $xy$ plane. }}\BibitemShut {NoStop}%
\bibitem [{\citenamefont {Blaha}\ \emph {et~al.}(2020)\citenamefont {Blaha}, \citenamefont {Schwarz}, \citenamefont {Tran}, \citenamefont {Laskowski}, \citenamefont {Madsen},\ and\ \citenamefont {Marks}}]{wien2k2020}%
  \BibitemOpen
  \bibfield  {author} {\bibinfo {author} {\bibfnamefont {Peter}\ \bibnamefont {Blaha}}, \bibinfo {author} {\bibfnamefont {Karlheinz}\ \bibnamefont {Schwarz}}, \bibinfo {author} {\bibfnamefont {Fabien}\ \bibnamefont {Tran}}, \bibinfo {author} {\bibfnamefont {Robert}\ \bibnamefont {Laskowski}}, \bibinfo {author} {\bibfnamefont {Georg K.~H.}\ \bibnamefont {Madsen}}, \ and\ \bibinfo {author} {\bibfnamefont {Laurence~D.}\ \bibnamefont {Marks}},\ }\bibfield  {title} {\enquote {\bibinfo {title} {{WIEN2k: An APW+lo program for calculating the properties of solids}},}\ }\href {\doibase 10.1063/1.5143061} {\bibfield  {journal} {\bibinfo  {journal} {The Journal of Chemical Physics}\ }\textbf {\bibinfo {volume} {152}},\ \bibinfo {pages} {074101} (\bibinfo {year} {2020})}\BibitemShut {NoStop}%
\bibitem [{\citenamefont {Blaha}\ \emph {et~al.}(2018)\citenamefont {Blaha}, \citenamefont {Schwarz}, \citenamefont {Madsen}, \citenamefont {Kvasnicka}, \citenamefont {Luitz}, \citenamefont {Laskowski}, \citenamefont {Tran},\ and\ \citenamefont {Marks}}]{wien2k}%
  \BibitemOpen
  \bibfield  {author} {\bibinfo {author} {\bibfnamefont {P.}~\bibnamefont {Blaha}}, \bibinfo {author} {\bibfnamefont {K.}~\bibnamefont {Schwarz}}, \bibinfo {author} {\bibfnamefont {G.K.H.}\ \bibnamefont {Madsen}}, \bibinfo {author} {\bibfnamefont {D.}~\bibnamefont {Kvasnicka}}, \bibinfo {author} {\bibfnamefont {J.}~\bibnamefont {Luitz}}, \bibinfo {author} {\bibfnamefont {R.}~\bibnamefont {Laskowski}}, \bibinfo {author} {\bibfnamefont {F.}~\bibnamefont {Tran}}, \ and\ \bibinfo {author} {\bibfnamefont {L.D.}\ \bibnamefont {Marks}},\ }\href@noop {} {\emph {\bibinfo {title} {WIEN2k, An Augmented Plane Wave + Local Orbitals Program for Calculating Crystal Properties (Karlheinz Schwarz, Vienna University of Technology, Austria)}}}\ (\bibinfo {year} {2018})\BibitemShut {NoStop}%
\bibitem [{\citenamefont {Perdew}\ and\ \citenamefont {Wang}(1992)}]{perdew_1992}%
  \BibitemOpen
  \bibfield  {author} {\bibinfo {author} {\bibfnamefont {John~P.}\ \bibnamefont {Perdew}}\ and\ \bibinfo {author} {\bibfnamefont {Yue}\ \bibnamefont {Wang}},\ }\bibfield  {title} {\enquote {\bibinfo {title} {Accurate and simple analytic representation of the electron-gas correlation energy},}\ }\href {\doibase 10.1103/PhysRevB.45.13244} {\bibfield  {journal} {\bibinfo  {journal} {Phys. Rev. B}\ }\textbf {\bibinfo {volume} {45}},\ \bibinfo {pages} {13244--13249} (\bibinfo {year} {1992})}\BibitemShut {NoStop}%
\bibitem [{\citenamefont {Kaprzyk}\ and\ \citenamefont {Bansil}(1990)}]{kaprzyk_1990}%
  \BibitemOpen
  \bibfield  {author} {\bibinfo {author} {\bibfnamefont {S.}~\bibnamefont {Kaprzyk}}\ and\ \bibinfo {author} {\bibfnamefont {A.}~\bibnamefont {Bansil}},\ }\bibfield  {title} {\enquote {\bibinfo {title} {{Green's function and a generalized {Lloyd} formula for the density of states in disordered muffin-tin alloys}},}\ }\href {\doibase 10.1103/PhysRevB.42.7358} {\bibfield  {journal} {\bibinfo  {journal} {Phys. Rev. B}\ }\textbf {\bibinfo {volume} {42}},\ \bibinfo {pages} {7358--7362} (\bibinfo {year} {1990})}\BibitemShut {NoStop}%
\bibitem [{\citenamefont {Gaspari}\ and\ \citenamefont {Gyorffy}(1972)}]{gaspari_1972}%
  \BibitemOpen
  \bibfield  {author} {\bibinfo {author} {\bibfnamefont {G.~D.}\ \bibnamefont {Gaspari}}\ and\ \bibinfo {author} {\bibfnamefont {B.~L.}\ \bibnamefont {Gyorffy}},\ }\bibfield  {title} {\enquote {\bibinfo {title} {{Electron-Phonon Interactions, $d$ Resonances, and Superconductivity in Transition Metals}},}\ }\href {\doibase 10.1103/PhysRevLett.28.801} {\bibfield  {journal} {\bibinfo  {journal} {Phys. Rev. Lett.}\ }\textbf {\bibinfo {volume} {28}},\ \bibinfo {pages} {801--805} (\bibinfo {year} {1972})}\BibitemShut {NoStop}%
\bibitem [{\citenamefont {Gomersall}\ and\ \citenamefont {Gyorffy}(1974)}]{gomersall_1974}%
  \BibitemOpen
  \bibfield  {author} {\bibinfo {author} {\bibfnamefont {I~R}\ \bibnamefont {Gomersall}}\ and\ \bibinfo {author} {\bibfnamefont {B~L}\ \bibnamefont {Gyorffy}},\ }\bibfield  {title} {\enquote {\bibinfo {title} {A simple theory of the electron-phonon mass enhancement in transition metal compounds},}\ }\href {\doibase 10.1088/0305-4608/4/8/015} {\bibfield  {journal} {\bibinfo  {journal} {Journal of Physics F: Metal Physics}\ }\textbf {\bibinfo {volume} {4}},\ \bibinfo {pages} {1204--1221} (\bibinfo {year} {1974})}\BibitemShut {NoStop}%
\bibitem [{\citenamefont {Mazin}\ \emph {et~al.}(1990)\citenamefont {Mazin}, \citenamefont {Rashkeev},\ and\ \citenamefont {Savrasov}}]{mazin_1990}%
  \BibitemOpen
  \bibfield  {author} {\bibinfo {author} {\bibfnamefont {I.~I.}\ \bibnamefont {Mazin}}, \bibinfo {author} {\bibfnamefont {S.~N.}\ \bibnamefont {Rashkeev}}, \ and\ \bibinfo {author} {\bibfnamefont {S.~Y.}\ \bibnamefont {Savrasov}},\ }\bibfield  {title} {\enquote {\bibinfo {title} {{Nonspherical rigid-muffin-tin calculations of electron-phonon coupling in high-{T}$_c$ perovskites}},}\ }\href {\doibase 10.1103/PhysRevB.42.366} {\bibfield  {journal} {\bibinfo  {journal} {Phys. Rev. B}\ }\textbf {\bibinfo {volume} {42}},\ \bibinfo {pages} {366--370} (\bibinfo {year} {1990})}\BibitemShut {NoStop}%
\bibitem [{\citenamefont {Rajput}\ \emph {et~al.}(1996)\citenamefont {Rajput}, \citenamefont {Prasad}, \citenamefont {Singru}, \citenamefont {Kaprzyk},\ and\ \citenamefont {Bansil}}]{kaprzyk_1996}%
  \BibitemOpen
  \bibfield  {author} {\bibinfo {author} {\bibfnamefont {S.~S.}\ \bibnamefont {Rajput}}, \bibinfo {author} {\bibfnamefont {R.}~\bibnamefont {Prasad}}, \bibinfo {author} {\bibfnamefont {R.~M.}\ \bibnamefont {Singru}}, \bibinfo {author} {\bibfnamefont {S.}~\bibnamefont {Kaprzyk}}, \ and\ \bibinfo {author} {\bibfnamefont {A.}~\bibnamefont {Bansil}},\ }\bibfield  {title} {\enquote {\bibinfo {title} {{Electronic structure of disordered {Nb} - {Mo} alloys studied using the charge-self-consistent {Korringa - Kohn - Rostoker} coherent potential approximation}},}\ }\href {\doibase 10.1088/0953-8984/8/17/006} {\bibfield  {journal} {\bibinfo  {journal} {Journal of Physics: Condensed Matter}\ }\textbf {\bibinfo {volume} {8}},\ \bibinfo {pages} {2929--2944} (\bibinfo {year} {1996})}\BibitemShut {NoStop}%
\bibitem [{\citenamefont {Wiendlocha}\ \emph {et~al.}(2006)\citenamefont {Wiendlocha}, \citenamefont {Tobola},\ and\ \citenamefont {Kaprzyk}}]{wiendlocha_2006}%
  \BibitemOpen
  \bibfield  {author} {\bibinfo {author} {\bibfnamefont {B.}~\bibnamefont {Wiendlocha}}, \bibinfo {author} {\bibfnamefont {J.}~\bibnamefont {Tobola}}, \ and\ \bibinfo {author} {\bibfnamefont {S.}~\bibnamefont {Kaprzyk}},\ }\bibfield  {title} {\enquote {\bibinfo {title} {{Search for {Sc}$_3${X}{B} ({X}={In},{Tl},{Ga},{Al}) perovskites superconductors and proximity of weak ferromagnetism}},}\ }\href {\doibase 10.1103/PhysRevB.73.134522} {\bibfield  {journal} {\bibinfo  {journal} {Phys. Rev. B}\ }\textbf {\bibinfo {volume} {73}},\ \bibinfo {pages} {134522} (\bibinfo {year} {2006})}\BibitemShut {NoStop}%
\bibitem [{\citenamefont {Waseda}\ \emph {et~al.}(1975)\citenamefont {Waseda}, \citenamefont {Hirata},\ and\ \citenamefont {Ohtani}}]{ta_lattice_const}%
  \BibitemOpen
  \bibfield  {author} {\bibinfo {author} {\bibfnamefont {Y}~\bibnamefont {Waseda}}, \bibinfo {author} {\bibfnamefont {K}~\bibnamefont {Hirata}}, \ and\ \bibinfo {author} {\bibfnamefont {M}~\bibnamefont {Ohtani}},\ }\bibfield  {title} {\enquote {\bibinfo {title} {{High-temperature thermal expansion of platinum, tantalum, molybdenum, and tungsten measured by X-ray diffraction}},}\ }\href@noop {} {\bibfield  {journal} {\bibinfo  {journal} {High Temperatures-High Pressures}\ }\textbf {\bibinfo {volume} {7}},\ \bibinfo {pages} {221--226} (\bibinfo {year} {1975})}\BibitemShut {NoStop}%
\bibitem [{\citenamefont {Straumanis}\ and\ \citenamefont {Zyszczynski}(1970)}]{nb_lattice_const}%
  \BibitemOpen
  \bibfield  {author} {\bibinfo {author} {\bibfnamefont {M.~E.}\ \bibnamefont {Straumanis}}\ and\ \bibinfo {author} {\bibfnamefont {S.}~\bibnamefont {Zyszczynski}},\ }\bibfield  {title} {\enquote {\bibinfo {title} {{Lattice parameters, thermal expansion coefficients and densities of Nb, and of solid solutions Nb–O and Nb–N–O and their defect structure}},}\ }\href {\doibase 10.1107/S002188987000554X} {\bibfield  {journal} {\bibinfo  {journal} {Journal of Applied Crystallography}\ }\textbf {\bibinfo {volume} {3}},\ \bibinfo {pages} {1--6} (\bibinfo {year} {1970})}\BibitemShut {NoStop}%
\bibitem [{\citenamefont {Russell}(2004)}]{hf_lattice_constant}%
  \BibitemOpen
  \bibfield  {author} {\bibinfo {author} {\bibfnamefont {R.~B.}\ \bibnamefont {Russell}},\ }\bibfield  {title} {\enquote {\bibinfo {title} {{On the Zr‐Hf System}},}\ }\href {\doibase 10.1063/1.1721254} {\bibfield  {journal} {\bibinfo  {journal} {Journal of Applied Physics}\ }\textbf {\bibinfo {volume} {24}},\ \bibinfo {pages} {232--233} (\bibinfo {year} {2004})}\BibitemShut {NoStop}%
\bibitem [{\citenamefont {Olinger}\ and\ \citenamefont {Jamieson}(1973)}]{zr_lattice_constant}%
  \BibitemOpen
  \bibfield  {author} {\bibinfo {author} {\bibfnamefont {B}~\bibnamefont {Olinger}}\ and\ \bibinfo {author} {\bibfnamefont {JC}~\bibnamefont {Jamieson}},\ }\bibfield  {title} {\enquote {\bibinfo {title} {Zirconium},}\ }\href@noop {} {\bibfield  {journal} {\bibinfo  {journal} {High Temperatures-High Pressures}\ }\textbf {\bibinfo {volume} {5}},\ \bibinfo {pages} {123--131} (\bibinfo {year} {1973})}\BibitemShut {NoStop}%
\bibitem [{\citenamefont {Pawar}\ and\ \citenamefont {Deshpande}(1968)}]{ti_lattice_constant}%
  \BibitemOpen
  \bibfield  {author} {\bibinfo {author} {\bibfnamefont {R.~R.}\ \bibnamefont {Pawar}}\ and\ \bibinfo {author} {\bibfnamefont {V.~T.}\ \bibnamefont {Deshpande}},\ }\bibfield  {title} {\enquote {\bibinfo {title} {{The anisotropy of the thermal expansion of {$\alpha$}-titanium}},}\ }\href {\doibase 10.1107/S0567739468000525} {\bibfield  {journal} {\bibinfo  {journal} {Acta Crystallographica Section A}\ }\textbf {\bibinfo {volume} {24}},\ \bibinfo {pages} {316--317} (\bibinfo {year} {1968})}\BibitemShut {NoStop}%
\bibitem [{\citenamefont {Stepanov}\ \emph {et~al.}(2018)\citenamefont {Stepanov}, \citenamefont {Yurchenko}, \citenamefont {Zherebtsov}, \citenamefont {Tikhonovsky},\ and\ \citenamefont {Salishchev}}]{stepanov_2018}%
  \BibitemOpen
  \bibfield  {author} {\bibinfo {author} {\bibfnamefont {N.D.}\ \bibnamefont {Stepanov}}, \bibinfo {author} {\bibfnamefont {N.Yu.}\ \bibnamefont {Yurchenko}}, \bibinfo {author} {\bibfnamefont {S.V.}\ \bibnamefont {Zherebtsov}}, \bibinfo {author} {\bibfnamefont {M.A.}\ \bibnamefont {Tikhonovsky}}, \ and\ \bibinfo {author} {\bibfnamefont {G.A.}\ \bibnamefont {Salishchev}},\ }\bibfield  {title} {\enquote {\bibinfo {title} {{Aging behavior of the {HfNbTaTiZr} high entropy alloy}},}\ }\href {\doibase 10.1016/j.matlet.2017.09.094} {\bibfield  {journal} {\bibinfo  {journal} {Materials Letters}\ }\textbf {\bibinfo {volume} {211}},\ \bibinfo {pages} {87--90} (\bibinfo {year} {2018})}\BibitemShut {NoStop}%
\bibitem [{\citenamefont {Chen}\ \emph {et~al.}(2019)\citenamefont {Chen}, \citenamefont {Tong}, \citenamefont {Tseng}, \citenamefont {Yeh}, \citenamefont {Poplawsky}, \citenamefont {Wen}, \citenamefont {Gao}, \citenamefont {Kim}, \citenamefont {Chen}, \citenamefont {Ren}, \citenamefont {Feng}, \citenamefont {Li},\ and\ \citenamefont {Liaw}}]{chen_2019}%
  \BibitemOpen
  \bibfield  {author} {\bibinfo {author} {\bibfnamefont {S.Y.}\ \bibnamefont {Chen}}, \bibinfo {author} {\bibfnamefont {Y.}~\bibnamefont {Tong}}, \bibinfo {author} {\bibfnamefont {K.-K.}\ \bibnamefont {Tseng}}, \bibinfo {author} {\bibfnamefont {J.-W.}\ \bibnamefont {Yeh}}, \bibinfo {author} {\bibfnamefont {J.D.}\ \bibnamefont {Poplawsky}}, \bibinfo {author} {\bibfnamefont {J.G.}\ \bibnamefont {Wen}}, \bibinfo {author} {\bibfnamefont {M.C.}\ \bibnamefont {Gao}}, \bibinfo {author} {\bibfnamefont {G.}~\bibnamefont {Kim}}, \bibinfo {author} {\bibfnamefont {W.}~\bibnamefont {Chen}}, \bibinfo {author} {\bibfnamefont {Y.}~\bibnamefont {Ren}}, \bibinfo {author} {\bibfnamefont {R.}~\bibnamefont {Feng}}, \bibinfo {author} {\bibfnamefont {W.D.}\ \bibnamefont {Li}}, \ and\ \bibinfo {author} {\bibfnamefont {P.K.}\ \bibnamefont {Liaw}},\ }\bibfield  {title} {\enquote {\bibinfo {title} {{Phase transformations of {HfNbTaTiZr} high entropy alloy at intermediate temperatures}},}\ }\href {\doibase
  10.1016/j.scriptamat.2018.08.032} {\bibfield  {journal} {\bibinfo  {journal} {Scripta Materialia}\ }\textbf {\bibinfo {volume} {158}},\ \bibinfo {pages} {50--56} (\bibinfo {year} {2019})}\BibitemShut {NoStop}%
\bibitem [{\citenamefont {McMillan}(1968)}]{mcmillan}%
  \BibitemOpen
  \bibfield  {author} {\bibinfo {author} {\bibfnamefont {W.~L.}\ \bibnamefont {McMillan}},\ }\bibfield  {title} {\enquote {\bibinfo {title} {{Transition Temperature of Strong-Coupled Superconductors}},}\ }\href {\doibase 10.1103/PhysRev.167.331} {\bibfield  {journal} {\bibinfo  {journal} {Phys. Rev.}\ }\textbf {\bibinfo {volume} {167}},\ \bibinfo {pages} {331--344} (\bibinfo {year} {1968})}\BibitemShut {NoStop}%
\bibitem [{\citenamefont {Anderson}(1959)}]{anderson_59}%
  \BibitemOpen
  \bibfield  {author} {\bibinfo {author} {\bibfnamefont {P.W.}\ \bibnamefont {Anderson}},\ }\bibfield  {title} {\enquote {\bibinfo {title} {Theory of dirty superconductors},}\ }\href {\doibase 10.1016/0022-3697(59)90036-8} {\bibfield  {journal} {\bibinfo  {journal} {Journal of Physics and Chemistry of Solids}\ }\textbf {\bibinfo {volume} {11}},\ \bibinfo {pages} {26--30} (\bibinfo {year} {1959})}\BibitemShut {NoStop}%
\bibitem [{\citenamefont {Anderson}\ \emph {et~al.}(1983)\citenamefont {Anderson}, \citenamefont {Muttalib},\ and\ \citenamefont {Ramakrishnan}}]{anderson_degradation}%
  \BibitemOpen
  \bibfield  {author} {\bibinfo {author} {\bibfnamefont {P.~W.}\ \bibnamefont {Anderson}}, \bibinfo {author} {\bibfnamefont {K.~A.}\ \bibnamefont {Muttalib}}, \ and\ \bibinfo {author} {\bibfnamefont {T.~V.}\ \bibnamefont {Ramakrishnan}},\ }\bibfield  {title} {\enquote {\bibinfo {title} {{Theory of the "universal" degradation of ${T}_{c}$ in high-temperature superconductors}},}\ }\href {\doibase 10.1103/PhysRevB.28.117} {\bibfield  {journal} {\bibinfo  {journal} {Phys. Rev. B}\ }\textbf {\bibinfo {volume} {28}},\ \bibinfo {pages} {117--120} (\bibinfo {year} {1983})}\BibitemShut {NoStop}%
\bibitem [{\citenamefont {Fukuyama}\ \emph {et~al.}(1984)\citenamefont {Fukuyama}, \citenamefont {Ebisawa},\ and\ \citenamefont {Maekawa}}]{weakly_localized_regime}%
  \BibitemOpen
  \bibfield  {author} {\bibinfo {author} {\bibfnamefont {Hidetoshi}\ \bibnamefont {Fukuyama}}, \bibinfo {author} {\bibfnamefont {Hiromichi}\ \bibnamefont {Ebisawa}}, \ and\ \bibinfo {author} {\bibfnamefont {Sadamichi}\ \bibnamefont {Maekawa}},\ }\bibfield  {title} {\enquote {\bibinfo {title} {{Bulk Superconductivity in Weakly Localized Regime}},}\ }\href {\doibase 10.1143/JPSJ.53.3560} {\bibfield  {journal} {\bibinfo  {journal} {Journal of the Physical Society of Japan}\ }\textbf {\bibinfo {volume} {53}},\ \bibinfo {pages} {3560--3567} (\bibinfo {year} {1984})}\BibitemShut {NoStop}%
\bibitem [{\citenamefont {Strongin}\ \emph {et~al.}(1970)\citenamefont {Strongin}, \citenamefont {Thompson}, \citenamefont {Kammerer},\ and\ \citenamefont {Crow}}]{destruction_films}%
  \BibitemOpen
  \bibfield  {author} {\bibinfo {author} {\bibfnamefont {Myron}\ \bibnamefont {Strongin}}, \bibinfo {author} {\bibfnamefont {R.~S.}\ \bibnamefont {Thompson}}, \bibinfo {author} {\bibfnamefont {O.~F.}\ \bibnamefont {Kammerer}}, \ and\ \bibinfo {author} {\bibfnamefont {J.~E.}\ \bibnamefont {Crow}},\ }\bibfield  {title} {\enquote {\bibinfo {title} {{Destruction of Superconductivity in Disordered Near-Monolayer Films}},}\ }\href {\doibase 10.1103/PhysRevB.1.1078} {\bibfield  {journal} {\bibinfo  {journal} {Phys. Rev. B}\ }\textbf {\bibinfo {volume} {1}},\ \bibinfo {pages} {1078--1091} (\bibinfo {year} {1970})}\BibitemShut {NoStop}%
\bibitem [{\citenamefont {Imry}\ and\ \citenamefont {Strongin}(1981)}]{destruction_granular}%
  \BibitemOpen
  \bibfield  {author} {\bibinfo {author} {\bibfnamefont {Yoseph}\ \bibnamefont {Imry}}\ and\ \bibinfo {author} {\bibfnamefont {Myron}\ \bibnamefont {Strongin}},\ }\bibfield  {title} {\enquote {\bibinfo {title} {Destruction of superconductivity in granular and highly disordered metals},}\ }\href {\doibase 10.1103/PhysRevB.24.6353} {\bibfield  {journal} {\bibinfo  {journal} {Phys. Rev. B}\ }\textbf {\bibinfo {volume} {24}},\ \bibinfo {pages} {6353--6360} (\bibinfo {year} {1981})}\BibitemShut {NoStop}%
\bibitem [{\citenamefont {Mishonov}\ \emph {et~al.}(2003)\citenamefont {Mishonov}, \citenamefont {Penev}, \citenamefont {Indekeu},\ and\ \citenamefont {Pokrovsky}}]{heat-prb}%
  \BibitemOpen
  \bibfield  {author} {\bibinfo {author} {\bibfnamefont {Todor~M.}\ \bibnamefont {Mishonov}}, \bibinfo {author} {\bibfnamefont {Evgeni~S.}\ \bibnamefont {Penev}}, \bibinfo {author} {\bibfnamefont {Joseph~O.}\ \bibnamefont {Indekeu}}, \ and\ \bibinfo {author} {\bibfnamefont {Valery~L.}\ \bibnamefont {Pokrovsky}},\ }\bibfield  {title} {\enquote {\bibinfo {title} {Specific-heat discontinuity in impure two-band superconductors},}\ }\href {\doibase 10.1103/PhysRevB.68.104517} {\bibfield  {journal} {\bibinfo  {journal} {Phys. Rev. B}\ }\textbf {\bibinfo {volume} {68}},\ \bibinfo {pages} {104517} (\bibinfo {year} {2003})}\BibitemShut {NoStop}%
\bibitem [{\citenamefont {Kasem}\ \emph {et~al.}(2021)\citenamefont {Kasem}, \citenamefont {Yamashita}, \citenamefont {Hatano}, \citenamefont {Sakurai}, \citenamefont {Oono-Hori}, \citenamefont {Goto}, \citenamefont {Miura},\ and\ \citenamefont {Mizuguchi}}]{Kasem_2021}%
  \BibitemOpen
  \bibfield  {author} {\bibinfo {author} {\bibfnamefont {Md~Riad}\ \bibnamefont {Kasem}}, \bibinfo {author} {\bibfnamefont {Aichi}\ \bibnamefont {Yamashita}}, \bibinfo {author} {\bibfnamefont {Taishi}\ \bibnamefont {Hatano}}, \bibinfo {author} {\bibfnamefont {Kota}\ \bibnamefont {Sakurai}}, \bibinfo {author} {\bibfnamefont {Naoko}\ \bibnamefont {Oono-Hori}}, \bibinfo {author} {\bibfnamefont {Yosuke}\ \bibnamefont {Goto}}, \bibinfo {author} {\bibfnamefont {Osuke}\ \bibnamefont {Miura}}, \ and\ \bibinfo {author} {\bibfnamefont {Yoshikazu}\ \bibnamefont {Mizuguchi}},\ }\bibfield  {title} {\enquote {\bibinfo {title} {{Anomalous broadening of specific heat jump at Tc in high-entropy-alloy-type superconductor TrZr$_2$}},}\ }\href {\doibase 10.1088/1361-6668/ac2554} {\bibfield  {journal} {\bibinfo  {journal} {Superconductor Science and Technology}\ }\textbf {\bibinfo {volume} {34}},\ \bibinfo {pages} {125001} (\bibinfo {year} {2021})}\BibitemShut {NoStop}%
\end{thebibliography}%

\newpage

\onecolumngrid

\section*{Supplemental Material}

\renewcommand{\thefigure}{{S\arabic{figure}}}
\setcounter{figure} 0

\vspace*{24pt}
\noindent
Supplemental Material contains:\\ \\
Table I with the atomic positions of all atoms in all 12 undistorted supercells.\\
Figures S1-S12 with the total and partial densities of states of the 12 supercells before and after relaxation and projections of the distorted supercells on the $xy$ plane.

\begin{table}[htb]
\begin{center}
\begin{tabular}{|c||c|c|c|c|c|c|c|c|c|c|c|c|c}
\hline
site & model 1 & model 2 & model 3 & model 4 & model 5 & model 6 & model 7 & model 8 & model 9 & model 10 & model 11 & model 12  \\
\hline
(0.0; 0.0; 0.0)	& Ta(1)	& Ta(1)	& Ta(1)	& Nb(1) & Nb(1) & Nb(1) & Nb(1) & Nb(1) & Ta(1) & Ti(1) & Ti(1) & Ti(1) \\ 	
(0.0; 1/3; 0.0)	& Ta(2)	& Ta(2)	& Ta(2)	& Nb(2) & Nb(2) & Nb(2) & Nb(2) & Nb(2) & Ta(2) & Ti(2) & Ta(1) & Ta(1) \\
(2/3; 1/3; 0.0)	& Ta(3)	& Ta(3)	& Ta(3) & Nb(3) & Nb(3) & Nb(3) & Nb(3) & Nb(3) & Ta(3) & Zr(2) & Zr(2) & Zr(2) \\
(0.0; 2/3; 0.0)	& Ta(4)	& Ta(4)	& Ta(4) & Ta(1) & Ta(1) & Ta(1) & Ta(1) & Ta(1) & Hf(1) & Zr(1) & Zr(1) & Hf(2)\\	
(1/3; 2/3; 0.0)	& Ta(5)	& Ta(5)	& Ta(5) & Ta(2) & Ta(2) & Ta(2) & Ta(2) & Ta(2) & Hf(2) & Nb(1) & Nb(1) & Nb(1)\\
(2/3; 2/3; 0.0)	& Ta(6)	& Ta(6)	& Ta(6) & Ta(3) & Ta(3) & Ta(3) & Ta(3) & Ta(3) & Zr(1) & Nb(4) & Nb(4) & Nb(4) \\    
(1/6; 1/6; 0.5)	& Nb(1)	& Nb(1)	& Nb(1) & Nb(4) & Nb(4) & Nb(4) & Nb(4) & Nb(4) & Nb(1) & Nb(2) & Nb(2) & Nb(2) \\
(0.5; 1/6; 0.5)	& Nb(2)	& Nb(2)	& Nb(2) & Nb(5) & Nb(5) & Nb(5) & Nb(5) & Nb(5) & Nb(2) & Nb(3) & Nb(3) & Nb(3) \\
(0.5; 5/6; 0.5)	& Nb(3)	& Nb(3)	& Nb(3) & Nb(6) & Nb(6) & Nb(6) & Nb(6) & Nb(6) & Nb(3) & Ta(4) & Ta(4) & Ta(4) \\
(5/6; 1/6; 0.5)	& Nb(4) & Nb(4)	& Nb(4) & Ta(4) & Ta(4) & Ta(4) & Ta(4) & Ta(4) & Zr(2) & Ta(1) & Hf(2) & Zr(1)  \\
(5/6; 0.5; 0.5)	& Nb(5)	& Nb(5)	& Nb(5) & Ta(5) & Ta(5) & Ta(5) & Ta(5) & Ta(5) & Ti(1) & Ta(5) & Ta(5) & Ta(5) \\
(5/6; 5/6; 0.5)	& Nb(6) & Nb(6)	& Nb(6) & Ta(6) & Ta(6) & Ta(6) & Ta(6) & Ta(6) & Ti(2) & Ta(2) & Ta(2) & Ta(2)\\
(1/3; 0.0; 0.0)	& Hf(1)	& Zr(1)	& Hf(1) & Zr(1)	& Hf(1) & Hf(1) & Hf(1)	& Hf(1) & Ta(4) & Hf(1) & Hf(1) & Hf(1) \\
(1/3; 1/3; 0.0)	& Hf(2) & Zr(2)	& Zr(1)	& Zr(2) & Hf(2) & Zr(1) & Zr(1) & Zr(1) & Ta(5) & Ta(6) & Ta(6) & Ta(6)\\
(1/6; 0.5; 0.5)	& Zr(1)	& Hf(1) & Zr(2)	& Hf(1)	& Zr(1) & Zr(2) & Ti(1) & Zr(2) & Ta(6) & Nb(6) & Nb(6) & Nb(6)\\
(0.5; 0.5; 0.5)	& Zr(2)	& Hf(2) & Hf(2)	& Hf(2) & Zr(2) & Hf(2) & Hf(2) & Ti(1) & Nb(4) & Nb(5) & Nb(5) & Nb(5)\\
(2/3; 0.0; 0.0)	& Ti(1)	& Ti(1)	& Ti(1)	& Ti(1)	& Ti(1) & Ti(1) & Zr(2) & Hf(2) & Nb(5) & Ta(3) & Ta(3) & Ta(3)\\
(1/6; 5/6; 0.5)	& Ti(2)	& Ti(2)	& Ti(2) & Ti(2) & Ti(2) & Ti(2) & Ti(2) & Ti(2) & Nb(6) & Hf(2) & Ti(2) & Ti(2)\\
\hline
\end{tabular}
\caption{\label{tab:sc_331_coordinates}Positions of atoms in each supercell model.  }
\end{center}
\end{table}

\begin{figure}[h!]
\centering
	\includegraphics[width=\textwidth]{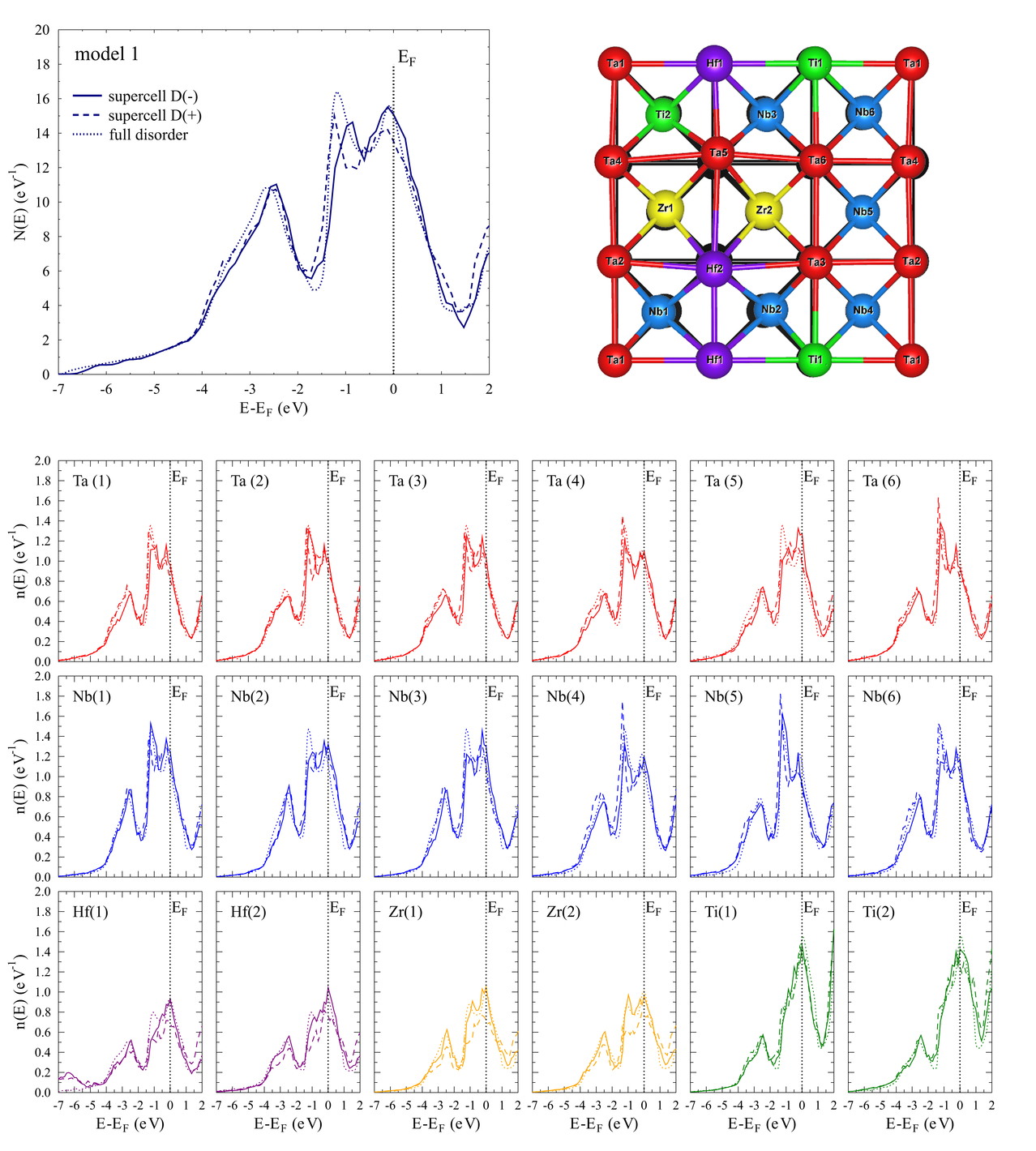}
	\caption{Model 1. Top left: total electronic densities of states, obtained for the supercell without distortions, labeled as D(-); the same supercell with distortions, labeled as  D(+); and KKR-CPA result for the full disorder with random site occupations, labeled as full disorder. Top right: projection of the distorted supercell on the $xy$ plane. In the background, the undistorted positions of the atoms are marked in black. Three bottom rows: atomic densities of states plotted in the same way as the total DOS in the top-left panel.}
	\label{fig_supp_es_1}
\end{figure}

\begin{figure}[h!]
\centering
	\includegraphics[width=\textwidth]{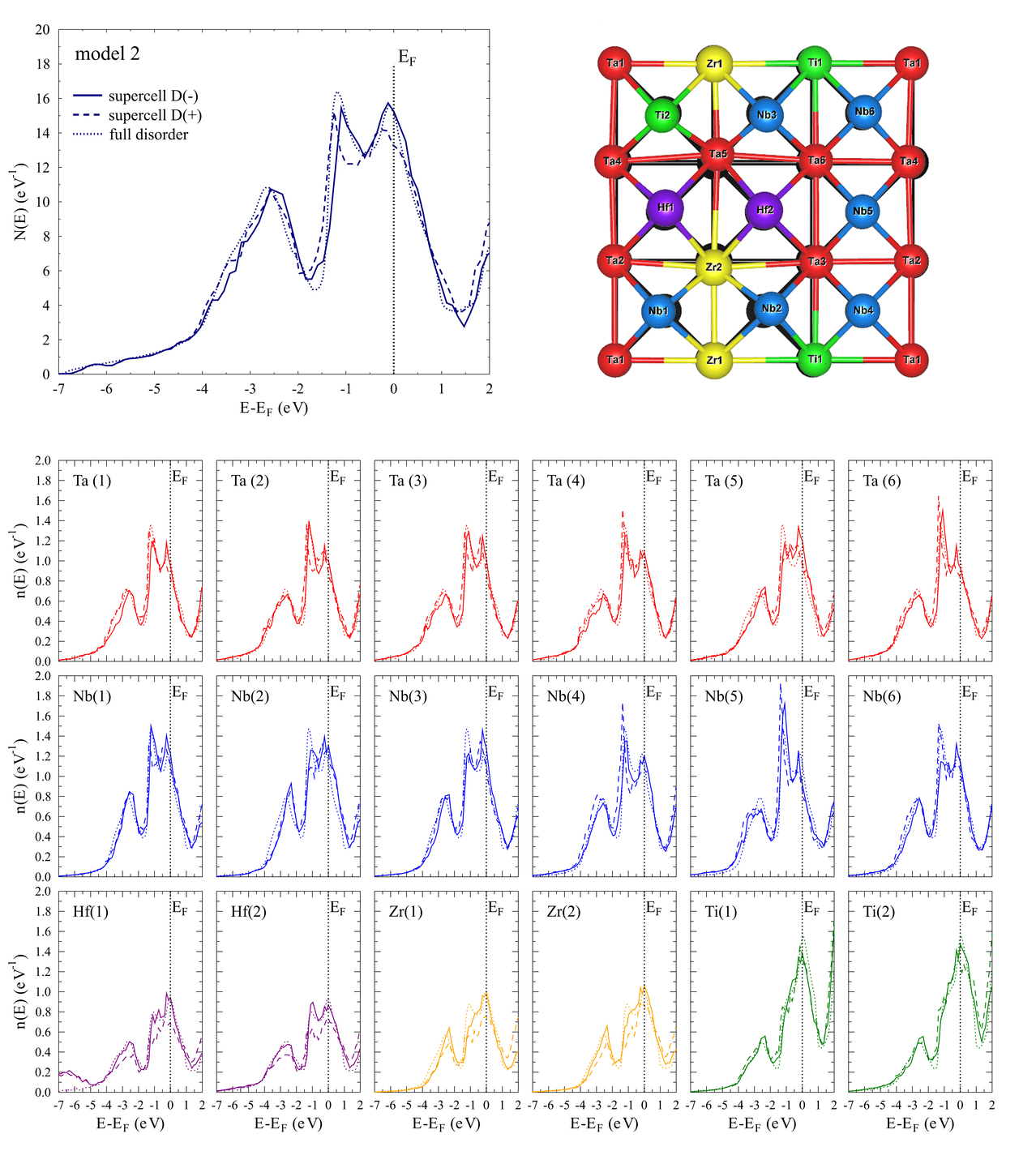}

	\caption{Model 2. Top left: total electronic densities of states, obtained for the supercell without distortions, labeled as D(-); the same supercell with distortions, labeled as  D(+); and KKR-CPA result for the full disorder with random site occupations, labeled as full disorder. Top right: projection of the distorted supercell on the $xy$ plane. In the background, the undistorted positions of the atoms are marked in black. Three bottom rows: atomic densities of states plotted in the same way as the total DOS in the top-left panel.}
	\label{fig_supp_es_2}
\end{figure}

\begin{figure}[h!]
\centering
	\includegraphics[width=\textwidth]{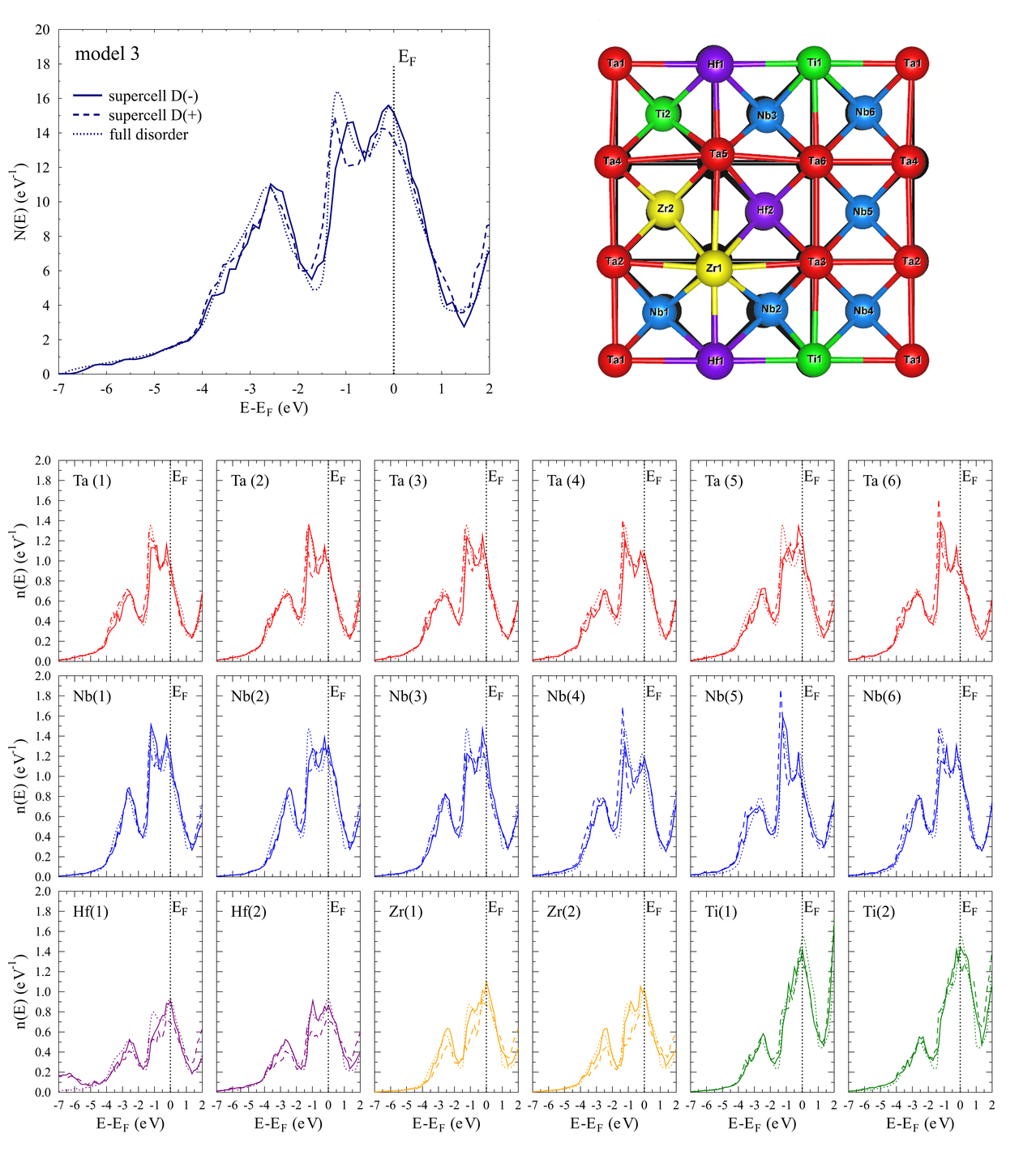}

	\caption{Model 3. Top left: total electronic densities of states, obtained for the supercell without distortions, labeled as D(-); the same supercell with distortions, labeled as  D(+); and KKR-CPA result for the full disorder with random site occupations, labeled as full disorder. Top right: projection of the distorted supercell on the $xy$ plane. In the background, the undistorted positions of the atoms are marked in black. Three bottom rows: atomic densities of states plotted in the same way as the total DOS in the top-left panel.}
	\label{fig_supp_es_3}
\end{figure}

\begin{figure}[h!]
\centering
	\includegraphics[width=\textwidth]{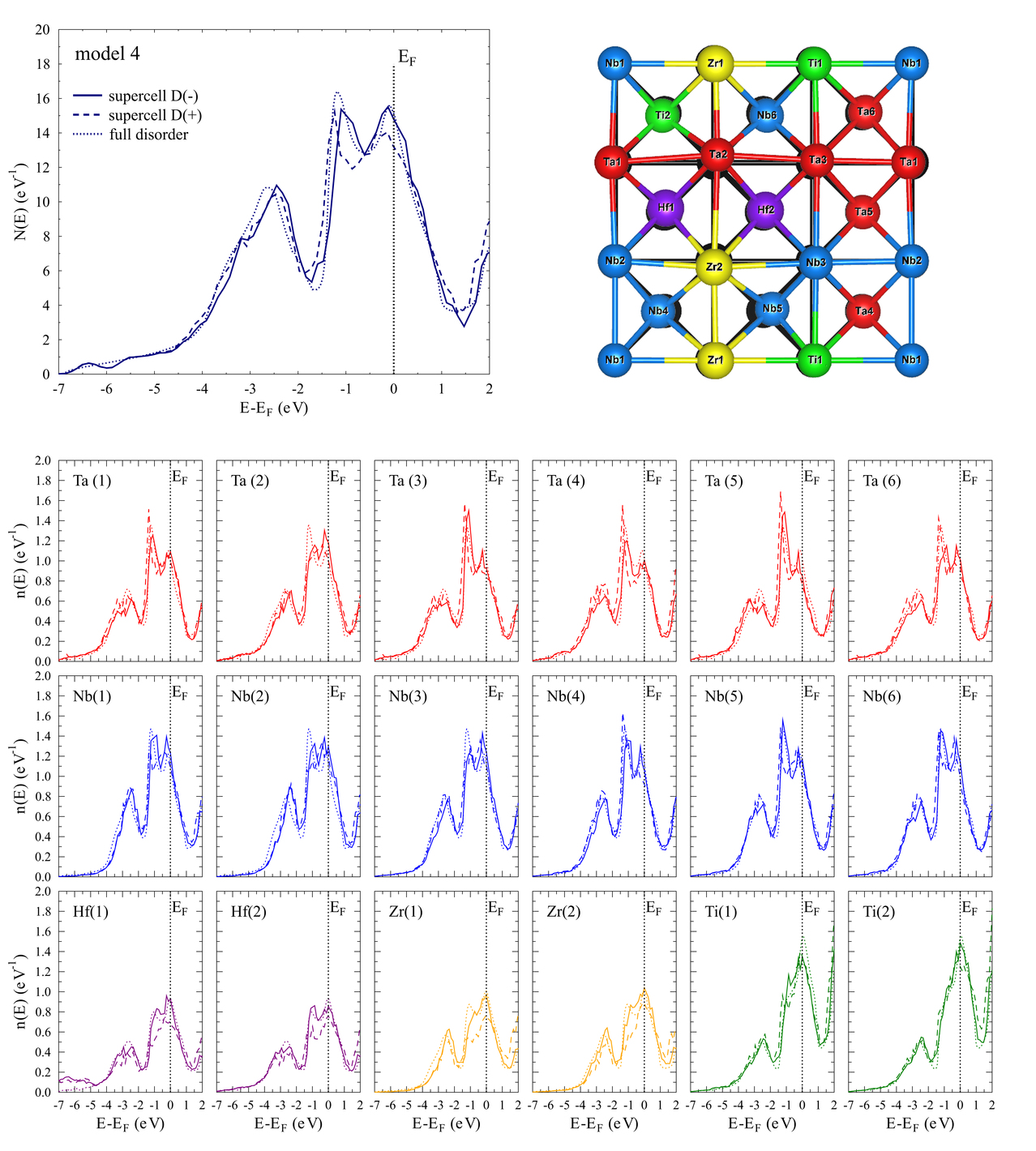}

	\caption{Model 4. Top left: total electronic densities of states, obtained for the supercell without distortions, labeled as D(-); the same supercell with distortions, labeled as  D(+); and KKR-CPA result for the full disorder with random site occupations, labeled as full disorder. Top right: projection of the distorted supercell on the $xy$ plane. In the background, the undistorted positions of the atoms are marked in black. Three bottom rows: atomic densities of states plotted in the same way as the total DOS in the top-left panel.}
	\label{fig_supp_es_4}
\end{figure}

\begin{figure}[h!]
\centering
	\includegraphics[width=\textwidth]{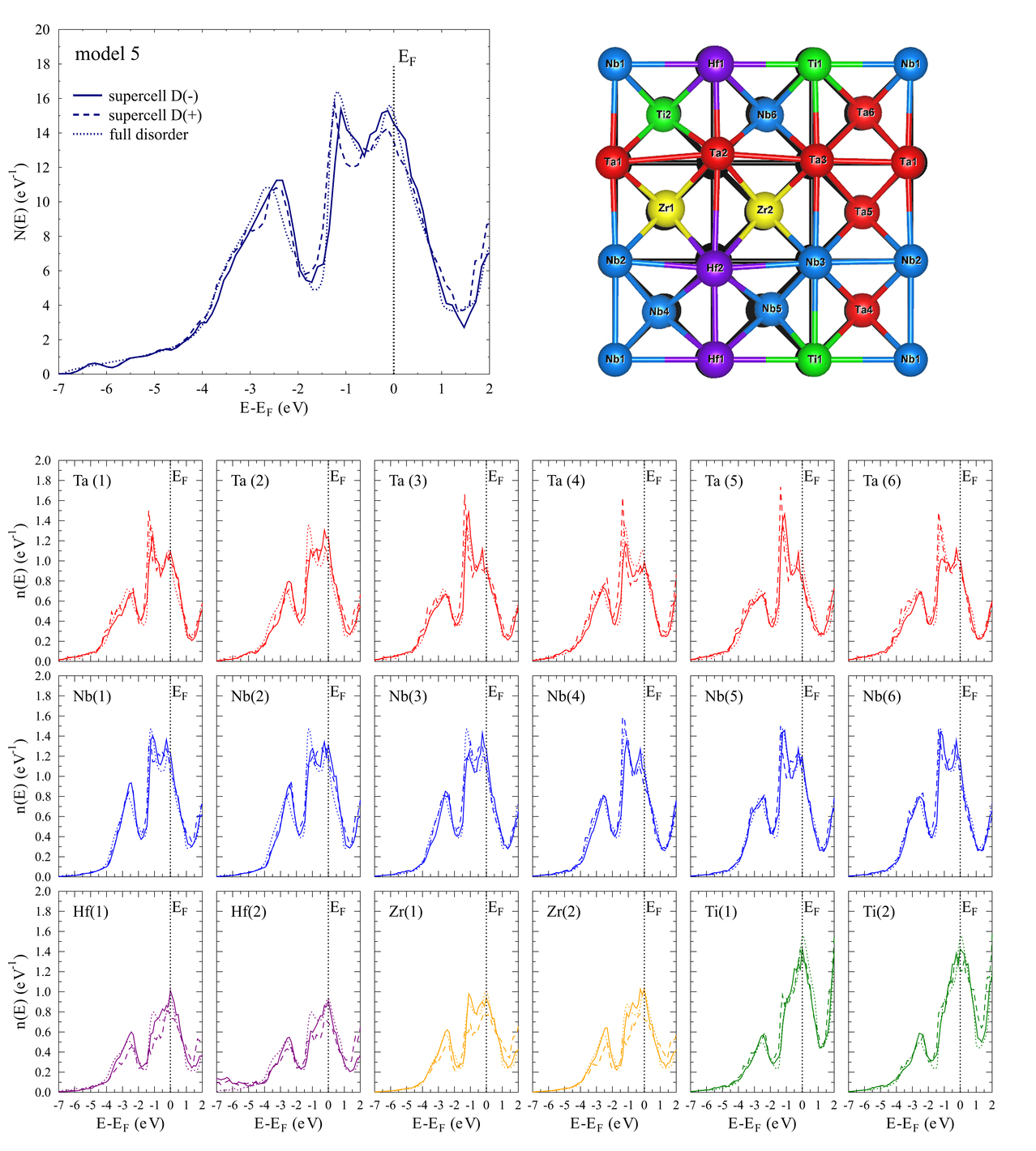}

	\caption{Model 5. Top left: total electronic densities of states, obtained for the supercell without distortions, labeled as D(-); the same supercell with distortions, labeled as  D(+); and KKR-CPA result for the full disorder with random site occupations, labeled as full disorder. Top right: projection of the distorted supercell on the $xy$ plane. In the background, the undistorted positions of the atoms are marked in black. Three bottom rows: atomic densities of states plotted in the same way as the total DOS in the top-left panel.}
	\label{fig_supp_es_5}
\end{figure}

\begin{figure}[h!]
\centering
		\includegraphics[width=\textwidth]{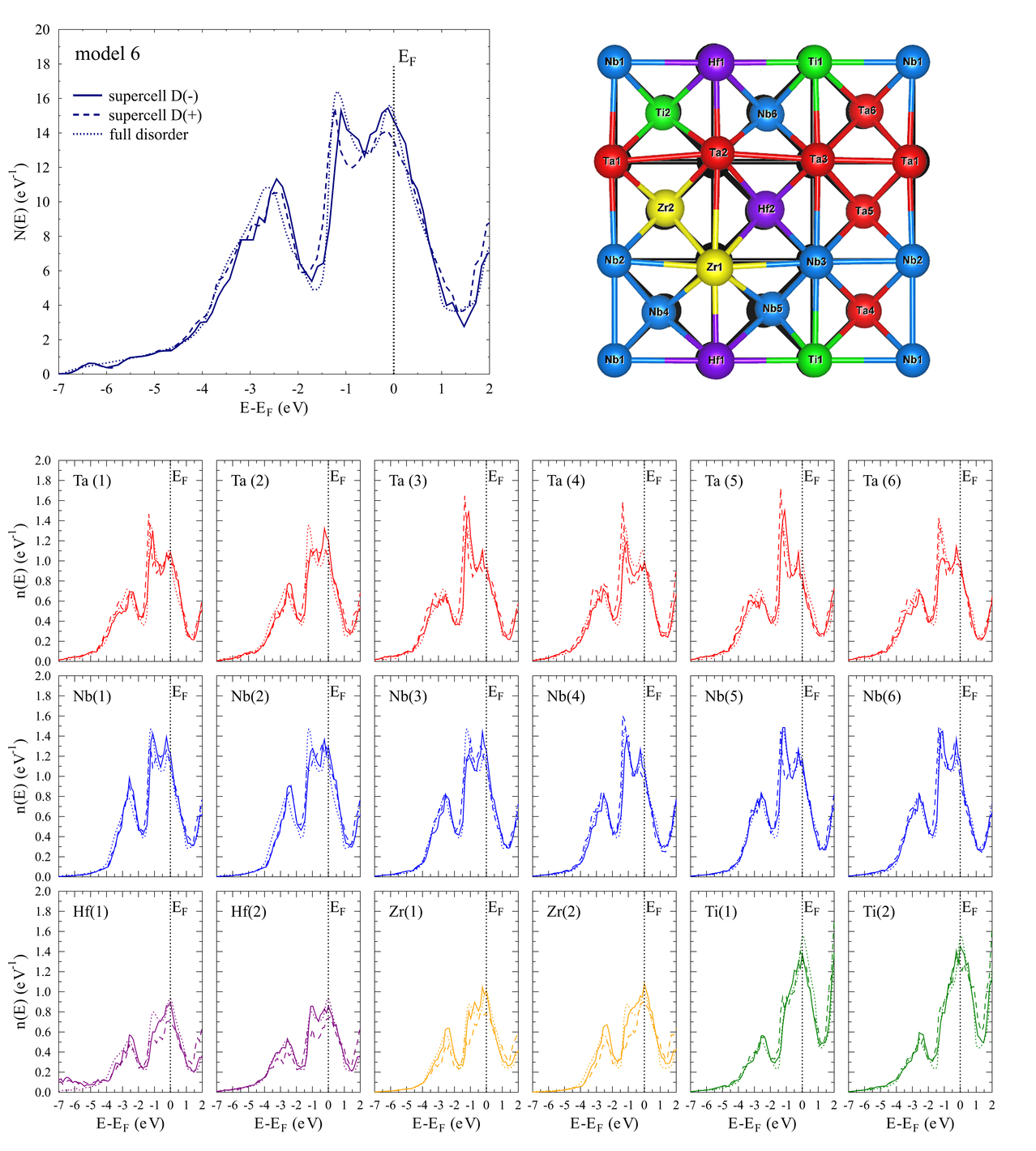}

	\caption{Model 6. Top left: total electronic densities of states, obtained for the supercell without distortions, labeled as D(-); the same supercell with distortions, labeled as  D(+); and KKR-CPA result for the full disorder with random site occupations, labeled as full disorder. Top right: projection of the distorted supercell on the $xy$ plane. In the background, the undistorted positions of the atoms are marked in black. Three bottom rows: atomic densities of states plotted in the same way as the total DOS in the top-left panel.}
	\label{fig_supp_es_6}
\end{figure}

\begin{figure}[h!]
\centering
	\includegraphics[width=\textwidth]{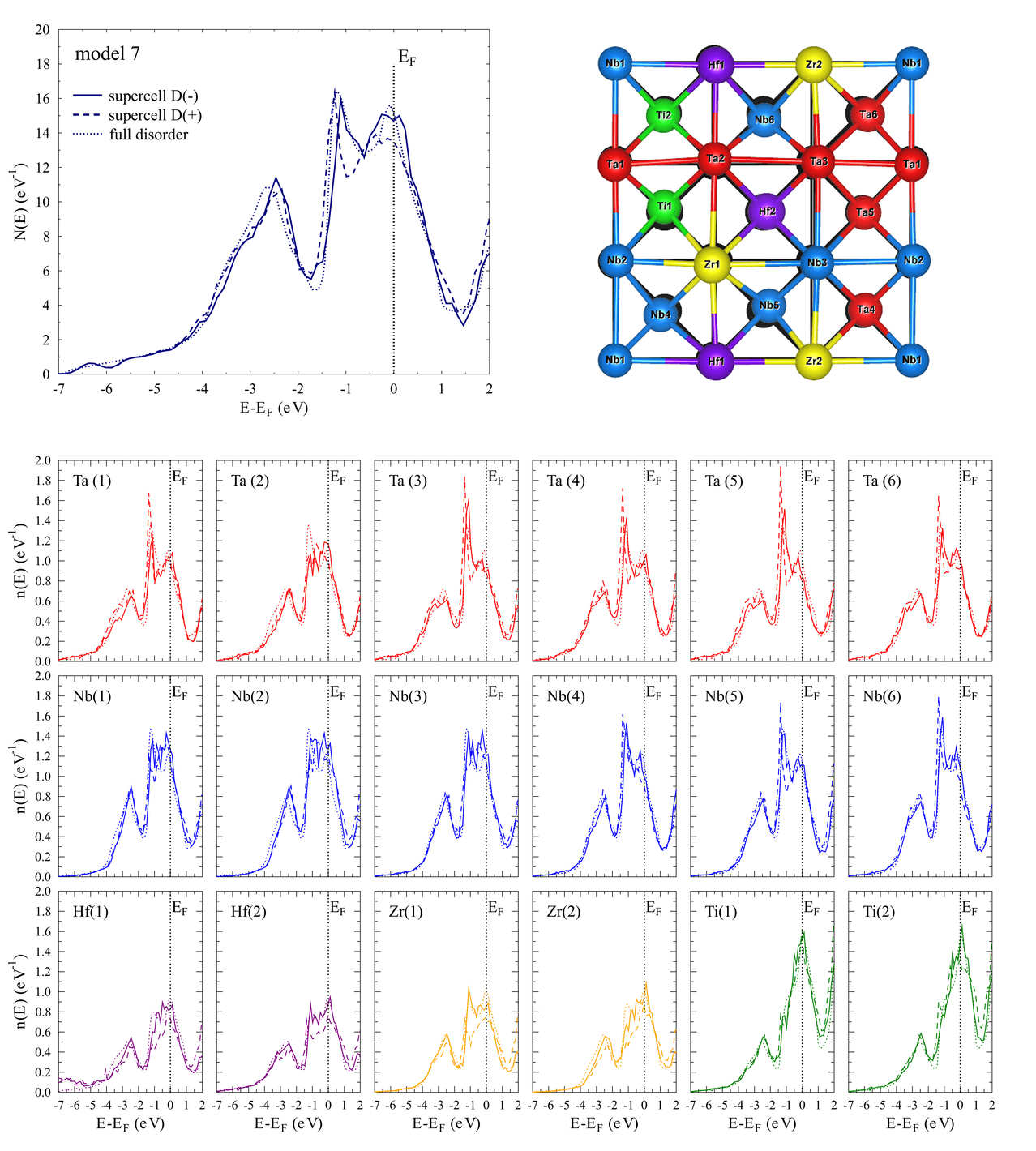}

	\caption{Model 7. Top left: total electronic densities of states, obtained for the supercell without distortions, labeled as D(-); the same supercell with distortions, labeled as  D(+); and KKR-CPA result for the full disorder with random site occupations, labeled as full disorder. Top right: projection of the distorted supercell on the $xy$ plane. In the background, the undistorted positions of the atoms are marked in black. Three bottom rows: atomic densities of states plotted in the same way as the total DOS in the top-left panel.}
	\label{fig_supp_es_7}
\end{figure}

\begin{figure}[h!]
\centering
	\includegraphics[width=\textwidth]{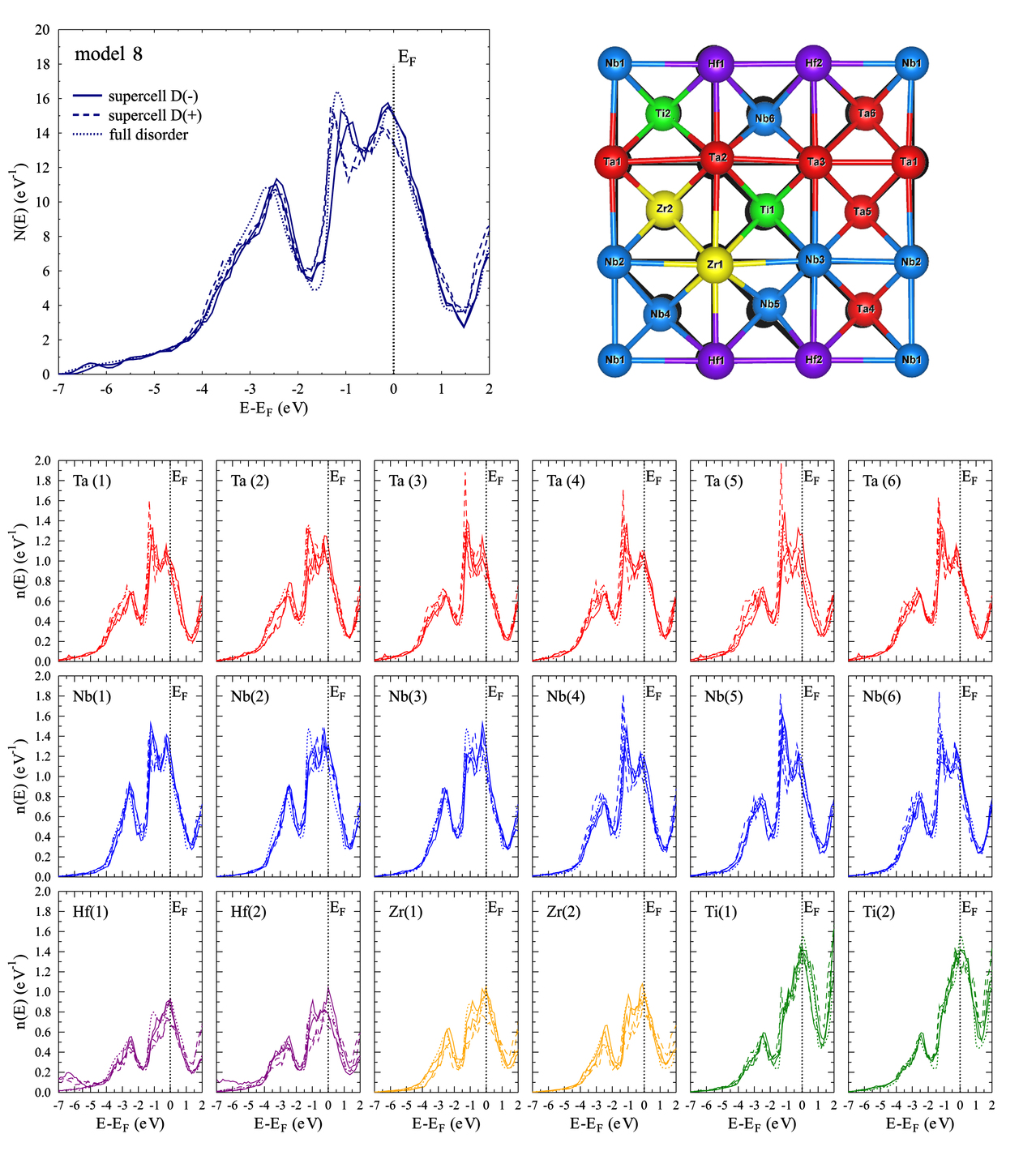}

	\caption{Model 8. Top left: total electronic densities of states, obtained for the supercell without distortions, labeled as D(-); the same supercell with distortions, labeled as  D(+); and KKR-CPA result for the full disorder with random site occupations, labeled as full disorder. Top right: projection of the distorted supercell on the $xy$ plane. In the background, the undistorted positions of the atoms are marked in black. Three bottom rows: atomic densities of states plotted in the same way as the total DOS in the top-left panel.}
	\label{fig_supp_es_8}
\end{figure}

\begin{figure}[h!]
\centering
	\includegraphics[width=\textwidth]{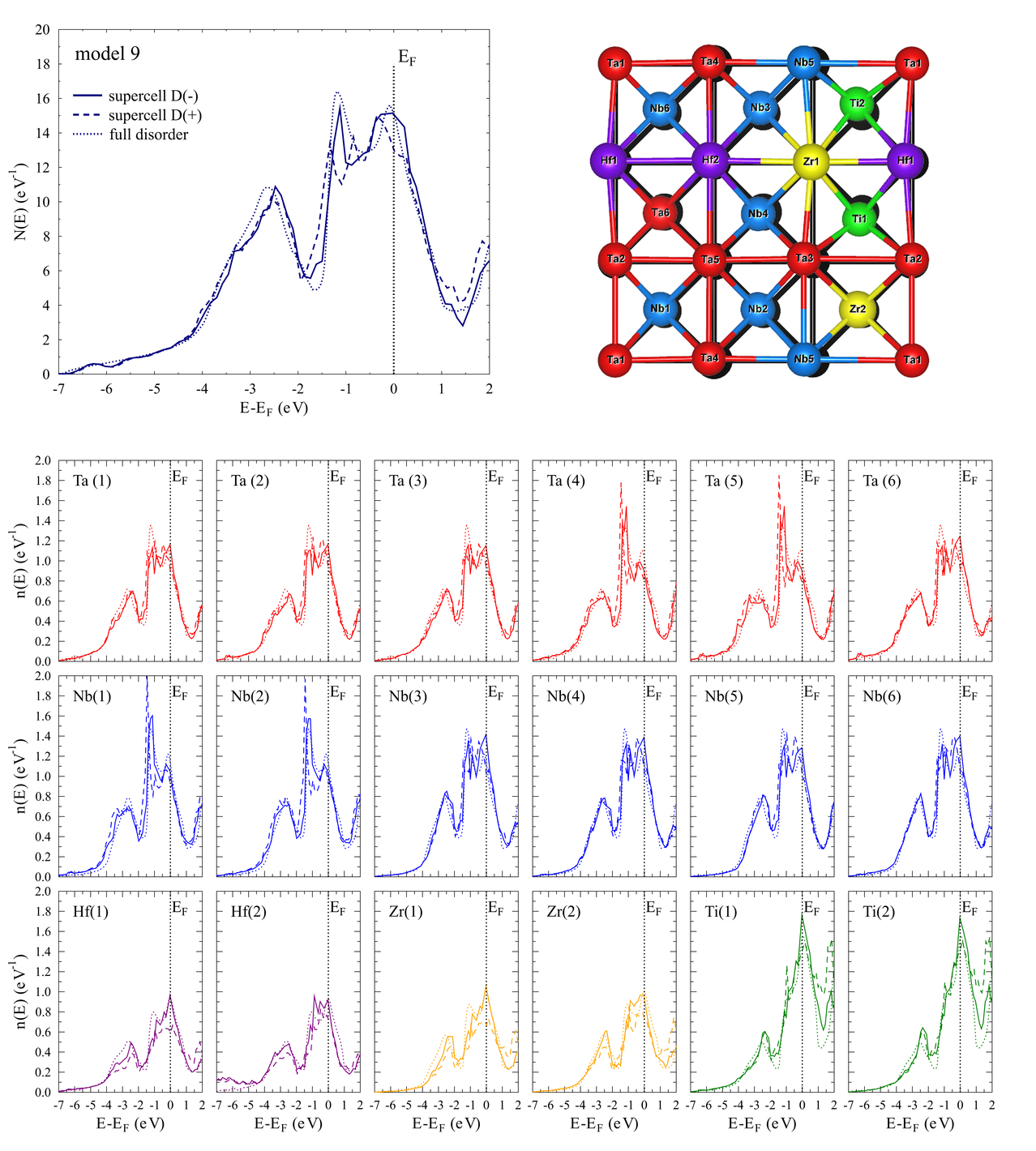}

	\caption{Model 9. Top left: total electronic densities of states, obtained for the supercell without distortions, labeled as D(-); the same supercell with distortions, labeled as  D(+); and KKR-CPA result for the full disorder with random site occupations, labeled as full disorder. Top right: projection of the distorted supercell on the $xy$ plane. In the background, the undistorted positions of the atoms are marked in black. Three bottom rows: atomic densities of states plotted in the same way as the total DOS in the top-left panel.}
	\label{fig_supp_es_9}
\end{figure}

\begin{figure}[h!]
\centering
	\includegraphics[width=\textwidth]{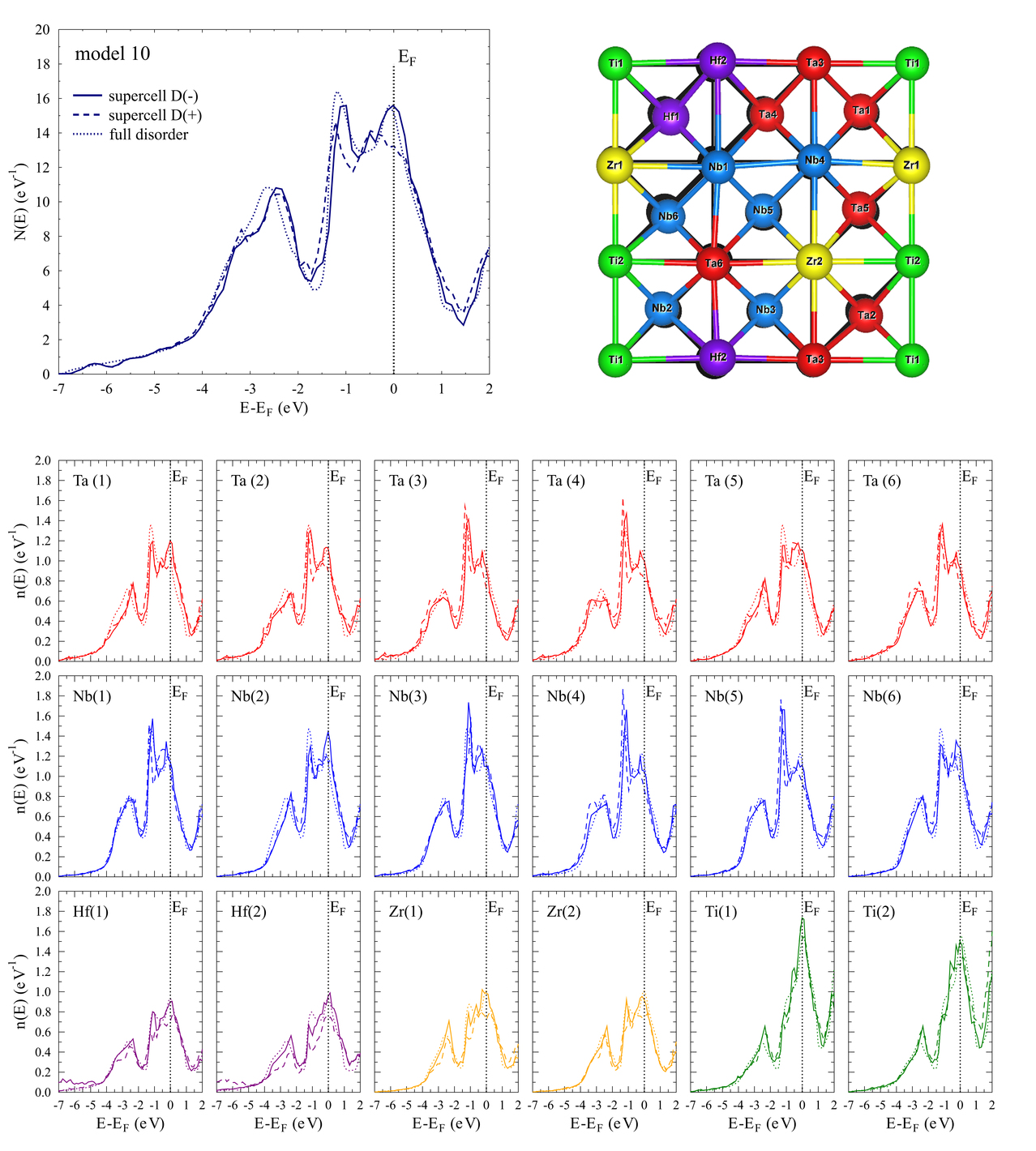}

	\caption{Model 10. Top left: total electronic densities of states, obtained for the supercell without distortions, labeled as D(-); the same supercell with distortions, labeled as  D(+); and KKR-CPA result for the full disorder with random site occupations, labeled as full disorder. Top right: projection of the distorted supercell on the $xy$ plane. In the background, the undistorted positions of the atoms are marked in black. Three bottom rows: atomic densities of states plotted in the same way as the total DOS in the top-left panel.}
	\label{fig_supp_es_10}
\end{figure}

\begin{figure}[h!]
\centering
	\includegraphics[width=\textwidth]{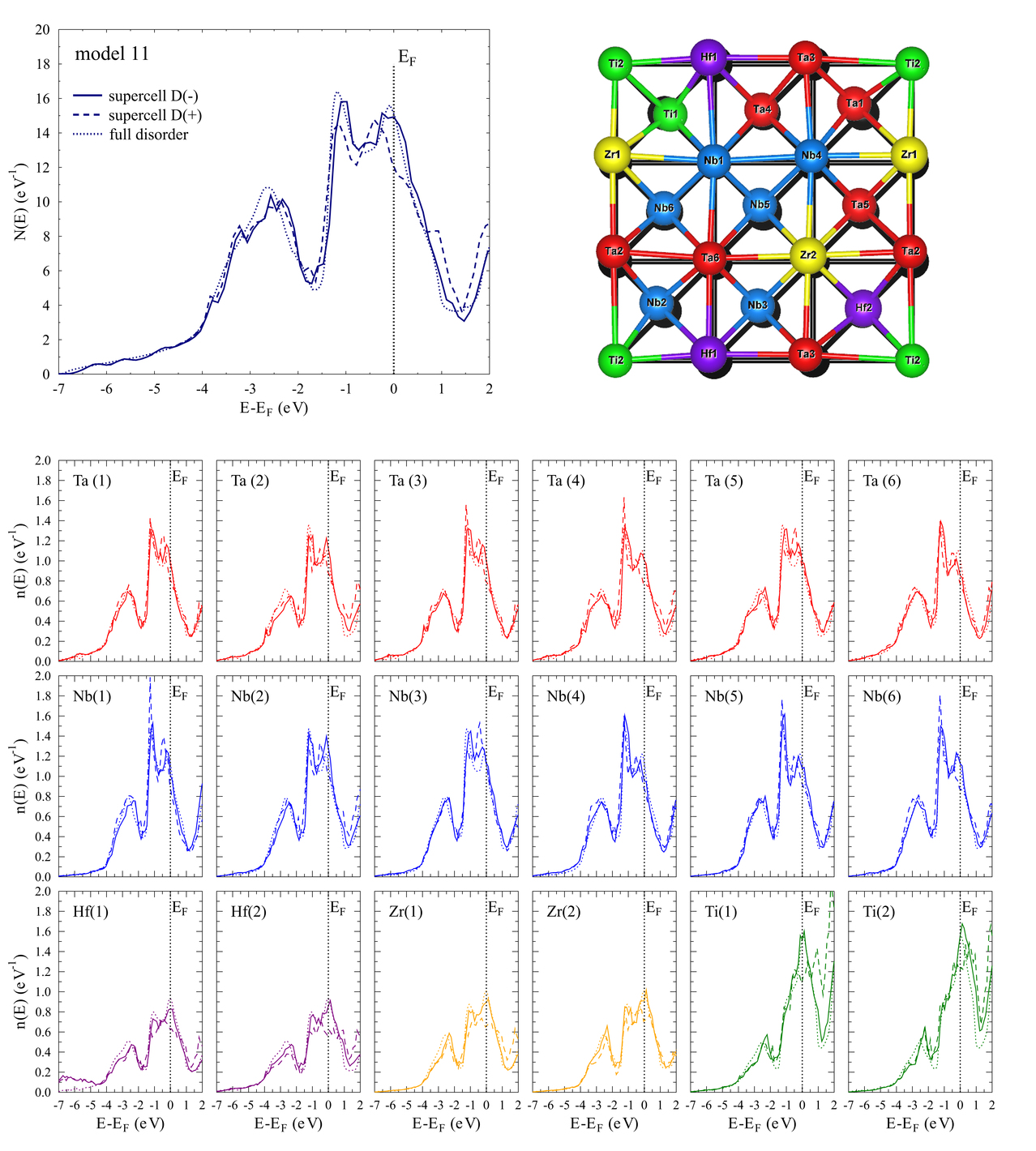}

	\caption{Model 11. Top left: total electronic densities of states, obtained for the supercell without distortions, labeled as D(-); the same supercell with distortions, labeled as  D(+); and KKR-CPA result for the full disorder with random site occupations, labeled as full disorder. Top right: projection of the distorted supercell on the $xy$ plane. In the background, the undistorted positions of the atoms are marked in black. Three bottom rows: atomic densities of states plotted in the same way as the total DOS in the top-left panel.}
	\label{fig_supp_es_11}
\end{figure}

\begin{figure}[h!]
\centering
		\includegraphics[width=\textwidth]{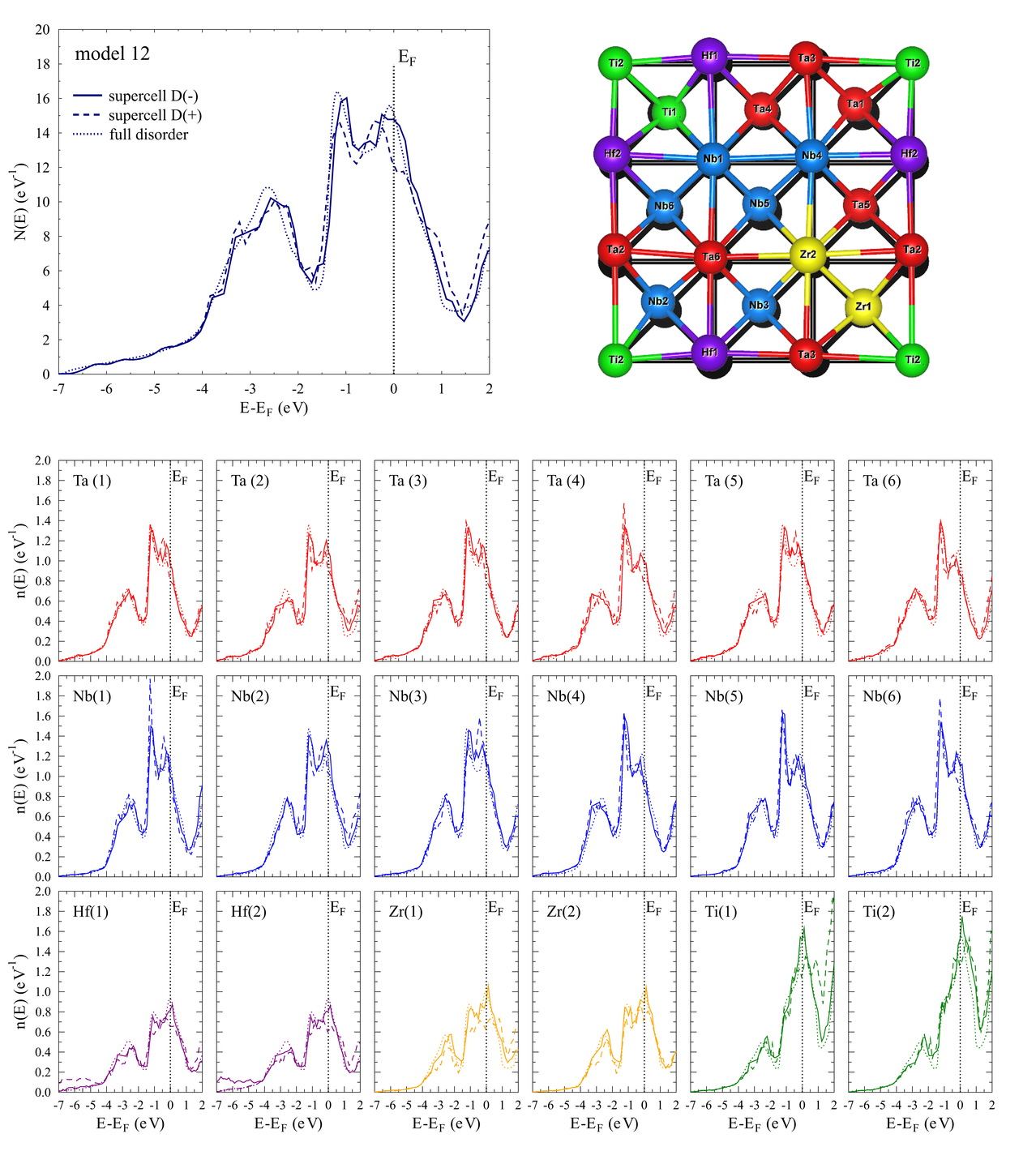}

	\caption{Model 12. Top left: total electronic densities of states, obtained for the supercell without distortions, labeled as D(-); the same supercell with distortions, labeled as  D(+); and KKR-CPA result for the full disorder with random site occupations, labeled as full disorder. Top right: projection of the distorted supercell on the $xy$ plane. In the background, the undistorted positions of the atoms are marked in black. Three bottom rows: atomic densities of states plotted in the same way as the total DOS in the top-left panel.}
	\label{fig_supp_es_12}
\end{figure}

\end{document}